\documentclass[11pt,a4paper]{article}

\usepackage[cp1251]{inputenc}

\usepackage[russian]{babel}
\usepackage{amsmath}
\usepackage{amssymb,amsfonts,amsmath,mathtext}
\usepackage{cite,enumerate,float,indentfirst}
\usepackage{graphicx}
\usepackage{subfig}
\usepackage{float}
\usepackage{xspace}
\usepackage{url}
\usepackage{xcolor}
\usepackage[normalem]{ulem}

\numberwithin{equation}{section}

\newcommand{\beq}{\begin{equation}}
\newcommand{\enq}{\end{equation}}
\newcommand{\be}{\begin{eqnarray}}
\newcommand{\ee}{\end{eqnarray}}
\newcommand{\rar}{\rightarrow}

\newcommand{\pd}[1]{\partial #1}

\newcommand{\hmpc}{\,h^{-1}\;{\rm Mpc}}

\def\Om{\mbox{$\Omega_{\rm M}$}}
\def\OL{\mbox{$\Omega_{\Lambda}$}}

\def\nifsx{${}^{56}$Ni\xspace}
\def\cofsx{${}^{56}$Co\xspace}
\def\fefsx{${}^{56}$Fe\xspace}

\let\jnlstyle=\rm
\def\refjnl#1{{\jnlstyle#1}}

\def\aj{\refjnl{Astron.~J.}}
\def\apj{\refjnl{Astroph.~J.}}
\def\apjl{\refjnl{Astroph.~J.~Lett.}}
\def\apjs{\refjnl{Astroph.~J.~Supp.~Ser.}}

\def\aap{\refjnl{Astron. Ap.}}

\def\azh{\refjnl{Astron.~Zh.}}

\def\sovast{\refjnl{Sov.~Ast.}}

\def\mnras{\refjnl{MNRAS}}

\def\prd{\refjnl{Phys.~Rev.~D}}

\def\rmp{\refjnl{Rev.~Mod.~Phys.}}
\def\pasp{\refjnl{Publ.~Astr.~Soc.~Pacific}}

\def\nat{\refjnl{Nature}}

\def\physrep{\refjnl{Phys.~Rep.}}

\newcommand{\bfomegamlcdm}{$0.295 \pm 0.034$\xspace}
\newcommand{\bfw}{$-1.018 \pm 0.057$\xspace}
\newcommand{\bfwall}{$-1.027 \pm 0.055$\xspace}
\newcommand{\LCDM}{\ensuremath{\Lambda}CDM\xspace}
\newcommand{\hoall}{$68.50 \pm 1.27$\xspace}

\begin{document}

\title{\bf  осмологическое ускорение}
\author{—.». Ѕлинников$^{a,b,c,d}$, ј.ƒ. ƒолгов$^{a,b}$}
\maketitle
\begin{center}
$^{a}$»нститут “еоретической и Ёкспериментальной ‘изики (Ќ»÷ `` урчатовский »нститут'' --- »“Ё‘), ћосква 117218\\
$^{b}$Ќовосибирский √осударственный ”ниверситет, Ќовосибирск 90\\
$^{c}$ Kavli IPMU (WPI), Tokyo University, Kashiwa,  277-8583, Japan \\
$^{d}$ ¬Ќ»»ј им Ќ.Ћ.ƒухова, ћосква 127055
\end{center}


%


\begin{abstract}
ƒан обзор современного состо€ни€ теории и наблюдений ускорени€ расширени€ наблюдаемой части
вселенной.
 \end{abstract}

\section{¬ведение}\label{s-intro}

ќдним из наиболее впечатл€ющих открытий в астрономии, сделанных за последние
два дес€тилети€, было обнаружение того факта, что космологическое расширение
отнюдь не замедл€етс€ со временем, как было бы естественно ожидать дл€ вещества,
движущегос€ в собственном гравитационном поле.
Ќапротив, скорость расширени€
возрастает, причем этот процесс началс€ в относительно недавнюю эпоху
космологической истории.

Ќобелевска€ ѕреми€ 2011 года по физике была присуждена трем астрономам
—олу ѕерлмуттеру, Ѕрайану Ўмидту и јдаму –иссу (Saul Perlmutter, Brian P. Schmidt,
Adam G. Riess) за открытие ускоренного расширени€ вселенной по наблюдением
отдаленных сверхновых
(``for the discovery of the accelerating expansion of the Universe through observations of
 distant supernovae'').
ќткрытие не было неожиданным, но чрезвычайно важным.
Oно поставило, если не окончательную точку в вопросе о характере космологического
 расширени€ в современную эпоху, то привело очень сильный аргумент
в пользу того, что расширение происходит с ускорением.
ƒалекие сверхновые оказались  более тусклыми, чем ожидалось.
точнее говор€, зарегистрированный от них поток излучени€ оказалс€ ниже
ожидаемого при измеренных красных смещени€х $z\sim 1$ и в предположении, что
светимoсть этих сверхновых при  $z\sim 1$ така€ же, как и при $z = 0$, т.е. что они
€вл€ютс€ стандартными свечами.

ќрсюда следует, что они дальше, чем мы думали, и следовательно вселенна€ расшир€етс€
быстрее, чем  предсказывала стандартна€ модель.
{—трого говор€, ни ускорение, ни даже расширение пр€мо не измерены,
в том смысле, что не хватает точности дл€ того, чтобы заметить изменение рассто€ний
между галактиками
от времени наблюдател€ $t$. –еально измер€ютс€ зависимости рассто€ни€ от красного
смещени€ галактик,
и вот эти зависимости оказываютс€ другими, чем  предсказывала стандартна€ модель.}
¬озможные подводные камни в интерпретации этих результатов и ошибки
наблюдений обсуждаютс€ ниже.

ќчень важно, что в пользу ускоренного расширени€ вселенной говор€т не только данные
по измерению потоков излучени€ и рассто€ний сверхновых, но и целый р€д других, совершенно независимых
астрономических наблюдений.
 ак мы обсудим ниже, сюда вход€т, в частности, данные о возрасте вселенной,
анализ ее крупномасштабной структуры, измерение угловых флуктуаций
космического микроволнового фона (CMB) и барионных акустических осцилл€ций (BAO).
¬ силу этого можно заключить, что ускоренное расширение вселенной вне вс€ких сомнений установлено.

„тобы оценить всю необычность этого открыти€, воспользуемс€ простой, хот€ и
не вполне точной, аналогией космологического расширени€ с движением камн€,
брошенного вертикально вверх, в поле т€жести «емли. ≈сли начальна€
кинетическа€ энерги€ камн€ будет меньше его потенциальной энергии, т.е.
$E_{\rm kin} < U$, то, достигнув определенной высоты, камень
на мгновенье остановитс€
и вернетс€ назад. ¬ противоположном случае, когда $E_{\rm kin} > U$, остановки не
будет и камень уйдет на бесконечность. ¬ промежуточном, весьма специальном
случае, когда $E_{\rm kin} = U$,
камень также уйдет бесконечно далеко, но на бесконечности его скорость обратитс€
в ноль.  ¬о всех случа€х скорость камн€ будет убывать по мере его движени€ {вверх}.

 ак полагали до недавнего времени, космологическое расширение происходит полностью
аналогично рассмотренным выше примерам.
–асширение вселенной можно понимать, как движение
по инерции, возникшее в результате некоего начального толчка, вызванного
гравитационным отталкиванием (антигравитацией)
в эпоху инфл€ции, которую мы обсуждаем ниже.
≈сли исходный толчок был не очень сильным, то
в  какой-то момент, вселенна€ прекратит расширение и
схлопнетс€ обратно к гор€чей и плотной сингул€рности. ѕри достаточно сильном
начальном толчке расширение будет идти вечно и гор€ча€ ``бан€''  нам не грозит.
 ак и  в примере с камнем, считалось, что скорость космологического расширени€
должна убывать со временем.

ѕоследние открыти€ показали, что это совсем не так.
¬ какой-то недавний по космологическим меркам момент нормальное расширение
 вселенной с уменьшением скорости сменилось ускоренным.
≈сли вернутьс€ к аналогии с камнем, то картина бы выгл€дела так, что вначале камень
летел бы, как обычно, постепенно тер€€ скорость, однако позже он начал
бы ускор€тьс€, как будто у него включилс€  реактивным двигатель или, что ближе
к космологической ситуации, гравитационное поле «емли на больших рассто€ни€х
стало бы антигравитационным,
вызывающим  отталкивание вместо прит€жени€.

¬ двух словах суть открыти€ ускоренного расширени€ вселенной состоит в том, что
на сравнительно поздней стадии космологической эволюции нормальное
гравитационное прит€жение, замедл€ющее скорость расширени€,
переходит в гравитационное отталкивание и скорость расширени€ начинает
расти со временем. «аметим сразу же, что подобна€ антигравитаци€ возможна  только
{в рел€тивистской теории гравитации, например,}
в теории Ёйнштейна --- в общей теории относительности (ќ“ќ), как обсуждаетс€
ниже в разделе \ref{s-friedman-eqs}.  ¬  теории Ќьютона возможно лишь
гравитационное прит€жение.
 стати, исходный толчок, который привел к созданию нашей астрономически
большой вселенной из микроскопически малого и тесного начального состо€ни€ и к
наблюдаемому сегодн€ расширению вселенной ``по инерции'', тоже был результатом
космической антигравитации. Ёта стади€ называетс€ инфл€ционной, и в течение
этой стадии расширение также происходило с ускорением.
ћожно сказать, что больша€, подход€ща€ дл€ жизни вселенна€ оказалась возможной
только благодар€ ќ“ќ. ¬последствии
первоначальна€ ``инфл€ционна€'' антигравитаци€  либо исчезла, либо стала
пренебрежимо малой почти во всей космологической истории. ѕричина, по которой
антигравитаци€ снова стала играть заметную роль в космологии на современном этапе,
остаЄтс€ непон€тной, тем более, что необходимости в ней, на первый взгл€д, нет.

¬о избежание недоразумени€ сразу отметим, что антигравитаци€ в ќ“ќ,
возможна только дл€ безграничных систем. Ћюбой конечный объект с положительной
плотностью энергии (а мы можем рассматривать лишь такие, чтобы избежать
неустойчивости мира) всегда создает
только нормальное гравитационное прит€жение.
ќднако два таких объекта, помещЄнных внутрь среды с отрицательным давлением, на
достаточно большом рассто€нии друг от друга уже не будут чувствовать ``нормальное''
прит€жение, а будут ускор€тьс€ в противоположные стороны.
ћожно показать, что дл€ двух галактик с массами пор€дка массы ћлечного ѕути
гравитационное прит€жение будет компенсироватьс€
гравитационным отталкиванием вакуумной энергии на рассто€ни€х около двух
мегапарсек (см.~ниже).

 ак считалось совсем недавно, окончательна€ судьба вселенной и геометри€
ее трехмерного пространства однозначно св€заны. «амкнута€ вселенна€, имеюща€
геометрию трехмерной сферы, не сможет расшир€тьс€ вечно. ¬ какой-то момент в
отдаленном будущем расширение остановитс€ и обратитс€ в сжатие, как в примере
с камнем с малой начальной скоростью.
–асширение открытой вселенной с геометрией трехмерного гиперболоида никогда не остановитс€.
¬ промежуточном случае плоского трехмерного пространства расширение также будет вечным.
ќстановка произойдет асимптотически при $t \rar \infty$.

ѕоследний случай представл€ет особый интерес, так как, согласно наблюдени€м,
геометри€ вселенной очень близка к плоской.

¬ случае ускоренного расширени€ наиболее веро€тным оказываетс€ вечное
расширение дл€ любой геометрии мира. «аметим, что этот результат пр€мо
противоположен утверждению инфл€ционной теории, что при неускоренном
(как считалось ранее)  расширении на современной и более поздних стади€х наиболее
веро€тна€ окончательна€ судьба вселенной --- сжатие обратно к сингул€рности.
“аким образом,  ускоренное расширение вселенной может спасти мир  от этой
печальной судьбы. Ќе в том ли его необходимость?

”тверждение о том, что достаточно длительна€ инфл€ци€ приведет к тому, что
наша вселенна€ закончит свою жизнь в гор€чей сингул€рности, требует по€снени€.
ƒело в том, что инфл€ци€ приводит к возмущени€м плотности, в частности,  на
масштабах космологического горизонта. Ёти возмущени€ €вл€ютс€ стохастическими
и знак флуктуации плотности, $\delta \rho$,
может быть как положительным, так и отрицательным. ѕоэтому естественно ожидать,
что когда-то  на масштабе горизонта  $\delta \rho$ окажетс€ положительной и тогда
этот кусок вселенной отцепитс€ от общего космологического расширени€ и
схлопнетс€. — точки зрени€ внешнего наблюдател€ мы превратимс€ в черную
дыру. ќднако, как 
уже отмечалось, при ускоренном расширении
така€ судьба нам не
грозит, если плотность тЄмной энергии, котора€, возможно, вызывает такое  расширение,
падает медленнее, чем квадрат масштабного
фактора, см. ниже, рассуждени€ после уравнени€ (\ref{a-vac}).

»сточник, вызывающий ускоренное расширение, неизвестен. √лавным образом
обсуждаютс€ два варианта. ¬о-первых, это так называема€ тЄмна€ энерги€, т.е.
кака€-то субстанци€, имеюща€ отрицательное давление, которое
по абсолютной величине превосходит одну треть от ее плотности энергии,
$|P| > \rho/3 $ (см. ниже).
ќдной из форм тЄмной энергии могла бы быть вакуумна€ энерги€
или, что то же, космологическа€ посто€нна€, дл€ которой $P  = - \rho $.
ƒругой вариант тЄмной энергии это  квазипосто€нное скал€рное поле, $\phi$,
аналогичное тому, которое, веро€тно, генерировало инфл€цию. ¬ таком случае
различие между экспоненциальным расширением на заре существовани€ мира
и сегодн€шним днЄм лишь в различии, правда, колоссальном, энергетических и
временных масштабов.

ѕлотность вакуумной энергии не мен€етс€ при расширении, см. ниже
уравнение (\ref{dot-rho}). —ледовательно, если тЄмна€ энерги€ €вл€етс€ вакуумной
 энергией, то ускоренное расширение будет продолжатьс€ вечно дл€ любой
трехмерной геометрии мира, как мы уже отмечали выше. ќднако если тЄмна€
энерги€ €вл€етс€ энергией очень легкого скал€рного пол€ или пол€ с почти плоским
потенциалом, то в весьма отдаленном будущем, когда параметр ’аббла станет
сравним с массой или наклоном потенциала этого пол€, темп расширени€ снова станет
замедл€тьс€, а само поле $\phi$ обратитс€ в нуль
за счет красного смещени€ и/или рождени€ безмассовых частиц.
¬ итоге судьба вселенной снова станет определ€тьс€ ее геометрией, как это было
в старой доброй фридмановской космологии.

  ускоренному расширению мог бы также приводить р€д модификаций
гравитации при малых кривизнах. ¬место обычной ќ“ќ, в которой лагранжиан
пропорционален скал€ру кривизны, $R$, рассматриваютс€ теории с дополнительным
нелинейным по кривизне членом, $R + f(R)$. ¬ принципе можно бы рассматривать
добавки, завис€щие от квадрата тензора –иччи или –имана или более сложных
инвариантов, однако пока детального анализа последних не проведено.
»з-за нелинейности действи€ по кривизне уравнени€ гравитационного пол€
оказываютс€, вообще говор€, более высокого пор€дка, чем обычного второго.
¬ результате могут возникнуть проблемы с духами, тахионами и с устойчивостью
теории. Ёто приводит к р€ду ограничений  на вид функции $f(R)$.

ѕроблема выбора между двум€ этими вариантами (тЄмной энергии или
модифицированной гравитации) €вл€етс€ одной из центральных
проблем в феноменологическом описании ускоренного расширени€ вселенной.
јналогичный вопрос существует и дл€ механизма инфл€ции, помимо скал€рного
пол€ (инфлатона), инфл€ци€ могла бы быть вызвана $R^2$-поправками
к действию ќ“ќ. “ут, однако, имеетс€ существенное отличие от механизма
ускорени€ на поздней космологической стадии, т.е. сейчас или несколько
раньше. ƒл€ $R^2$-инфл€ции требуетс€ модификаци€ гравитации при больших
кривизнах, котора€ довольно естественно возникает вследствие радиационных
поправок, а дл€ феноменологического описани€ ускоренного расширени€ сегодн€
требуетс€ модификаци€ гравитации при очень малых кривизнах, котора€ вводитс€
{\it ad hoc} без специальных на то теоретических оснований.

 роме того, имеетс€ гораздо более глубока€ фундаментальна€
теоретическа€ проблема, тесно
св€занна€ с ускоренным расширением  -- это проблема вакуумной энергии.
ƒело в том, что теори€ и, в каком-то смысле, даже эксперимент говор€т, что
вакуум не пустой, а обладает колоссальной плотностью энергии, на 50--100
пор€дков превышающей наблюдаемое ограничение или, может быть, измеренное
значение. ѕоэтому необходим какой-то механизм компенсации этих громадных
вкладов. Ќесмотр€ на многочисленные попытки, этот механизм до сих пор не
найден. ѕроблема компенсации вакуумной энергии, веро€тно, одна из наиболее
важных проблем современной фундаментальной физики.

ќбзор построен следующим образом. ¬ следующем разделе приведены (а в
ѕриложении ј даже выведены, как  на простом наивном уровне, так и вполне
строго) основные космологические уравнени€ --
уравнени€
‘ридмана, введено пон€тие космологической посто€нной (вакуумной энергии)
и по€сн€етс€,  как может возникнуть космическа€ антигравитаци€. ¬ этом же разделе
определены основные космологические параметры. ¬ разделе \ref{s-lambda}
описана проблема вакуумной энергии и возможные способы ее решени€.
¬ разделе \ref{s-observ} обсуждаютс€
астрономические данные, указывающие на ускоренное расширение вселенной.
¬ отдельный раздел \ref{s-SNBAO} вынесен анализ данных по сверхновым  и барионным акустическим осцилл€ци€м
в силу большой важности такого анализа и интересных перспектив на будущее.
‘еноменологическое описание ускоренного расширени€ вселенной
за счет тЄмной энергии
или модифицированной гравитации приведено соответственно в
разделах \ref{s-DE}  и \ref{s-grav-mdf}. ¬ приложении ј
приведены два варианта упрощенного вывода уравнений ‘ридмана (один из
ньютоновского предела, второй из вариационного принципа), а в приложении Ѕ
приведены
основные космологические параметры и обсуждаютс€ способы их измерени€.
ѕриложение ¬ посв€щено скал€рному полю в космологии.

\section{”равнени€ ‘ридмана и космологическое ускорение \label{s-friedman-eqs}}

¬ основе современной космологии лежат уравнени€ Ёйнштейна~\cite{ein-eq,ein-eq1},
св€зывающие кривизну пространства-времени с $T_{\mu\nu}$ -- тензором
энергии-импульса материи:
\be
R_{\mu\nu} - \frac{1}{2} g_{\mu\nu} R = \frac{8\pi}{m_{Pl}^2}\, T_{\mu\nu}\,.
\label{ein-eq}
\ee
«десь $g_{\mu\nu}$ -- метрический тензор, который определ€ет интервал в
четырЄхмерном пространстве-времени согласно
\be
ds^2 = g_{\mu\nu} dx^\mu dx^\nu\,,
\label{ds2}
\ee
а $R_{\mu\nu}$ и $R = g^{\mu\nu} R_{\mu\nu}$ -- тензор –иччи и скал€рна€ кривизна
соответственно. ќни известным образом выражаютс€ через метрический тензор
и его первые и вторые производные, как описываетс€ в любом учебнике по ќ“ќ или
–имановой геометрии.
¬еличина $m_{Pl} = 1.2\times 10^{19}$ √э¬ носит название массы ѕланка.
«десь и ниже мы
используем естественную систему единиц, где $c = k = h/(2\pi) =1$. ¬ этих единицах
ньютоновска€ гравитационна€ посто€нна€ равна $G_N = 1/m_{Pl}^2$.

”равнени€ Ёйнштейна были впервые применены к космологии реальной ¬селенной
јлександром јлександровичем ‘ридманом~\cite{Friedman1922}, хот€ уже сам Ёйнштейн
пыталс€ построить статическую модель вселенной (дл€ чего и ввЄл Ћ€мбда-член),
а де —иттер строил модели вакуумной вселенной.
‘ридман предположил, что вселенна€ однородна и изотропна (во вс€ком случае на больших масштабах) и,
следовательно,
пространство должно иметь посто€нную трехмерную кривизну с интервалом вида:
\be
 ds^2 = dt^2 - a^2[d\chi^2+\sin^2\chi^2(d\theta^2+\sin^2\theta\;d\varphi^2)] \ .
\label{ds2-friedman}
\ee
Ќесколько иной вид получаетс€ при замене
$\sin\chi=r$, тогда $d\chi^2=dr^2/(1-r^2)$, и можно записать
интервал в виде (ввод€ $k=1$)
\be
ds^2 = dt^2 - a^2(t)\left[{{dr^2}\over{1-kr^2}}
  + r^2(d\theta^2 +\sin^2\theta d\varphi^2)\right].
\label{ds2-cosm}
\ee
¬еличина $a(t)$ носит название масштабного фактора и определ€ет
рассто€ние между двум€ событи€ми в трехмерном пространстве.
Ёто форма метрики ‘ридмана из работы \cite{Friedman1922} и другой его метрики
из работы \cite{Friedman1924} в записи –о\-берт\-со\-на--”окера
\cite{Robertson1933,Walker1933} ({  Robertson-Walker}). —оответственно
метрика однородной и изотропной вселенной получила название метрики
‘ридмана-–обертсона-”окера (FRW metric).

ѕри $k=+1$ трехмерное пространство можно рассматривать как
трехмерную сферическую поверхность, вложенную в
плоское четырехмерное пространство.
¬ этом случае говор€т о замкнутой вселенной. ¬ остальных случа€х вселенна€ будет открыта€.
ѕри $k= - 1$ трехмерное пространство представл€ет собой
гиперболоид, а  при $k=0$ наше пространство €вл€етс€ трехмерно плоским, описываемым обычной (изучаемой в средней школе) геометрией Ёвклида.

ѕодстановка метрического тензора, отвечающего интервалу (\ref{ds2-cosm}),
в уравнени€ (\ref{ein-eq}) приводит к следующим уравнени€м, определ€ющим
эволюцию масштабного фактора от ``сотворени€'' мира почти до наших
дней, пока вселенную можно считать однородной и изотропной:
\be
\left( \frac{\dot a}{a}\right)^2 = \frac{8\pi}{3}\,\frac{\rho}{m_{Pl}^2} - \frac{k}{a^2}\, ,
\label{H2}
\ee
где $\rho$ -- плотность энергии вещества, т.е. $T_{tt}$ (или $T_{00}$) --
компонента тензора энергии-импульса. «аметим, что в соответствии с
гипотезой об однородности и изотропии вселенной
предполагаетс€, что последний имеет
диагональный вид с пространственными компонентами, пропорциональными
плотности давлени€, $T_{i}^{j} = -\delta_{i}^{j} P$.

¬торое уравнение ‘ридмана выражает ускорение при расширении через плотность энергии и давлени€ вещества:
\be
\frac{\ddot a}{a} = -\frac{4\pi}{3 m_{Pl}^2}(\rho+3P) \,.
\label{ddot-a}
\ee
—разу же отметим, что при $P < -\rho/3$, ускорение оказываетс€ положительным, т.е. гравитаци€ становитс€ отталкивающей (антигравитацией).  ак  упоминалось во
введении,  с этим св€зан первоначальный толчок, приведший к расширению вселенной,
а также очень веро€тно, что
это же  €вл€етс€ причиной наблюдаемого сейчас ускоренного расширени€ вселенной.

ѕриведем еще уравнение, описывающее эволюцию плотности энергии во фридмановской
космологии:
\be
\dot \rho = - 3H (\rho + P)\,.
\label{dot-rho}
\ee
Ёто уравнение представл€ет собой ковариантный закон сохранени€ тензора энергии-импульса,
\be
D_\mu T_\mu^\nu = 0\,,
\label{D-T}
\ee
где $D_\mu$ --  ковариантна€ производна€ в гравитационном поле.
”равнение (\ref{dot-rho})
€вл€етс€ следствием двух уравнений ‘ридмана (\ref{H2})  и (\ref{ddot-a}), но выделено тут в силу его важности.

¬ ѕриложении A приведен элементарный (хот€ и слегка жульнический)
вывод этих уравнений практически без использовани€ ќ“ќ. ќн позвол€ет
лучше почувствовать их физический смысл.
“ам же дан и более строгий, но нетрадиционный вывод из вариационного принципа ќ“ќ.

ƒва независимых уравнени€ из вышеприведенных
трех (\ref{H2},\ref{ddot-a},\ref{dot-rho})
содержат три неизвестных функции, $a(t)$, $\rho(t)$ и $P(t)$. „тобы полностью
определить систему необходимо еще одно уравнение. ќбычно этим уравнением
€вл€етс€ уравнение состо€ни€, выражающее плотность давлени€ как функцию
плотности энергии, $ P = P(\rho)$. ¬о многих практически интересных случа€х
справедливо линейное соотношение:
\be
P = w \rho\, ,
\label{P-w-rho}
\ee
где $w$--обычно некоторый посто€нный параметр. ѕоследнее, однако, не
об€зательно и нередко обсуждаютс€ версии теории с $w=w(t)$. “акое предположение
естественно делать дл€ феноменологического описани€ и анализа наблюдательных
данных при разных красных смещени€х (т.е. в разные моменты космологической
эволюции). “акже $w(t)$ естественно возникает, например, в случае, когда
гравитирующа€
матери€ оказываетс€ каким-то динамическим полем. “огда $w(t)$ определ€етс€
уравнени€ми движени€ этого пол€, см. ниже раздел \ref{s-DE}. «аметим, что
соотношение $P = P(\rho)$, строго говор€, не всегда справедливо. ƒавление
может выражатьс€ через плотность энергии через ее производные или интеграл по
времени, или может зависеть от других термодинамических переменных (температура,
удельна€ энтропи€ и т.д.).
Ќо разумеетс€, соотношение~(\ref{P-w-rho}) формально всегда справедливо:
$w(t) = P(t) /\rho(t)$.

¬ более простых, но практически интересных случа€х $w=\mbox{const}$. Ќапример,
дл€ нерел€тивистского вещества $P\ll \rho$, поэтому полагают $w=0$.
ƒл€ рел€тивистского вещества, как известно, $w= 1/3$.
«акон космологического расширени€ имеет особенно простой вид при $k=0$,
см. уравнение  (\ref{H2}), когда трехмерное пространство €вл€етс€ плоским, эвклидовым.
—огласно наблюдени€м именно это с хорошей точностью и реализуетс€ в нашем мире.
¬ нерел€тивистском случае масштабный фактор растет по закону:
\be
a_{\rm NR} (t) \sim t^{2/3}\,.
\label{a-NR}
\ee
¬ рел€тивистском случае закон расширени€ имеет вид:
\be
a_{\rm rel} (t) \sim t^{1/2}\,.
\label{a-rel}
\ee

‘ридманом были найдены решени€ уравнений (\ref{H2},\ref{ddot-a},\ref{dot-rho})
дл€ различных законов расширени€
и {\it предсказано} расширение вселенной, впоследствии обнаруженное в
астрономических наблюдени€х. ќбычно это открытие приписывают
’абблу~\cite{hubble}, однако оно было обнаружено ранее
Ћеметром~\cite{lemaitre}. ѕоэтому закон расширени€ вселенной
было бы уместно называть законом ‘ридмана-Ћеметра-’аббла.

ѕо какой-то причине Ёйнштейн был противником нестационарной вселенной
и длительное врем€ (до работы ’аббла) не признавал решений ‘ридмана.
ѕыта€сь применить уравнени€ ќ“ќ к космологии, Ёйнштейн обнаружил, что
стационарные решени€ отсутствуют и дл€ ``спасени€'' стационарной вселенной
предложил ввести  в уравнени€ (\ref{ein-eq}) дополнительное слагаемое,
так называемый Ћ€мбда-член или, что то же, космологическую посто€нную \cite{ein-lambda}:
\be
R_{\mu\nu} - \frac{1}{2} g_{\mu\nu} R  - \Lambda g_{\mu\nu} =
\frac{8\pi}{m_{Pl}^2}\, T_{\mu\nu}\,.
\label{Lambda}
\ee
¬ведение этого члена не нарушает ковариантного сохранени€ правой и левой
части этого уравнени€. “ензор Ёйнштейна автоматически сохран€етс€ в
метрической теории:
\be
D_\mu \left( R^\mu_\nu - \frac{1}{2} g^\mu_\nu R \right) \equiv 0\,.
\label{ricci}
\ee
Ёто так называемое свЄрнутое тождество Ѕь€нки,
которое автоматически 
выполн€етс€ в
римановой геометрии. «ануление производной тензора Ёйнштейна напоминает
автоматическое обращение в ноль производной левой части уравнений ћаксвeлла,
$\partial_\mu \partial_\nu F^{\mu\nu} \equiv 0$.
ѕрава€ часть уравнени€ (\ref{Lambda}) ковариантно сохран€етс€, см. (\ref{D-T}),
в силу общей ковариантности, т.е. инвариантности теории относительно выбора
системы координат.  овариантна€ производна€ метрического тензора
равна нулю по построению теории. Ёто легко проверить,
%
%
если найти ковариантную производную метрики:
\be
&g_{ik;m}&=\partial_m g_{ik} - \Gamma^{j}_{im}g_{jk} - \Gamma^{j}_{km}g_{ij}
 = \partial_m g_{ik} - \Gamma_{kim} - \Gamma_{ikm} \cr
&=& \uline{\partial_m g_{ik}}
-{1\over 2} \left(
   \uline{\partial_m g_{ik}} + \uwave{\partial_i g_{mk}}-\uuline{\partial_k g_{{im}}}\right)
-{1\over 2} \left(
   \uline{\partial_m g_{ki}} + \uuline{\partial_k g_{mi}}-\uwave{\partial_i g_{{km}}}\right)=0.
\label{dgcov}
\ee
«десь члены, подчЄркнутые одинаковыми символами, взаимно уничтожаютс€ при
учЄте симметрии $g_{ik}=g_{ki}$.
%
% ...
%
ѕоэтому ковариантное сохранение всех слагаемых в уравнении (\ref{Lambda}) не будет нарушено,
если  $\Lambda = \mbox{const}$. ќтсюда и им€ ---  космологическа€ посто€нна€.

ѕо мысли Ёйнштейна антигравитаци€, создаваема€ Ћ€мбда-членом, могла
бы компенсировать гравитационное прит€жение обычного вещества и
обеспечить стационарность вселенной. ѕодобное стационарное решение
действительно существует, но оно 
неустойчиво и, что главное,
противоречит позднее открытому расширению вселенной.
ѕоэтому впоследствии
Ёйнштейн отказалс€ от гипотезы о существовании Ћ€мбда-члена, полага€ его
величайшей  ошибкой (blunder) своей жизни (со слов √. √амова в книге ``ћо€ мирова€
лини€'',  впрочем, точна€ цитата нам неизвестна).

¬последствии стало €сно, что космологическа€ посто€нна€ эквивалентна энергии
вакуума с тензором энергии-импульса:
\be
T_{\mu\nu}^{\rm vac} = g_{\mu\nu} \rho^{\rm vac}
\label{T-vac}
\ee
и с уравнением состо€ни€ $P= -\rho$ (ссылки см., например, в обзоре \cite{Chernin2008}).
ќчевидно,  что при доминантности
вакуумной энергии вселенна€ должна расшир€тьс€ ускоренно с экспоненциально
растущим масштабным фактором:
\be
a^{\rm vac} (t) \sim \exp \left[ H_{\rm vac} t\right]\,,
\label{a-vac}
\ee
где $H_{\rm vac}^2 = 8\pi \rho^{\rm vac}/ 3 m_{Pl}^2 \approx \mbox{const}$.

«аметим еще, что согласно уравнению (\ref{dot-rho}) плотность энергии
рел€тивистского вещества убывает, как $1/a^4$, а нерел€тивистского, как
$1/a^3$. ¬ эпоху доминантности как того, так и другого, космологическа€
плотность энергии падает со временем, как $1/t^2$. — другой стороны,
плотность энергии вакуума не измен€етс€ ни при расширении,
ни при сжатии,  $\rho^{\rm vac} =\mbox{const}$.

“акое поведение космологической плотности энергии объ€сн€ет св€зь геометрии
пространства и окончательной судьбы вселенной.
ќсобенно просто эту судьбу вы€снить при отсутствии вакуумной или
тЄмной энергии.
—огласно уравнению (\ref{H2}) при
$k>0$, т.е. в замкнутой вселенной, член пропорциональный кривизне, $k/a^2$,
рано или поздно окажетс€ больше, чем слагаемое, пропорциональное плотности
энергии материи, которое убывает, по крайней мере, как $1/a^3$. —ледовательно,
$H$ обратитс€ в ноль, и затем расширение сменитс€ сжатием.

 осмологические решени€ в теории с Ћ€мбда-членом исследовались в
работах ‘ридмана~\cite{Friedman1922,Friedman1924}, а также в
работах~\cite{lambda-OK,lambda-OK1,lambda-OK2,lambda-OK3}.
јвторы последних работ считали,  вопреки Ёйнштейну, что добавление
космологической посто€нной в уравнени€ (\ref{Lambda}) было  чрезвычайно
важным обобщением ќ“ќ.  — другой стороны, многие очень хорошие физики
относились к космологической посто€нной крайне отрицательно. ¬ частности
√амов в своей автобиографической книге ``ћо€ мирова€ лини€''  по поводу астрономических данных, указывающих на ненулевое значение последней в
середине 1960-ыx годов,
писал: ``Ћ€мбда оп€ть поднимает свою безобразную голову''. ¬прочем,
эти указани€ впоследствии было исчезли, но позднее по€вились снова -
см. раздел~\ref{ss-history}

ќднако, как мы увидим в следующем разделе, квантова€ теори€ пол€
не только требует, чтобы
космологическа€ посто€нна€ (или, что то же, энерги€ вакуума) была
отлична от нул€,  но и предсказывает громадную величину
последней.

\section{ѕроблема вакуумной энергии \label{s-lambda} }

ѕроблема вакуумной энергии, по-видимому, наиболее серьезна€ проблема
современной фундаментальной физики. “еоретические оценки различных
вкладов в вакуумную энергию привод€т к фантастически громадному результату.
Ќе будет большим преувеличением сказать, что теори€ предсказывает
$\rho^{\rm vac} \approx \infty$. –азногласие между теорией и астрономическими
данными составл€ет величину пор€дка $10^{50}-10^{100}$. –азброс этих
значений отвечает различным физическим источникам вакуумной энергии.
√ромадна€ величина теоретических вкладов и ничтожно малый суммарный результат
заставл€ют вспомнить высказывание ‘ейнмана о радиационных поправках в
квантовой элетродинамике: ``ѕоправки бесконечны, но малы.''

¬акуумом в квантовой теории называетс€ состо€ние с низшей энергией и,
вообще говор€, эта энерги€ не об€зана быть равной нулю. ¬спомним хорошо
известный пример квантового осцилл€тора, дл€ которого энерги€ основного
уровн€ равна $\omega/2$, где  $\omega$-частота осцилл€тора.

≈сли вакуумна€ энерги€ отлична от нул€, то отвечающий ей тензор
энергии-импульса должен быть пропорционален метрическому тензору,
см. уравнение (\ref{T-vac}), так как это единственный симметричный
``инвариантный'' тензор второго пор€дка, т.е. тензор, который не измен€етс€ при
переходе из одной системы координат в другую, что естественно ожидать дл€
вакуума.

 вантовое поле представл€ет собой бесконечный набор осцилл€торов
со всеми возможными частотами. —оответственно плотность энергии
вакуумных квантовых флуктуаций какого-либо бозонного пол€ оказываетс€
бесконечно большой (см. приложение ¬):
\be
\rho^{\rm vac}_b = \langle {\cal H}_b \rangle_{\rm vac} = \int \frac{d^3 k}{(2\pi)^3}\,
\frac{\omega_k}{2}
\langle a^\dagger_k a_k + b_k b^\dagger_k \rangle_{\rm vac}
=\int \frac{d^3k}{2\, (2\pi)^3}\,\omega_k = \infty^4\,.
\label{h-b}
\label{rhobos}
\ee
«десь $\langle {\cal H}_b \rangle_{\rm vac}$ среднее вакуумное значение гамильтониана  этого бозонного
пол€, $\omega_k = (k^2 + m^2)^{1/2}$ -- энерги€ кванта с импульсом  $k$ и массой $m$, а $a_k$, $a_k^\dagger$ и
$b_k$, $b_k^\dagger$ - операторы уничтожени€ и рождени€
частиц и античастиц соответственно.

≈сли вычислить давление квантовых флуктуаций вакуума, то оно также
окажетс€ бесконечным. ≈сли при этом воспользоватьс€ выражени€ми (\ref{rho-P}) из ѕриложени€ ¬,
 формально обреза€ интегралы дл€ $\rho$ и $P$ на одном и том же верхнем пределе, то
 вычисленные таким
образом вакуумные средние не будут удовлетвор€ть условию
$\rho^{\rm vac} = - P^{\rm vac}$.
Ќа этом основании в работе~\cite{prokopec} был сделан вывод, что вакуумные
флуктуации нарушают Ћоренц-инвариантность вакуума, а так как вакуум должен
быть Ћоренц-инвариантен, то надо
потребовать, чтобы наиболее расход€щиес€ члены, которые €кобы нарушают Ћоренц
инвариантность, обращались бы в ноль. ќднако обращени€ с бесконечно
расход€щимис€
интегралами требует осторожности и, в частности, сравнение формально равных
выражений
дл€ разных величин вовсе не требует того, что они и в самом деле равны, см. ниже
пример с сокращением бозонныx и фермионных вкладов  в суперсимметричном
мире, когда получающийс€ конечный
(с точностъю до логарифмической расходимости) результат удовлетвор€ет
необходимому Ћоренц-инвариантному условию $ \rho^{\rm vac} = -P^{\rm vac}$.

ќчевидно, что жизнь в мире с бесконечно большой плотностью энергии невозможна.
ќн либо расширитс€ с ``бесконечно''  большой скоростью, если плотность энергии
положительна, либо мгновенно схлопнетс€ в сингул€рность, если она отрицательна.
  счастью, из-за замены коммутаторов на антикоммутаторы при квантовании,
фермионные пол€ внос€т точно такой же по величине,
но противоположный по знаку вклад:
\be
\rho_{\rm vac}^{(f)} \equiv \langle {\cal H}_f \rangle_{\rm vac} = \int \frac{d^3 k}{(2\pi)^3}\,
\frac{\omega_k}{2}
\langle a^\dagger_k a_k - b_k b^\dagger_k \rangle_{\rm vac} \,
= -\int \frac{d^3k}{(2\pi)^3}\,\omega_k = -\infty^4\,.
\label{h-f}
\ee

ѕоэтому, если бы дл€ каждого бозона существовал
бы фермионный партнер с точно равной по величине массой, то вакуумна€ энерги€
квантовых флуктуаций обратилась бы в ноль. Ётот факт был впервые отмечен
ѕаули \cite{PauliFQ}
и независимо «ельдовичем~\cite{zeld-vac},
см. также \cite{Visser2016}, где подробно обсуждаетс€ работа ѕаули.
»нтересно, что иде€ о сокращении вакуумных энергий бозонов и фермионов
по€вилась еще до по€влени€ работ, где была предложена
теори€ суперсимметрии~\cite{golfand}, при наличии которой это сокращение естественно
происходит.
ќднако суперсимметри€, если она и существует, не €вл€етс€ точной, и массы бозонов и фермионов должны сильно различатьс€. ѕри этом расходимости четвертой степени
будут все же сокращатьс€ независимо от масс. ѕри так называемом м€гком нарушении
суперсимметрии должны сократитьс€ и квадратичные расходимости, но конечный
остаток, пропорциональный разности масс обычных частиц и их суперпартнЄров, оказываетс€
колоссальным в космологических масштабах:
\be
\rho^{\rm vac}_{\rm SUSY}  \sim m^4_{\rm SUSY} > 10^{55} \rho_{c} \,.
\label{rho-sus}
\ee
где $\rho_c \approx 4\times 10^{-47} \; \mbox{√э¬}^4$ -- космологическа€ плотность
энергии в сегодн€шней вселенной (см. ѕриложение Ѕ), а
$m_{\rm SUSY} $ -- массовый масштаб нарушени€ суперсимметрии, который согласно
имеющимс€ экспериментальным ограничени€м должен быть выше 100 √э¬.
ќтсутствие сигнала от суперсимметричных частиц на большом адронном коллайдре
(LHC) в ÷≈–Ќ, по-видимому, говорит, либо, что $m_{\rm SUSY} $ должна быть заметно выше,
либо, что суперсимметрии вообще не существует.

ќтметим, что суммарный вклад бозонов и фермионов с массами, которые подобраны
так, чтобы квартично и квадратично расход€щиес€ вклады сокращались, дает логарифмически
расход€щийс€ результат, удовлетвор€ющий условию $\rho = -P$, так что Ћоренц-инвариантность, как уже отмечалось,
не нарушаетс€. ѕри этом €сно, что ответ расходитс€, как втора€ и даже четверта€
степень массы при разности масс, стрем€щейс€ к бесконечности.

≈сли точна€ суперсимметри€ приводит к нулевой вакуумной энергии, то в
нарушенной суперсимметрии это не об€зательно. Ѕолее того,
так называема€ глобальна€ суперсимметри€ при м€гком
ее нарушении, которое не портит перенормируемости теории, требует ненулевой
величины вакуумной энергии, близкой по величине к указанной выше $m_{\rm SUSY}^4 $.
ќднако, если расширить теорию, включив в нее гравитацию, т.е. рассмотреть
теорию супергравитации (SUGRA), то утверждение об об€зательности ненулевой вакуумной
энергии снимаетс€ и за счет очень тонкой подгонки параметров можно получить
$\rho^{\rm vac}_{\rm SUGRA} = 0$. ќднако естественное значение
вакуумной энергии в такой
теории составл€ет $\rho^{\rm vac}\sim m_{Pl}^4$, так что необходима€ подгонка
дл€ сокращени€ вакуумной энергии
должна иметь точность около $10^{-123}$, что выгл€дит крайне неестественным.

»нтересно отметить, что в ходе космологической эволюции вакуумна€ энерги€
претерпевала колоссальные скачки при фазовых переходах от симметричной
фазы к фазе с нарушенной симметрией. ƒело в том, что
в основе современной теории элементарных
частиц лежит пон€тие спонтанно нарушенной (калибровочной) симметрии. ѕри охлаждении
вселенной измен€етс€ вакуумное состо€ние в такой теории.
ѕри этом скачок в плотности энергии
составл€ет $\delta \rho^{\rm vac} \sim 10^{60}\, \; \mbox{√э¬}^4$ дл€ фазового перехода
в модели большого объединени€ (GUT),
$\delta \rho^{\rm vac} \sim 10^{8} \; \mbox{√э¬}^4$ -- в электрослабой теории и
$\delta \rho^{\rm vac} \sim 10^{-2} \; \mbox{√э¬}^4$ -- в квантовой хромодинамике ( ’ƒ)
при переходе от фазы конфайнмента к деконфайнменту.

ѕроблема вакуумной энергии приобретает особую серьезность, если рассмотреть
структуру вакуума квантовой хромодинамики.  ак известно, протон состоит из
трех легких кварков: $p \sim uud $, масса каждого из которых, грубо говор€, 5 ћэ¬.
ѕосему следует ожидать, что масса протона должна быть весьма мала,
$ m_p \sim $ (15 ћэ¬ $- E_B) < 15 $ ћэ¬, где $E_B$ -- энерги€ св€зи кварков в
протоне. ѕолученный результат, по крайней мере,  в 60 раз меньше массы протона.
Ќедостающий вклад в массу беретс€
из нетривиальных свойств вакуума  ’ƒ. ¬опреки интуитивному ожиданию этот
вакуум оказываетс€ не пустым, а заполненным
конденсатом кварковых~\cite{q-cond} и глюонных~\cite{g-cond} полей:
\be
\langle \bar q q  \rangle \neq 0\,, \,\,\,\,\,
 \langle { G_{\mu\nu} G^{\mu\nu} }\rangle  \neq 0\,.
\label{QCD-cnd}
\ee
ѕлотность энергии этих конденсатов отрицательна и составл€ет примерно
1 √э¬$^4$. Ёто разумна€ величина по меркам физики элементарных частиц и
громадна€ по космологическим стандартам современной вселенной:
\be
\rho^{\rm vac}_{\rm QCD} \approx\, - 10^{45} \rho_c.
\label{rho-QCD}
\ee

¬нутри протона кварки уничтожают глюонный конденсат, отсутствие которого
поднимает массу протона до нужной величины:
\be
m_p = 2m_u + m_d - \rho^{\rm vac} l_p^3 \sim 1\,\,\; \mbox{√э¬}\,,
\label{m-p}
\ee
где $l_p$--размер протона.

¬озникает почти мистическа€ ситуаци€.
ѕрекрасно установленна€, согласующа€с€ с
экспериментом теори€ говорит, что вакуум не пустой. ¬ нем ``живут'' кварковые и
глюонные пол€, плотность энергии которых на 45 пор€дков превышает космологическую.
“ем не менее полна€ плотность вакуумной энергии равна космологической. „то-то
еще ``живет'' в вакууме и это ``что-то'', возможно, какое-то новое поле, компенсирует
$\rho^{\rm vac}_{\rm QCD} $ на 45 пор€дков. «аметим, что это новое поле должно быть лЄгким,
иначе оно не могло бы действовать на космологических масштабах.  роме того, оно
ничего на знает ни о кварках, ни о глюонах. ¬ противном случае, если бы это поле заметно
взаимодействовало с ними, то оно было бы обнаружено на эксперименте.
ќчевидно, вакуумна€ энерги€ описываетс€ одним параметром, и все описанные вклады можно
компенсировать одной единственной вычитательной константой, но такое решение выгл€дит
весьма неестественным.

ѕроблема усугубл€етс€ еще и тем, что  вакуумна€ (или вакуумно-подобна€) энерги€
немножко не нулева€ и по какой-то не€сной причине близка по величине к плотности
энергии обычной материи (включа€ сюда тЄмную материю), хот€ законы космологической эволюции $\rho^{\rm vac}$ и $\rho_{m}$ совершенно различны.  ак мы отмечали выше,
$\rho^{\rm vac}$ остаетс€ посто€нной при расширении вселенной, а $\rho_{m}$ падает, как
куб масштабного фактора $a$ дл€ нерел€тивистского вещества и как $1/a^4$ -- дл€
рел€тивистского.
¬ насто€щее врем€ имеетс€ р€д попыток решени€ этой проблемы или, лучше
сказать, этих проблем, но ни одну из них нельз€ назвать до конца удачной.
Ќиже мы их кратко опишем.
Ѕолее детальные обзоры по этой теме содержатс€ в работах~\cite{rho-vac-rev} --\cite{rho-vac-rev13}. 

ѕервое, самое простое и ужасно некрасивое решение, уже упом€нутое несколькими строками выше, состоит в том, что
все эти вклады в вакуумную энергию компенсируютс€ некоторой исключительно точно подобранной вычитательной константой, котора€
равна сумме всех физических
вкладов в вакуумную энергию с точностью $10^{-45}-10^{-127}$. Ёто по существу
представл€ет собой выбор нул€ в отсчете энергии. “ак как нам надо объ€снить величину лишь одного числа, то исключить такую
возможность никак нельз€, но она представл€етс€
весьма неестественной.  роме того, хотелось бы, чтобы все вклады в вакуумную
энергию  происходили от физических полей, а не от какого-то ``богом данного''
числа.

¬ несколько лучшем положении €вл€етс€ предложение решени€ проблемы вакуумной
энергии на основе антропного принципа, который гласит, что вопрос о величине тех или иных значений фундаментальных посто€нных задаетс€ наблюдателем,
наход€щимс€ во вселенной, где имеютс€ подход€щие услови€ дл€ существовани€
этого наблюдател€. —писок ранних работ, видимо не полный, по антропному принципу
мы даЄм в ссылках~\cite{anthrop-early,anthrop-early1,anthrop-early2,anthrop-early3};
более подробное обсуждени€ может быть найдено, например, в книгах~\cite{book-anthrop,book-anthrop1,book-anthrop2,book-anthrop3}.

ƒл€ естественной реализации антропного принципа необходимо, чтобы
существовало близкое к бесконечному (или даже бесконечное) число вселенных с
различными значени€ми фундаментальных посто€нных, в частности, с различными
затравочными значени€ми вакуумной энергии (т.е. с различными вычитательными константами). ¬есьма естественно космологически большие
вселенные с различными физическими свойствами возникают в инфл€ционной
модели~\cite{vilenkin-infl-83}. ќсобенно хороша в этом смысле
хаотическа€ инфл€ци€~\cite{linde-chaotic}.
ѕо-видимому, впервые иде€ о большом числе вселенных
в применении к решению проблемы вакуумной энергии была высказана в
работе~\cite{sakharov-lambda}.

»де€ о множественности вселенных получила сильную поддержку в
контексте теории суперструн. Ёти теории формулируютс€ в 10-мерном пространстве,
но предполагаетс€, что дополнительные к нашему миру 6 измерений €вл€ютс€ маленькими,
компактными. –еалистический механизм компактификации дополнительных
измерений совместимый с инфл€цией
был предложен в работе~\cite{kklt}. ¬скоре было осознано, что существует
колоссальное количество способов компактификации этого 6-мерного пространства.
—огласно оценке работы~\cite{douglas}
количество возможных типов компактификации оказалось около $10^{{500}}$ или,
возможно, даже заметно больше. јналогичные идеи о существовании большого
числа различных вселенных развивались в работе~\cite{suskind-landscape}.

—реди этого множества
вселенных лишь относительно немногие могут оказатьс€ пригодными дл€ жизни
(а могут и не оказатьс€). ¬ пригодных дл€  жизни вселенных вакуумна€ энерги€,
как положительна€, так и отрицательна€, не может быть слишком большой по
абсолютной величине. ѕри большой отрицательной $\rho^{\rm vac}$ вселенна€ схлопнетс€
еще до образовани€ звезд и планет, а при большой положительной $\rho^{\rm vac}$
расширение будет настолько быстрым, что никакие небесные тела не успеют возникнуть,
так как плотность материи очень быстро станет пренебрежимо малой.
Ётим аргументам был придан количественный характер в
работах~\cite{carter-83,weinberg-anthrop,linde-87}.
Ѕыло показано, что
в случае $\rho^{\rm vac}>0$ плотность
вакуумной энергии не должна отличатьс€ более, чем на 2-3 пор€дка  от
космологической плотности массы обычной материи ~\cite{weinberg-anthrop}.
¬ работе\cite{tegmark-rees-98} антропный принцип был использован дл€  вывода
ограничений на значени€ различных космологических параметров, включа€
вакуумную энергию.

Ѕолее детально вопрос об антропном ограничении на величину вакуумной
энергии рассмотрен в работе~\cite{vilenkin-2004}. јвтор делает существенно
более сильное утверждение, что
антропный принцип предсказывает ненулевую вакуумную энергию, причем
довольно близкую по величине к наблюдаемой. јнтропное ограничение
на величину (положительной) вакуумной энергии может быть получено
из требовани€, что вакуумна€ энерги€ могла начать доминировать
лишь позже эпохи образовани€ галактик. “ак как плотность энергии обычной материи убывает при
космологическом расширении, как куб масштабного фактора, а $\rho^{\rm vac}$
остаетс€ посто€нной, то должно выполн€тьс€ условие:
\be
\rho^{\rm vac} < (1+z_{\rm max})^3 \rho_m^0\,,
\label{rho-vac-rho-m}
\ee
где $\rho_m^0$-плотность материи в сегодн€шней вселенной, а
$z_{\rm max}$-максимальное красное смещение, когда начинают образовыватьс€
галактики.


—огласно работе~\cite{vilenkin-2004} при $z_{\rm max}\sim 10$ антропное ограничение на
плотность вакуумной энергии составл€ет $\rho^{\rm vac} < 4000 \rho_m^0$.
Ёто ограничение
было улучшено в работе~\cite{weinberg-anthrop}
(см также~\cite{Carriga2000}):
$\rho^{\rm vac} < 100 \rho_m^0$,
где было впервые отмечено, что не все значени€ $\Lambda$ совместимы с наличием во
¬селенной  разумной жизни.

¬ работе~\cite{Carriga1999}
антропные соображени€ примен€лись
к трем временным масштабам: времени образовани€ галактик, моменту доминантности космологической
посто€нной в плотности энергии ¬селенной и возрасту ¬селенной и был сделан вывод, что
плотность вакуумной энергии  близка к наблюдаемой.

Ѕлизкое утверждение о малости вакуумной энергии сделано в работе~\cite{Mersini2008},
однако, по словам авторов, нулева€ вакуумна€ энерги€ намного более веро€тна, чем
наблюдаема€ величина тЄмной энергии.

 ƒетальный анализ антропного подхода к веро€тности того или иного значени€ вакуумной
энергии был проведен в работах~\cite{Hong2012,Hartle2013}.
’от€ авторы и  сход€тс€  во мнении, что антропные соображени€ говор€т в пользу наибольшей веро€тности  малой вакуумной энергии,
близкой к критической плотности энергии, всЄ же их результаты слегка различаютс€.
”читыва€ неопределенность оценок, можно говорить о хорошем их согласии.

ќтрицательна€ плотность энергии вакуума ведет к гравитационному прит€жению
и, чтобы избежать раннего сжати€ вселенной до того, как она состарилась до
современного возраста, необходимо выполнение услови€ $|\rho^{\rm vac}|< \rho_m^0$.

Ќесмотр€ на очевидные успехи в понимании возможности естественной
реализации антропного принципа, остаетс€ чувство неудовлетворенности теорией,
в которой нет возможности динамического вычислени€ основных параметров.
¬ этой св€зи хочетс€ вспомнить о проблемах фридмановской космологии, дл€
решени€ которых пришлось бы примен€ть антропный принцип, если бы не
было найдено красивое и экономное инфл€ционное решение, давшее чЄткое
количественное предсказание о спектре возмущений плотности.  ритику
антропного подхода можно найти в статье~\cite{kane-anti-anthrop}.
Ќо в физике любой подход имеет право на существование, пока и поскольку
он не противоречит опыту и хорошо установленным теори€м в области их
применени€.

 акое-то врем€ назад возлагались надежды на существование симметрии,
котора€ требовала бы нулевого значени€ Ћ€мбда-члена, как это происходит в
ненарушенной суперсимметрии. ѕохоже, эти надежды оказались тщетными.
—имметри€ должна реализовыватьс€ при достаточно низких энерги€х, наверн€ка,
ниже 100 ћэ¬, иначе будет невозможно компенсировать вакуумную энергию
в  ’ƒ~(\ref{rho-QCD}),
а при этих энерги€х физика хорошо известна и никакой
подход€щей симметрии там наверн€ка не существует. Ѕолее того, чтобы
получить наблюдаемое значение вакуумной (или вакуумо-подобной) энергии на уровне
$(10^{-12}$  √э¬$)^4$, мы должны сдвинутьс€ в область гораздо более низких,
чем 100 ћэ¬, энергий.

— точки зрени€, по крайней мере, одного из авторов этого обзора наибoлее
интересным представл€етс€ механизм динамической компенсации вакуумной
энергии, предложенный в работе~\cite{ad-nuff} в 1982 году. ѕредполагаетс€, что существует
новое лЄгкое или безмассовое поле,  $\phi$, таким образом
св€занное с гравитацией, что при наличии
вакуумной энергии возникает конденсат этого пол€, энерги€ которого компенсирует
исходную вакуумную энергию. Ётот механизм напоминает известное решение проблемы
нарушени€ CP-инвариантности аксионным полем. —амо предположение,
что обратна€ реакци€ системы на внешнее воздействие стремитс€ уменьшить это
воздействие, €вл€етс€ совершенно
общим. ќно известно в химии с XIX века и носит название принципа Ће Ўателье
(Le Chatelier).

»сходна€ иде€~\cite{ad-nuff} и множество последующих
предложений (см. обсуждени€ в обзорах~\cite{rho-vac-rev} --\cite{rho-vac-rev13})
опирались на использование скал€рного пол€, однако не исключены и пол€ высших спинов,
например, векторное~\cite{ad-vector}  или тензорное~\cite{ad-tensor}.
Ќезависимо от конкретной
реализации, эти предсказани€ имеют общие и весьма привлекательные черты. ¬о-первых, вклад
энергии-импульса $\phi$ в космологическую динамику приводит к тому, что исходное
экспоненциальное
расширение переходит в степенное. ¬о-вторых, компенсаци€ вакуумной энергии полем
$\phi$ не €вл€етс€ полной, а лишь с точностью до величины пор€дка ${\rho_c (t)}$. ¬-третьих,
``недокомпенсированное'' значение вакуумной энергии может иметь необычное уравнение
состо€ни€, в частности привод€щее к ускоренному расширению. “аким образом, подобный механизм
в принципе решает не только проблему компенсации вакуумной энергии от гигантской величины,
типичной дл€ физики элементарных частиц, до космологически малых значений, но и позвол€ет
решить так называемую проблему совпадени€, т.е.
объ€снить близость величины нескомпенсированной тЄмной энергии  $\rho_{\rm DE}$
к полной, завис€щей от времени плотности энергии во вселенной, ${\rho_c (t)}$ в полном
соответствии с тем, что мы видим на небе. ¬ этом смысле наличие космологической тЄмной
энергии было предсказано в 1982 году~\cite{ad-nuff} задолго до ее астрономического
наблюдени€.   сожалению, до насто€щего времени, несмотр€ на многочисленные попытки,
не удалось найти реалистической космологической модели, включающей этот механизм.

ѕо-видимому, впервые предположение, что вакуумна€ энерги€ может зависеть от времени
и быть близкой по величине к космологической плотности энергии, было высказано
ћатвеем Ѕронштейном~\cite{mb}. ќднако, вз€тые буквально модели с
${{ \Lambda = \Lambda (t)}}$
далеко не безобидны, так как ковариантное сохранение левой части уравнений Ёйнштейна
требует,
как известно, $\Lambda = \mbox{const}$, см. обсуждение сразу после уравнени€~(\ref{Lambda}).
„тобы компенсировать
несохранение вакуумной энергии, ввод€т равнoе по величине, но противоположное по знаку
несохранение  тензора энергии-импульса материи, $T_{\mu\nu}$. ќднако при этом тер€етс€
предсказательна€ сила теории, так как подобные модели говор€т лишь о том, чему
должна быть равна ковариантна€ дивергенци€:
\be
D_\mu T^\mu_\nu = -( m_{Pl}^2/8\pi) \,\partial_\nu \Lambda \,,
\label{DT}
\ee
но ничего не могут сказать о величине  $T_{\mu\nu}$. ѕо подобной причине работа Ѕронштейна была
подвергнута Ћандау довольно жЄсткой критике.

Ќапомним, что в OTO тензор энергии-импульса вычисл€етс€ как функциональна€
производна€ действи€ материи по метрике, так что он автоматически сохран€етс€ в силу принципа
общей ковариантности. ќтказ от последнего привел бы, вообще говор€, к ненулевой массе
гравитона в противоречии с наблюдени€ми. (—м. работу \cite{KoyamaSakstein} об анализе модифицированных
теорий гравитации на масштабах звЄзд и галактик  с учЄтом возможного нарушени€ механизма
¬айнштейна \cite{Vainshtein}, который мог бы помочь избежать проблем с конечной массой гравитона.)
ƒруга€ возможность -- это революционна€ модификаци€
теории гравитации, например, обращение к неметрическим теори€м.

ћожно, однако, избежать всех этих проблем, если ввести в теорию новое
легкое или безмассовое поле, которое имеет приближенное уравнение состо€ни€ ${P \approx -\rho}$,
отвечающее квазипосто€нной вакуумо-подобной плoтности энергии.

ѕерва€ модель динамического сокращени€ вакуумной энергии, предложенна€ в
работе~\cite{ad-nuff}, опиралась на безмассовое скал€рное поле, неминимально
св€занное с гравитацией и удовлетвор€ющее уравнению движени€ вида:
\be
\ddot \phi + 3H \dot \phi +U'(\phi, R) = 0.
\label{ddot-phi-R}
\ee
ѕотенциал $U(\phi, R)$ был выбран в простейшей форме, ${ U = \xi R \phi^2/2}$.
Ћегко видеть, что при ${\xi R<0}$ это уравнение в пространстве де —иттера имеет неустойчивые,
экспоненциально растущие со временем, решени€, так как при посто€нной кривизне $R$ квадрат
эффективной массы пол€ $\phi$ отрицателен. јналогичное положение имеет место в модели
’иггса, когда длинноволновые состо€ни€ с $\phi = 0$ оказываютс€ неустойчивыми.

— ростом $\phi$ уже нельз€ пренебрегать его воздействием на космологическую эволюцию и,
как легко проверить, исходно экспоненциальное расширение, $a(t) \sim \exp (H_v t)$,
асимптотически перейдет в степенное:
\be
\phi \sim t\,, \,\,\,\, a(t) \sim t^\beta\,,
\ee
где $\beta$ - некотора€ посто€нна€, выражающа€с€ через параметры модели.
“аким образом, обратна€ реакци€ $\phi$ на космологическое расширение приводит к
превращению экспоненциального закона расширени€ во фридмановский,
несмотр€ на наличие исходно ненулевой вакуумной энергии.

  числу недостатков этой простой модели, основанной на скал€рном поле, следует отнести
то, что тензор энергии-импульса  этого пол€ не пропорционален
метрическому тензору,
\be
T_{\mu\nu} (\phi) \neq \tilde\Lambda g_{\mu\nu}\,.
\label{T-mu-nu-of-g}
\ee
и потому вакуумна€ энерги€ не занул€етс€ даже и асимптотически.
»зменение режима расширени€ достигаетс€ благодар€ ослаблению гравитационного
взаимодействи€, константа которого падает со временем, в  начале экспоненциально, а
 затем, как квадрат времени:
\be
G_N \sim 1/t^2 .
\label{G-of-t}
\ee
≈сли бы подобное изменение $G_N$ имело место в ранней вселенной, а позже как-то
стабилизировалось, то этот механизм мог бы объ€снить иерархию гравитационного и
электрослабого масштабов~\cite{notari}.

”спешные предсказани€ теории первичного нуклеосинтеза требуют, чтобы $G_N$
изменилась не более, чем на  ~10\% от момента, когда возраст вселенной был около
одной секунды, до нашего времени \cite{Rothman1982,Accetta1990,Copi2004,Bambi2005}.
≈ще более сильные ограничени€ на переменность  $G_N$ следуют из анализа прихода
сигнала от пульсара J1713+0747 (метод так называемого пульсарного тайминга -- pulsar timing):
$\dot G_N/G_N = (-0.6\pm 1.1)\cdot 10^{-12} $  лет$^{-1}$  (99.7\%CL),
что по крайней мере
в 30 раз медленнее, чем скорость расширени€ ¬селенной~\cite{Zhu2015}. “ак что
 стабилизаци€ изменени€ $G_N$ со временем €вно необходима.
ќднако имеютс€ указани€ на некоторую мистическую переменность $G_N (t)$ \cite{Anderson2015,Schlamminger2015}.

  насто€щему времени опубликовано несколько дес€тков работ, где обсуждаютс€
различные механизмы динамической компенсации вакуумной энергии скал€рным полем.
–анние работы приведены в~\cite{fujii} --\cite{fujii10}, а также
в~\cite{earlier-papers,earlier-papers1,earlier-papers2,earlier-papers3}, где
в основном речь идет о феноменологическом описании тЄмной энергии скал€рным
полем. —сылки на более поздние статьи можно найти
в обзорах~\cite{rho-vac-rev} --\cite{rho-vac-rev13}.   сожалению, ни один из предложенных механизмов
нельз€ считать полностью удовлетворительным.

—. ¬айнбергом была сформулирована теорема о невозможности (no-go theorem)
естественной компенсации
вакуумной энергии скал€рным полем (см. статью~\cite{rho-vac-rev}),
основанна€ на
том, что дл€ компенсации необходимо выполненине двух условий: равенства  нулю
суммы $\rho_{\rm tot}=\rho^{\rm vac} + \rho_\phi$ и занулени€ в этой точке
производной потенциала $U'(\phi, R)$.
ѕоследнее условие означает, что то значение $\phi$, где исчезает полна€ вакуумо-подобна€ энерги€,
$\rho_{\rm tot} = 0$, €вл€eтс€ одновременно и решением уравнени€ движени€ (\ref{ddot-phi-R}) с
$\phi = \mbox{const}$. ќднако, как это нередко бывает в физике, теоремы о невозможности удаетс€ обойти,
изменив услови€ задачи. ¬ частности, можно обратитьс€ к пол€м с ненулевым спином (см. ниже) или
же, модифицировав форму взаимодействи€ $\phi$ с гравитацией, как это
было сделано в
работе~\cite{muk-rand} и более детально рассмотрено в работах~\cite{ad-kawa-2005,ad-kawa-2003}. ќсновна€ иде€ этих работ
состоит в том, чтобы модифицировать кинетический член скал€рного пол€, введ€ в него коэффициент
обратно пропорциональный {квадрату кривизны}:
\be
  A= \int d^4 x\sqrt{g} \left[ -\frac{1}{2} (R+2\Lambda)
  + F_1(R)\right.   + {{\left.
 { \frac{D_\mu \phi D^\mu \phi} { 2\,R^2}}\,
- U(\phi, R) \right] }}\,.
\label{A-phi-R}
\ee
«десь используетс€ система единиц, в которой $m_{Pl}^2/8\pi =1$.

 —оответствующее уравнение движени€ дл€ ${\phi}$ имеет вид:
\be
D_\mu\left[ D^\mu \phi\,\left(\frac{1}{R}\right)^2\right] + U'(\phi) = 0,
\label{D-mu2-phi}
\ee
которое  в FRW-метрике дл€ пространственно однородного пол€ $\phi (t)$ сводитс€ к:
\be
\left( \frac{d}{dt} + 3 H \right)\left(\frac {\dot \phi}{R^2}
\right) + U' (\phi) =0,
\label{ddot-phi-U}
\ee
а уравнени€ Ёйнштейна приобретают, в частности, дополнительное слагаемое пропорциональное
высшим производным пол€ $\phi$:
\be &&
 R_{\mu\nu} - \frac{1}{2}  g_{\mu\nu}  R
 - 4C_1 \left( R_{\mu\nu} -\frac{1}{4} g_{\mu\nu} R + g_{\mu\nu} D^2  -  D_\mu D_\nu \ \right) R
 -\frac {D_\mu\phi D_\nu \phi}{R^2} + \nonumber\\
&& \frac{\left( D_\alpha \phi \right)^2}{2 R^2}
\left({g_{\mu\nu} } + \frac{4 R_{\mu\nu}}{R}\right)
-g_{\mu\nu} \left[ U(\phi) +\rho_{vac} \right]
+ 2\left(g_{\mu\nu} D^2 -D_\mu D_\nu\right)
 \frac{\left( D_\alpha \phi\right)^2}{R^3} =
 T_{\mu\nu},
\label{ein-eq}
\ee
где $T_{\mu\nu}$  - тензор энергии-импульса вещества и
$(D \phi)^2 \equiv D_\alpha \phi D^\alpha \phi$  и дл€ простоты мы положили $F_1(R) = C_1 R^2$.

¬з€в след по индексам $\mu$ и $\nu$, получим
\be
- R + 3 \left( \frac{1}{R}\right)^2 \left( D_\alpha \phi \right)^2
- 4\left[ U(\phi) +\rho^{\rm vac}\right]
- 6 D^2 \left[ 2C_1 R - \left(\frac{1}{R}\right)^2
\frac{\left( D_\alpha \phi\right)^2}{R}\right] =
T^{\mu}_{\mu}.
\label{trace1}
\ee
¬ пространственно однородном случае ковариантный оператор ƒ'јламбера, дeйствующий на скал€р
имеет вид $D^2 = d^2/dt^2 + 3 H d/dt $, причем параметр ’аббла св€зан со скал€ром кривизны соотношением
\be
R = -6 \left( 2 H^2 +\dot H \right).
\label{r-of-h1}
\ee
”равнени€ (\ref{ddot-phi-U},\ref{trace1},\ref{r-of-h1}) полностью описывают нашу систему в отсутствии обычного вещества.
 ак показано в работax~\cite{ad-kawa-2005,ad-kawa-2003}
эти уравнени€ даже при исходно ненулевой вакуумной энергии
имеют асимптотическое решение: \\
\be
H &=& h/t, \nonumber \\
R &=& r/t^2 , \nonumber \\
\left[ U (\phi) - \rho_{vac} \right]  &\sim & 1/t^2 .
\label{U-rho-vac}
\ee
јмплитуда пол€ $\phi$ стремитс€  к значению, при котором потенциал $U(\phi)$ компенсирует исходно ненулевую вакуумную
энергию и экспонециальный закон расширени€ переходит во фридмановский.
—ледовательно, это решение компенсирует вакуумную энергию, позвол€€ избежать запрет ¬айнберга.
–ешение обладает и другими привлекательными свойствами, в частности, можно показать, что отвечающий ему
параметр ’аббла равен ${ H=1/(2t)}$, как это имеет место в
реальной космологии на стадии доминантности рел€тивистского вещества. ќднако это значение $H$ никак не св€зано
с тем, какое вещество содержитс€ во вселенной, тогда как в обычной космологии $H\sim \sqrt{\rho}$.

 роме того, оказываетс€, что такие ``хорошие'' решени€ неустойчивы~\cite{ad-kawa-2003}.
ƒл€ приведенного выше  решени€ (\ref{U-rho-vac}) последнее слагаемое в правой части уравнени€  (\ref{trace1}),
содержащее $D^2$, оказываетс€  малым, несущественным.  ќднако при изучении устойчивости решений того или
иного уравнени€ члены с высшими производными должны быть сохранены, так как они коренным образом вли€ют на
устойчивость. ¬ цитированной выше работе было показано, что именно члены  с высшими производными
ведут к сингул€рному решению с $R$ и $H$  обращающимис€ в бесконечность за конечное и короткое врем€. ¬ общем
решении знак кривизны измен€етс€ с исходно отрицательного на положительный, причем параметр ’аббла
стремитс€ к минус бесконечности, привод€ к коллапсу ¬селенной, независимо от наличи€ обычного вещества.

Ѕолее общее действие вида
\be
A = \int d^4 x \sqrt{-g}
\left[-\frac{m_{Pl}^2}{16\pi}(R+2\Lambda )
+ F_2 (\phi, R) D_\mu \phi D^\mu \phi
+ F_3 (\phi, R) D_\mu \phi D^\mu R - U(\phi, R) \right]
\label{A-gnrl}
\ee
пока не исследовано.  роме того, в принципе в действие можно бы включить члены,
завис€щие от ${R_{\mu\nu}}$ или ${R_{\mu\nu\alpha\beta}}$ в надежде получить
реалистическую
модель. ќднако, наличие высших тензоров кривизны, помимо скал€ра, вообще говор€,
должно приводить к тахионам или духам в теории гравитации, которых счастливо
избегает ќ“ќ или теории с действием, завис€щим лишь только от $R$, несмотр€ на
наличие в действии вторых производных.

“еорему ¬айнберга удаетс€ обойти также в модели
динамической компенсации с двум€ скал€рными
пол€ми~\cite{two-scalars-fujii,two-scalars-fujii1,two-scalars-fujii2,two-scalars-rub,two-scalars-hebecker}. ќдно из
двух скал€рных полей $\phi$, рассмотренных в модели работы~\cite{two-scalars-rub}, имеет обычный кинетичeский член, а второе - сингул€рный при $\phi \rar 0$:
\be
L_{\rm kin} =  \frac{1}{2} \partial_\mu \phi \partial^\mu \phi +
\frac{\mu^{2p}}{2\phi^{2p}} \partial_\mu \chi \partial^\mu \chi
\label{L-kin-phi-chi}
\ee
и в этом смысле напоминает более позднюю работу~\cite{muk-rand}, заставл€€
релаксировать поле $\phi$ вблизи нул€. “очнее, надо бы сказать наоборот: в
работе~\cite{muk-rand} используетс€ эта иде€ работы~\cite{two-scalars-rub}.
Ёто приводит к довольно заметному сходству обеих работ как в достоинствах, так
и недостатках.

¬озможно, компенсирующие пол€ с ненулевым спином смогут решить проблему вакуумной
энергии успешнее, чем скал€рное поле. ћодель компенсирующего векторного пол€, ${V_\mu}$,
была предложена в работе ~\cite{ad-vector} с лагранжианом, выбранном в форме:
\be
{ L_1} = \eta \left[ F^{\mu\nu} F_{\mu\nu} /4 + (V^\mu_{;\mu})^2\right]
+\xi R m^2 \ln \left( 1 +\frac{V^2}{m^2}\right)\,.
\label{L1-of-V}
\ee
¬ этой теории временна€ компонента  пол€, $V_t$, оказываетс€ неустойчивой в пространстве
де —иттера и растет со временем:
\be
V_t \sim t + c/t\,.
\label{V-of-t}
\ee
ƒоминирующий член в тензоре энергии-импульса, отвечающего этому решению,  пропорционален
метрическому тензору и имеет вид:
\be
T_{\mu\nu} (V_t) = -\rho^{\rm vac} g_{\mu\nu} + \delta T_{\mu\nu}\,,
\label{T-mu-nu-vector}
\ee
где $\delta T_{\mu\nu} $ асимптотически стремитс€ к нулю.

¬ этой модели гравитационна€ посто€нна€ зависит от времени, но лишь логарифмически, что могло
бы и не противоречить имеющимс€ ограничени€м на вариацию $G_N$. ќднако схема
обладает типичным
недостатком известных компенсационных механизмов, а именно отсутствием св€зи между
темпом космологического расширени€, $H$, и плотностью эенергии материи
во вселенной, как это имеет место в обычной космологии, см. (\ref{H2}).

ƒругую интересную возможность открывает безмассовое тензорное поле второго
ранга, $ S_{\mu\nu}$~\cite{ad-tensor},  с лагранжианом,
который содержит лишь кинетические члены:
\be
{\cal L}_2 =
\eta_1 S_{\alpha\beta;\gamma} S^{\alpha\gamma;\beta}
+\eta_2 S^\alpha_{\beta;\alpha} S^{\gamma\beta}_{\> \>;\gamma}
+\eta_3 S^\alpha_{\alpha;\beta} S_\gamma^{\gamma;\beta}\,.
\label{L2}
\ee
»ными словами, это свободное поле с минимальным гравитационным взаимодействием.
 ак известно, безмассовое поле должно взаимодействовать
с сохран€ющимс€ источником во избежание инфракрасной катастрофы. — другой стороны,
единственным сохран€ющимс€ источником тензорного пол€ второго ранга €вл€етс€ тензор
энергии-импульса, см. например книгу~\cite{weinberg-2}. ќтсюда делаетс€ вывод, что единственным
безмассовым полем спина два может быть лишь только гравитон. ќднако дл€ свободных полей
это ограничение неприменимо,  и безмассовое тензорное поле второго ранга не исключено.
ќтметим еще, что  лагранжиан (\ref{L2}) зависит только от производных пол€ и поэтому
инвариантен при сдвиге на ковариантно посто€нный тензор второго ранга. ¬ силу этого квантовые
радиационные поправки не должны генерировать массу дл€ $ S_{\alpha\beta}$. ¬ этом состоит
важное отличие от скал€рного пол€, дл€ которого, вообще говор€, масса может возникнуть в результате
радиационных поправок.

¬ плоской метрике ‘ридмана-–обертсона-”окера компоненты пол€ $ S_{\alpha\beta}$
удовлетвор€ют следующим уравнени€м:
\be
 (\partial_t^2 + 3H\partial_t - {6H^2}) S_{tt} -2H^2 s_{jj} = 0\,, \\
 \label{dt2tt}
 (\partial_t^2 + 3H\partial_t -{6H^2}) s_{tj} = 0\,, \\
\label{dt2tj}
 (\partial_t^2 + 3H\partial_t -{2H^2}) s_{ij} -2H^2 \delta_{ij} S_{tt} = 0\,,
\label{dt2ij}
\ee
где  ${ s_{tj} = S_{tj}/a(t)}$ and ${ s_{ij} = S_{ij}/a^2(t)}$.

Ћегко видеть, что временно-временна€ компонента ${ S_{tt}}$, и изотропна€
пространственно-пространственна€ компонента, ${ S_{ij}\sim \delta_{ij}}$, нестабильны
и растут со временем в метрике де —иттера. “ензор энергии-импульса этих компонент
стремитс€ к посто€нной величине, пропорциональной метрическому тензору, и
компенсирует исходную вакуумную энергию с точностью до членов пор€дка
$m_{Pl}^2/t^2$. ќднако, хот€ это немедленно и не очевидно, константа св€зи  гравитации
с веществом мен€етс€ со временем, согласно аргументам, представленным в
работе~\cite{rub-G-of-t}.

«десь уместно сделать также следующее общее замечание. ќбычно при построении теории полей
со спином накладываютс€ дополнительные услови€ занулени€ компонент с низшими спинами.
“ак например, в теории векторного пол€ устран€етс€ временна€ компонента вектора, котора€
€вл€етс€ трехмерным скал€ром. ¬ теории тензорного пол€ второго ранга устран€етс€ не только
трехмерный скал€р, но и трехмерный вектор. ќднако в рассмотренном выше примере ситуаци€
противоположна€, когда трехмерно скал€рные компоненты четырехмерного тензора остаютс€
физическими, а устранение высших спинов €вл€етс€ необходимым. Ќе€сно, можно ли
непротиворечиво сформулировать подобную теорию, но в литературе имеетс€ пример
калибровочной теории скал€рного пол€, которое описываетс€ временной компонентой
вектора ${{ V_t}}$, а компоненты со спином единица исключены~\cite{notoph}.

Ќедавно в серии работ~\cite{klink,klink1,klink2,klink3} были получены многообещающие результаты
дл€ механизма динамической компенсации вакуумной энергии в теории нескольких
св€занных векторных полей, временна€ компонента которых растет пропорционально
времени, компенсиру€ вакуумную энергию. ¬ такой модели, по-видимому, удаЄтс€
избежать отмеченных выше проблем с ростом гравитационной посто€нной, хот€
``вечна€'' проблема отсутстви€ канонической зависимости параметра ’аббла от
плотности энергии обычной материи в ранние времена, например, в период
первичного нуклеосинтеза, по-видимому, остаЄтс€.

ќписанные выше механизмы динамической компенсации вакуумной энергии рассматривались большей
частью в рамках инфл€ции, котора€, как мы уже говорили, дала успешное
количественное предсказание спектра возмущений плотности.
“ем не менее, пока нельз€ говорить об инфл€ции как о факте, это всЄ ещЄ
гипотеза, кроме того сама инфл€ци€ порождаетс€ разными способами, как например, в описанном выше механизме~\cite{two-scalars-rub}.
ѕомимо инфл€ции развиваютс€ и другие сценарии, не столь всеобъемлющие, но которые претендуют на объ€снение малой величины
сегодн€шней космологической константы, см., например, \cite{2006Sci...312.1180S}, где динамический механизм
включаетс€ в сценарий ``циклической вселенной''.

”пом€нем в заключение еще одну космологическую проблему, св€занную с вакуумной
энергией. »менно, современное значение $\rho^{\rm vac}$ очень близко к величине
плотности энергии материи, отлича€сь от него лишь в два раза, несмотр€ на
существенное различие их эволюции в ходе расширени€ вселенной. ≈стественного
понимани€ этой близости мы не имеем, как, впрочем, нет понимани€ близости
плотностей энергии барионов и тЄмной материи.

¬ заключение этого раздела подчеркнЄм, что представл€етс€ весьма естественным
наличие глубокой св€зи между проблемами компенсации вакуумной энергии и
физикой тЄмной энергии. —корее всего, решение этих проблем возможно лишь
в совокупности.


\section{ƒанные в пользу космического ускорени€ \label{s-observ}}

\subsection{»стори€ \label{ss-history}}
ѕримерно в 1920-е годы физики стали всерьЄз обсуждать модели вселенных на основе общей
теории относительности (ќ“ќ), в том числе и модели дл€ вселенной, конечной по объЄму.
ѕервыми общерел€тивистские модели вселенной  создали Ёйнштейн, де —иттер, ‘ридман.
«абавно, что в это врем€ ещЄ не были известны даже по пор€дку величины
размеры ћлечного ѕути.
Ќе было даже твЄрдо установлено, существуют ли вне ћлечного ѕути другие
галактики 
\cite{GreatDebate}.

„то знали астрономы о размерах √алактики и о
нашем положении в ней к 1920-м годам?
”же в 19-м веке начали аккуратно измер€ть рассто€ни€ и обнаруживать собственные
движени€ близких звЄзд (пон€тие о сфере неподвижных звЄзд давно исчезло).
ѕо подсчЄтам слабых звЄзд в разных направлени€х сложилось представление,
что наше —олнце близко к центру √алактики -- ведь плотности звЄзд вдоль
полосы ћлечного ѕути примерно одинаковы.
» только к 1920-му году ’арлоу Ўепли правильно указал на центр √алактики
в направлении созвезди€ —трельца -- вокруг него скучивались шаровые звЄздные
скоплени€.
Ўепли далеко отодвинул —олнце от центра √алактики. —ейчас мы знаем, что до центра
примерно 8~кпк (около 25 тыс€ч световых лет)~\cite{Boehle2016}, а Ўепли даже сильно завысил это рассто€ние.
Ѕольшинство же астрономов придерживались точки зрени€  Єртиса, который давал такую оценку:
``Ќаша √алактика, веро€тно, не более, чем 30 тыс€ч световых лет в диаметре'',
причЄм —олнце полагали вблизи центра √алактики.

26 апрел€ 1920 года в Ќациональной академии наук —Ўј в ¬ашингтоне состо€лс€
исторический диспут между Ўепли и 
 Єртисом -- ``¬еликий спор''.
 Єртис отстаивал малую шкалу рассто€ний в √алактике, а Ўепли большую.

 ¬ этих дебатах  Єртис приводит некоторые оценки,
вз€тые из статьи Ћундмарка:
``¬ольф -- около 14000 световых лет в диаметре;
Ёддингтон -- около 15000 световых лет;
Ўепли (1915) -- около 20000 световых лет;
Ќьюком -- не меньше 7000 световых лет, а позднее -- возможно 30000 световых лет в диаметре
и 5000 по толщине;
 аптейн -- около 60000 световых лет''.

ќценка же Ўепли на 1920 год такова: ``√алактика имеет примерно 300 тыс€ч световых лет в диаметре.''

¬ тот момент никто никого не убедил.
“олько в 1930-е годы стало €сно, что Ўепли в вопросе o положении —олнца
вдали от центра был прав.
{(—овременные данные по поводу размера √алактики: у  Єртиса было $\sim$30 тыс. св. лет,
у Ўепли $\sim$300 тыс. св. лет. Ќасто€щее значение $\sim$100 тыс. св. лет $\approx 30$~кпк.
ќни ошиблись примерно одинаково, но в разные стороны)}.
ѕравоту Ўепли удалось установить, только когда астрономы (Ћиндблад, ќорт)
разобрались в асимметрии звЄздных движений вокруг нас, во вращении √алактики.

Ќо Ўепли ошибс€ в другом -- он упорно считал спиральные туманности, такие
как ћ31 (јндромеда), маленькими газовыми образовани€ми, лежащими внутри
√алактики или совсем р€дом с ней.
ј  Єртис упорно доказывал, что эти спирали -- такие же звЄздные миры как ћлечный
ѕуть. » оказалс€ прав (только размер галактик занижал).

“еперь можно сопоставить картину той маленькой стационарной ¬селенной, модель которой
пыталс€ строить Ёйнштейн ещЄ до 1920-го года, с реальностью, о которой
мы знаем теперь.
»зменились не только наблюдаемые масштабы (в миллионы раз!).
–одилось пон€тие о нестационарной ¬селенной.

√рафик зависимости красных смещений объектов от их
рассто€ний называют диаграммой ’аббла, по его работе 1929 года~\cite{Hubble1929}.
Ќа основе своей диаграммы ’аббл установил закон роста рассто€ни€ до галактики с красным смещением.

Ќа самом деле, первым этот закон открыл за восемь лет до того немецкий астроном
 арл ¬ирц (Carl Wirtz, 1876---1939), которого 
јлан —ендидж (Alan Sandage) назвал ``≈вропейским ’абблом без телескопа'' \cite{Sandage1995}
{(—ендидж был PhD студентом Ѕааде, на ’аббла он работал с 1950 по 1953 год)}.
¬ирц был насто€щим пионером наблюдательной космологии и в 1921 г.
по данным дл€ 29 спиральных галактик обнаружил, что чем
дальше галактика, тем больше еЄ красное смещение \cite{Wirtz1922}
(опубликовано в июне 1922~г.). 
  сожалению, из-за т€жЄлой жизни в √ермании в то врем€
его исследовани€ не получили поддержки.
¬ 1936~г., когда ’аббл  был уже прославлен на весь мир,
¬ирц сделал отча€нную попытку напомнить о своЄм приоритете в журнале
Zeitschrift f\"ur Astrophysik \cite{Wirtz1936},  
но всЄ равно был практически забыт.
“ем не менее, авторы обзора~\cite{SeitterDuerbeck1999} назвали ¬ирца ``пионером космических размеров''.

ќбъективно и  кратко истори€ открыти€ расширени€ ¬селенной изложена в статье~\cite{vandenBergh2011}.
–оль таких наблюдателей как ¬есто —лайфер освещена в~\cite{Raifeartaigh2012}.

¬ан ден Ѕерг \cite{vandenBergh2011} пр€мо пишет:
``ћиф о том, что расширение ¬селенной было обнаружено ’абблом, впервые был распространен ’ьюмасоном (1931) \cite{Humason1931}.
»стинна€ природа этого открыти€ оказалась более сложной и интересной.''

—огласно исследованию истории этого открыти€~\cite{Pruzh2015,Pruzh2016} в 1927 году,
∆орж јнри-ƒжозеф Ёдуард Ћеметр (1894-1966) опубликовал работу
под названием ``ќднородна€ ¬селенна€ посто€нной массы с растущим
радиусом, с учЄтом лучевой скорости внегалактических туманностей''
в јнналах Ќаучного общества Ѕрюссел€~\cite{Lemaitre1927}.
(Aнглийский перевод французского оригинала по€вилс€ позже в~\cite{Lemaitre1931}.)
¬ этой работе он нашел 
линейное соотношение между скоростью удалени€ галактики и рассто€нием до неЄ.
Ћеметр был первым, кто получил численное значение параметра ``’аббла'' в этом соотношении,
св€зывающем  скорость ($ v $) и рассто€ние до галактики ($ D $).
—егодн€ этот параметр известен как $ H_0 $ в равенстве $ v = H_0 D $.
»спользу€ измерени€ красного смещени€, проведЄнные ¬есто ћелвиным —лайфером (1875-1969) и √уставом —тромбергом (1882-1962), опубликованные в 1925 году,
и оценки рассто€ний ’аббла (1926)~\cite{Hubble1926} дл€ 42 галактик, Ћеметр получил значение $H_0=625$~км/с/ ћпк.


 огда ¬ирц, Ћеметр и ’аббл обнаружили своЄ соотношение красное смещение -- рассто€ние,
они не имели надЄжного способа измерить рассто€ни€ до галактик.
ќни только предположили зависимость рассто€ни€ от размеров и светового потока галактик,
причЄм ¬ирц сформулировал закон только качественно,
а Ћеметр и ’аббл недооценили фактические значени€ удалени€
 галактик почти в 10 раз.
Ќапомним, что ¬ирцу ещЄ было вовсе не €сно -- вне или внутри ћлечного ѕути
наход€тс€ спиральные туманности, он мог ошибатьс€ в рассто€нии ещЄ сильнее.
“ем не менее, факт роста красного смещени€ с рассто€нием ¬ирцем был установлен верно.

\subsubsection{»стори€ по€влени€ намЄков на л€мбда-член в наблюдени€х}

ƒо работ по сверхновым Ia дл€ всех трЄх указанных выше типов вселенных
обычно всерьЄз  рассматривали только те варианты,
где сейчас происходит торможение, а не  ускорение  под действием взаимного т€готени€ вещества во
¬селенной.
Ѕольшинство астрономов не  ожидало увидеть данные, которые можно истолковать как ускорение
расширени€, хот€ такие модели вселенной давно уже были построены  де —иттером (дл€ пустоты)
и ‘ридманом (дл€ непустой вселенной) с Ћ€мбда-членом.
‘ридман \cite{Friedman1922} уже в 1922 году во всех детал€х рассмотрел судьбу
ускоренно расшир€ющейс€ вселенной при положительном $\Lambda$.
(¬ той же статье он рассматривал и космологические модели с нулевым и отрицательным
$\Lambda$.)

»ногда, при по€влении особенностей, например, в распределении квазаров
по $z$ возникала мода на модели с Ћ€мбда-членом. Ёти модели выдвигались
».—.Ўкловским \cite{Shklovsky1967} и Ќ.—. ардашЄвым \cite{Kardashev1967} в 1967 г.
Ќо потом особенности ``рассасывались'' и мода проходила.
—ерьЄзные аргументы в пользу реальности Ћ€мбда-члена приводила
Ѕеатриса “инсли с коллегами в 1975-1978 гг. на основе сопоставлени€ возраста
¬селенной и шаровых звЄздных скоплений \cite{GunnTinsley1975,Tinsley1978}.
“огда казалось, что  шаровые скоплени€ старше ¬селенной в модел€х без ускорени€.
Ќо это противоречие тоже ``рассосалось''.

¬ажную роль сыграли статьи M.Fukugita с соавторами
\cite{Fukugita1990,FukugitaHogan1990,Fukugita1993}.
ќни подчеркнули, во-первых, что параметр $H_0$ не может быть столь малым как
50~км/с/ћпк, что получили Sandage и Tamman \cite{Sandage1974}, а тогда нужный возраст
¬селенной может дать положительное $\Lambda$.
 роме того в работе \cite{Fukugita1990} было показано, что подсчЄты слабых
галактик требуют ``ощутимого'' (sizable) значени€ $\Lambda$.

ѕоначалу эти утверждени€ про $H_0$ и $\Lambda$ многие астрономы-профессионалы
считали еретическими, но в 1998 г. диаграмма ’аббла дл€ SN~Ia показала, что многие
из них наход€тс€ от нас чуть
дальше, чем в простых модел€х.
”скорение расширени€ ¬селенной (например, за счЄт Ћ€мбда-члена)
могло бы объ€снить увеличение рассто€ни€, и это объ€снение стало общеприн€тым.
¬ следующем разделе рассмотрим, как измер€ют рассто€ни€ с помощью сверхновых.

\section{ ƒанные по сверхновым  и барионным акустическим осцилл€ци€м \label{s-SNBAO} }

\subsection{Ёлементарна€  осмографи€: рассто€ни€ во ¬селенной}

≈сли правильно отождествить
длины волн различных спектральных линий в спектрах
далЄких объектов, и если известны их лабораторные значени€, то
можно измерить красное смещение, которое определ€етс€ как отоношение
сдвига  частот: $ z = \omega_1/\omega_0 -1$,
јстрономы умеют делать это с очень высокой точностью дл€ близких объектов и с точностью
в несколько процентов дл€ далЄких.
«начит, можно сказать, насколько длины волн линий
изменились с момента испускани€ $t_1$, до
момента наблюдени€ $t_0$.
ќтсюда и из
\begin{equation}
   \omega a(t) = \mbox{const} \; .
\label{oma_eq}
\end{equation}
получаем отношение масштабных факторов  в эти два момента:
$$
\frac{a(t_0)}{a(t_1)}= 1+z \; .
$$

 ак же измер€ть рассто€ни€? Ёта задача сложнее.

ћожно дать формальное определение \textit{собственного} (по-английски \textit{proper})
рассто€ни€.
ѕусть у нас координата $r=0$, у далекой галактики $r=r_1$.
“огда
\begin{equation}
  D_{\rm prop}(t) = \int_{0}^{r_1}\sqrt{g_{rr}}dr= a(t) \int_{0}^{r_1} \frac{dr}{\sqrt{1-r^2}}  \; ,
\label{dprop}
\end{equation}
-- соответствует рассто€нию, которые измерили бы наблюдатели, сид€щие достаточно
плотно в расшир€ющейс€ вселенной между $r=0$ и $r_1$, в \emph{один момент
космического времени~$t$}.
ƒл€ того, чтобы осуществить такое измерение, им пришлось бы заранее сговоритьс€.
ѕоэтому непосредственно от $D_{\rm prop}$ не очень много пользы.

¬ведЄм вспомогательную величину
\begin{equation}
   D_{\rm flat}(t) \equiv a(t) r  \; .
\label{dflat}
\end{equation}
Ќазовем $D_{\rm flat}$ `плоским' (flat) рассто€нием.
“олько при нулевой кривизне $k=0$ эта величина $D_{\rm flat}$ была бы измерена  в момент координатного
времени $t$ набором наблюдателей с жЄсткими линейками между точками с радиальными
координатами $0$ и $r$ и совпала бы с собственным рассто€нием. 
(Ќапомним, что $r$ безразмерно!)
ћы увидим ниже, что $D_{\rm flat}(t)$ войдет в измеримые рассто€ни€ и в общем случае
$k \ne 0$.

¬ теоретической космологии важным €вл€етс€ определение
\emph{сопутствующего} (co\-mo\-ving) рассто€ни€
$$
D_{\rm com}=a(t_0)\chi \ ,
$$
где $\chi$ --- фридмановска€ радиальна€ координата, вход€ща€
в выражение (\ref{ds2-friedman}).
“ам ‘ридман использовал $\chi$ дл€ замкнутого мира, когда $k=1$.
¬ общем случае можно написать:
\beq
  r= \left\{
  \begin{array}{ll}
    \sin \chi \quad &\mbox{при} \quad k=1, \\
    \chi \quad  &\mbox{при} \quad k=0, \\
    \sh \chi  \quad &\mbox{при} \quad k=-1 .
 \end{array}
 \right.
\enq

Mы не можем непосредственно измерить ``рассто€ни€'' $D_{\rm prop}$,  $D_{\rm flat}$ и
$D_{\rm com}$  до далЄких объектов в расшир€ющейс€ вселенной с помощью линеек
или радиолокации.
¬место этого в космографии ввод€т различные определени€ рассто€ний, завис€щие
от тех измерений, которые можно реально провести (например, рассто€ние углового размера  ---
по измерению углового поперечника стандартной линейки).
ѕодробности прекрасно изложены ¬ейнбергом \cite{Weinberg1972,Weinberg2008}.
ћы следуем ему, а также тексту  эррола \cite{Carroll2004}.

\subsubsection{–ассто€ние углового размера}

«десь дадим только набросок определени€. ѕодробнее см.
\cite{Weinberg1972,Weinberg2008,GorbunovRubakov2006}.

≈сли имеетс€ источник с известным размером (стандартна€ линейка), то его
можно использовать дл€ измерени€ рассто€ни€, измер€€ его угловой размер и
использу€ очевидные тригонометрические формулы дл€ эвклидова пространства.
Ќапример, если луч зрени€ перпендикул€рен стандартной линейке, то €сно, что
при уменьшении углового размера вдвое рассто€ние увеличилось вдвое.

ѕри больших рассто€ни€х $D$ углы малы, и дл€ линейки длины $R$
в плоском пространстве мы видим линейку под углом
$\theta=R/D$.
Ёто позвол€ет определить \textit{рассто€ние углового размера}, или
\textit{рассто€ние по угловому диаметру}
(по-английски \emph{angular diameter distance})
\begin{equation}
D_A \equiv R/\theta,
\label{dA}
\end{equation}
(дл€ малых $\theta$).  

Ќетрудно показать, что в расшир€ющейс€ вселенной (не только  в пространственно
плоской, но и в искривленной):
\beq
 D_A = \frac{ a_0 r}{1+z} ,
\label{daz}
\enq
где $r$ -- безразмерна€ радиальна€ координата в метрике FRW (\ref{ds2-cosm}),
поскольку свет был испущен, когда линейка была в $(1+z)$ раз ближе, чем в момент
приЄма, когда масштабный фактор равен $a_0$.

–ассто€ние углового размера в космологии используютс€ в основном при анализе
анизотропии реликтового излучени€ ---
cosmic microwave background (CMB).
¬ микроволновом диапазоне наблюдаетс€ почти идеальный спектр чЄрного тела,
который мы принимаем с поверхности ``последнего рассе€ни€''
(``Surface of Last  Scattering'').
ћалые отклонени€ от средней температуры,  $T\approx 2.7$~K, измер€емые как
функци€ углового размера, говор€т об акустических колебани€х первичной плазмы
в этот период. ћаксимальна€ величина флуктуации $\delta T$, наблюдаема€ на
угловом разделении примерно в один градус, отвечает ``последней'' звуковой волне,
вошедшей под горизонт как раз в момент рекомбинации. ƒлина этой волны
нам известна и по величине угла можно определить
рассто€ние до поверхности последнего рассе€ни€.
ќтсюда, в частности, получают, что наша ¬селенна€ пространственно плоска€.
ƒетали см. \cite{Challinor,GorbunovRubakov2006,RubVlasov2012}. 
ƒругое приложение рассто€ни€ углового размера даЄт анализ BAO -- барионных акустических осцилл€ций -- см. раздел \ref{BAO} ниже.

\subsection {‘отометрическое –ассто€ние}
\label{phDist}

Ѕолее ценным дл€ приложений сверхновых к космографии €вл€етс€ так называемое фотометрическое рассто€ние (photometric distance):
\begin{equation}
  D_{\rm ph} = \left({{L} \over {4\pi F}}\right)^{1/2} \,,
  \label{dph}
\end{equation}
где $L$ --- абсолютна€ светимость (т.е. светова€ мощность) источника --
\textit{стандартной свечи} (\emph{standard candle}), а $F$
--- поток, измеренный наблюдателем (энерги€, приход€ща€ в единицу времени на
единичную площадь приЄмника). »ногда это рассто€ние обозначают $D_L$ (от
``luminosity distance'').
Ёто определение соответствует тому, что в плоском пространстве
поток  источника на рассто€нии $D$ есть $F=L/(4\pi D^2)$.

¬о фридмановской вселенной,
однако, нельз€ сюда просто подставить  $ D_{\rm flat}=a_0 r$ из (\ref{dflat}), где $a_0$ -
масштабный фактор в момент,
когда фотоны были наблюдены на сопутствующей координате $r$ от источника.
ƒело в том, что с ростом $r$  поток уменьшаетс€ не только из-за обычной дилюции
(поверхность сферы растЄт как $4\pi D_{\rm flat}^2$), а ещЄ
из-за двух эффектов:
индивидуальные фотоны испытывают красное смещение на  фактор $(1+z)$, и
темп прихода фотонов тоже снижаетс€ в  $(1+z)$ раз.
ѕоэтому мы имеем
\begin{equation*}
  {F} = {L\over{4\pi a_0^2 r^2 (1+z)^2}}\ ,
\end{equation*}
или
\begin{equation}
  D_{\rm ph} = a_0 r (1+z) = D_{\rm flat}(t_0) (1+z) \; .
\label{dphdflat}
\end{equation}

»так, получаем, что рассто€ни€, введЄнные в выражени€х (\ref{dA}) и (\ref{dph}) не
равны друг другу
в общем случае даже в случае плоского пространства, если метрика нестационарна.
–авенство $D_A =  D_{\rm ph}$ выполн€етс€ только в статическом плоском пространстве,
а в расшир€ющейс€ вселенной
$D_{\rm ph} =D_A (1+z)^2$.
“.е. дл€ сгустков гор€чей плазмы, которую мы видим на момент рекомбинации,
когда $z \approx 1000$, различие в \emph{миллион раз}!

‘отометрическое рассто€ние $D_{\rm ph}$ --- это нечто доступное
измерению, если у нас есть астрофизический источник,  чь€
абсолютна€ светимость $L$ известна (``стандартна€ свеча'').
Ќо $r$ в выражении (\ref{dphdflat}) не наблюдаемо непосредственно, так что мы должны
избавитьс€ от него.
Ќа нулевой геодезической, по которой к наблюдателю распростран€ютс€ фотоны от
далЄкого объекта, можно положить $d\theta=d\varphi=0$  и тогда получим:
\begin{equation}
  0 =ds^2 = c^2dt^2 - {{a^2}\over {1-k\bar r^2}}d \bar r^2\ ,
\end{equation}
или
\begin{equation}
  \int\limits_{t_1}^{t_0} {{cdt}\over{a(t)}}
  = \int\limits_{0}^{r} {{d\bar r}\over{(1-k\bar r^2)^{1/2}}}\ .
\label{intNullGeod}
\end{equation}
»нтеграл в правой части (\ref{intNullGeod}) элементарный:
\beq
  \int\limits_{0}^{r} {{d\bar r}\over{(1-k\bar r^2)^{1/2}}}
  = \left\{
  \begin{array}{ll}
    \arcsin(r) \quad &\mbox{при} \quad k=1, \\
    r \quad  &\mbox{при} \quad k=0, \\
   \mbox{Arsh}(r)  \quad &\mbox{при} \quad k=-1 .
 \end{array}
 \right.
\enq

ƒл€ левой части (\ref{intNullGeod}) сделаем следующие преобразовани€
$$
 \int\limits_{t_1}^{t_0} {{dt}\over{a(t)}}=\int\limits_{a_1}^{a_0}
{{dt}\over{da}}
 \frac{da}{a} = -\int\limits_{a_0/a_1}^1 \frac{a}{a_0} {{dt}\over{da}}
 d\left(\frac{a_0}{a}\right)
 = \int\limits_1^{a_0/a_1} \frac{a}{a_0} {{dt}\over{da}}
 d\left(\frac{a_0}{a}\right) \; .
$$
ќтсюда
$$
a_0\int_{0}^{r} {{d\bar r}\over{(1-k\bar r^2)^{1/2}}}=c\int_1^{z+1}\frac{d(\bar z+1)}{H}=
c\int_0^{z}\frac{d\bar z}{H}
$$
и все свелось к наблюдаемым величинам типа $1/H$ (равного $a/\dot a$),
$z$ (с помощью $a_0/a_1=z+1$) и т.п.
“аким путЄм мы можем избавитьс€ от $r$ в выражении (\ref{dphdflat}) дл€
$D_{\rm ph}$, если выразим $r$ через элементарный интеграл $\int_{0}^{r} (1-k\bar r^2)^{-1/2}{d\bar r}$ и $\int_0^{z}{d\bar z}/{H}$.

„тобы сделать это, воспользуемс€ уравнением ‘ридмана (\ref{H2}) и
учтЄм при этом  возможность ненулевой энергии вакуума:
$$
  H^2 = \frac{8\pi G {\rho}}{3} - \frac{k}{a^2},
$$
что эквивалентно
\begin{equation}
  H^2 = H_0^2[\Omega_{\rm m}(1+z)^3 + \Omega_\Lambda
  + (1-\Omega_{\rm m}-\Omega_\Lambda)(1+z)^2] ,
\label{HzLamDM}
\end{equation}
где $\Omega_{\rm m} = { {\cal E}_{\rm m}/\rho_c c^2} $ и
$\Omega_\Lambda  = { {\cal E}_{\rm DE}/\rho_c c^2}$ ---
введЄнные выше параметры плотности дл€ нерел€тивистского вещества (плотность
которого ${\rho}_{\rm m}$ мен€етс€ обратно пропорционально сопутствующему
объему, т.е. как $(1+z)^{3}$)
и тЄмной энергии соответственно (эта плотность
посто€нна в простейшем случае $\Lambda$-члена).
«аметим, что ${\rho }_{\rm m}$ содержит энергию как обычного, так и
невидимого нерел€тивистского вещества (Dark Matter, сокращЄнно DM).
Ќиже мы будем использовать обычные сокращени€ дл€ космологических моделей,
например, CDM -- модель с холодной тЄмной материей, $\Lambda$CDM -- та же модель с
учЄтом $\Lambda$-члена, и т.п.

ѕодставив $H$ в выражение
$\int_0^{z}{dz_1}/{H}$, и выразив
$r$ через  $\int_{0}^{r} (1-k\bar r^2)^{-1/2}{d\bar r}$ только дл€ случа€
$k=-1$ (остальные случаи $k$ получатс€ автоматически аналитическим продолжением), мы
находим рабочую формулу дл€ фотометрического рассто€ни€:
\beq
D_{\rm ph}(z)={c\over H_0}(1+z){1\over\sqrt{\Omega_k}}
\sh\left\{
          \sqrt{\Omega_k}\int\limits_0^z \left[
             \Omega_{\rm m} (1+z)^3  +\Omega_\Lambda
            +\Omega_k (1+z)^2\right]^{-1/2} {\rm d}z \right\}.
\label{dL}
\enq
«десь $\Omega_k \equiv 1-\Omega_{\rm m}-\Omega_\Lambda$,
и при $\Omega_k<0$ гиперболический синус ($\sh$)
переходит в тригонометрический  ($\sin$), а
$\sqrt{\Omega_k}$ в $\sqrt{|\Omega_k|}$.
ѕри $\Omega_k \rightarrow 0$ легко берЄтс€ предел,
причЄм  $\sh$ исчезает из выражени€ дл€  $D_{\rm ph}$, а остаЄтс€ только интеграл
$\int_0^z [\ldots]^{-1/2} dz$.

“еперь зависимость $D_{\rm ph}(z)$ выражена через космологические
параметры.
ћы видим из (\ref{dL}), что $D_{\rm ph}$ очень просто св€зано и
с `плоским' $D_{\rm flat}$  из (\ref{dflat}) и с фридмановской радиальной
координатой $\chi$. 
Ќетрудно найти аналогичное выражение дл€ $D_{\rm ph}(z)$ в случае
непосто€нной плотности тЄмной энергии, когда $P=w(z) {\cal E}$.

≈сли $P_{\rm DE} =w{\cal E_{\rm DE}}$ (DE $\equiv $ Dark Energy), имеем:
\begin{equation}
  H(z) = H_0[\Omega_{\rm m}(1+z)^3 + v(z) \Omega_{\rm DE}  +
\Omega_k (1+z)^2]^{1/2} \; ,
\label{Hzw}
\end{equation}
где $\Omega_{\rm DE}  = { {\cal E}_{\rm DE}/\rho_c }$.
“еперь $H^{-1}(z)$ из (\ref{Hzw}) нужно подставить на место членов в квадратных
скобках в (\ref{dL}).
ѕлотность ${\cal E}_{\rm DE}$ посто€нна в простейшем случае $\Lambda$-члена
(когда $w=-1$).
‘ункци€ {$v(z)$} определ€етс€ уравнением состо€ни€ тЄмной энергии
{$P(z) =w(z){\cal E_{\rm DE}}(z)$}:
\begin{equation}
  v(z) = \exp{\left[3 \int\limits_0^z{\frac{1 + w(z)}{1 + z} dz}\right]} \; ,
\label{vz}
\end{equation}
и равна единице дл€ посто€нного $\Lambda$-члена, т.е. дл€ $w=-1$.

—ледует обратить внимание, что формулы дл€
фотометрического рассто€ни€ $D_{\rm ph}$, в которые входит
$D_{\rm flat}$, типа \ref{dphdflat} применимы дл€ любых $k$, а не только дл€ $k=0$.
–ассто€ние $D_{\rm ph}$ св€зывает $L$ и $F$ через площадь сферы, через которую
вс€ мощность $L$ и проходит.
Ќо у нас в метриках типа FRW площадь сферы св€зана только с
$g_{22} \equiv g_{\theta\theta}$ и $g_{33}\equiv g_{\phi\phi}$), т.е. в неЄ не входит $g_{11}=g_{rr}$, где и ``сидит''
параметр 3D кривизны $k$.
¬сегда получаем дл€ площади сферы $4\pi a^2 r^2$, а наше $D_{\rm flat}$ радиус
этой сферы и измер€ет.
ѕоэтому мы получаем формулу $D_{\rm ph}$, верную и в кривом 3D пространстве.
ј ``истинное'' рассто€ние $D_{\rm prop}$ содержит в себе $\int \sqrt{g_{rr}}\, dr$.

 онечно, наша формула не будет работать в анизотропном кривом пространстве, где
$g_{22}$ и $g_{33}$ не имеют эвклидова вида.

\subsection{Ћестница космологических рассто€ний. —верхновые}

Ћестница космологических рассто€ний опираетс€ на \textit{первичные индикаторы
рассто€ний} (\emph{primary distance indicators}) такие как тригонометрические
параллаксы звЄзд, близкие скоплени€ звЄзд с общим движением,
пульсирующие звЄзды, в частности, цефеиды (когда к ним примен€ют метод
Ѕааде-¬еселинка).

»меетс€ очень много \textit{вторичных индикаторов рассто€ний} (\emph{secondary
distance indicators}), использующих различные калибровки дл€ рассто€ний,
которые установлены по предшествующим ступен€м лестницы
\cite{Keel2009}.
Ќапример, те же цефеиды €вл€ютс€ вторичными индикаторами рассто€ний, когда
используетс€ соотношение период-светимость, прокалиброванное по объектам с известным
рассто€нием.

—верхновые относ€тс€ к астрономическим объектам наивысшей светимости $L$, поэтому
они могут наблюдатьс€ на огромных рассто€ни€х и  играют очень важную роль
в проверках космологических моделей.
¬ этом разделе мы по€сним, как сверхновые используютс€ в качестве индикаторов
рассто€ний.

јстрономическа€ классификаци€ сверхновых
основана на их спектрах в видимом свете вблизи максимума блеска.
ѕрежде всего провер€ют, есть ли линии  \textit{водорода}.

1) —верхновые типа I: \emph{нет линий водорода} вблизи максимума блеска.
»з них выдел€ют подтип Ia с линией ионизованного кремни€.
“акие сверхновые €вл€ютс€ наибoлее €ркими (т.е. имеют очень высокую светимость
в максимуме).

2) —верхновые типа II: \emph{чЄткие линий водорода} в спектрах и в максимуме блеска
и долгое врем€ спуст€.

ƒл€ измерений потока $F$ и светимости $L$ астрономы по традиции используют
звЄздные величины (звЄздна€ вел., или mag), которые €вл€ютс€ логарифмической мерой потока.
–азность двух звЄздных величин есть по определению
\begin{equation}
    m_1-m_0 = - 2.5(\lg F_1  - \lg F_0) = - 2.5\lg (F_1 / F_0) .
\label{mags}
\end{equation}
 оэффициент -2.5 выбран потому, что по традиции звезда нулевой звЄздна€вел. имеет блеск
(т.е. даЄт световой поток) ровно в 100 раз больше, чем звезда 5-й звЄздна€вел.
Ќуль-пункт $m_0$ в (\ref{mags}) должен быть определЄн по стандартной звезде
(обычно определ€ют по ¬еге~$=\alpha$~Lyrae).
¬се буквы $m$ и $F$  в (\ref{mags}) могут иметь дополнительные индексы (например, дл€ частоты $\nu$ в случае монохроматических потоков, или $U$, $B$, $V$ и т.п. дл€ различных фильтров).

јбсолютна€ звЄздна€ величина ${\cal M}$ определ€етс€ как звЄздна€ величина на
стандартном рассто€нии $D$ в 10 парсек (пк).
“аким образом, ${\cal M}=-2.5 \lg L +\mbox{const}$ --- чем больше светимость $L$,
тем меньше  ${\cal M}$
%

\begin{figure}[H]
\begin{center}
\includegraphics[width=0.55\linewidth]{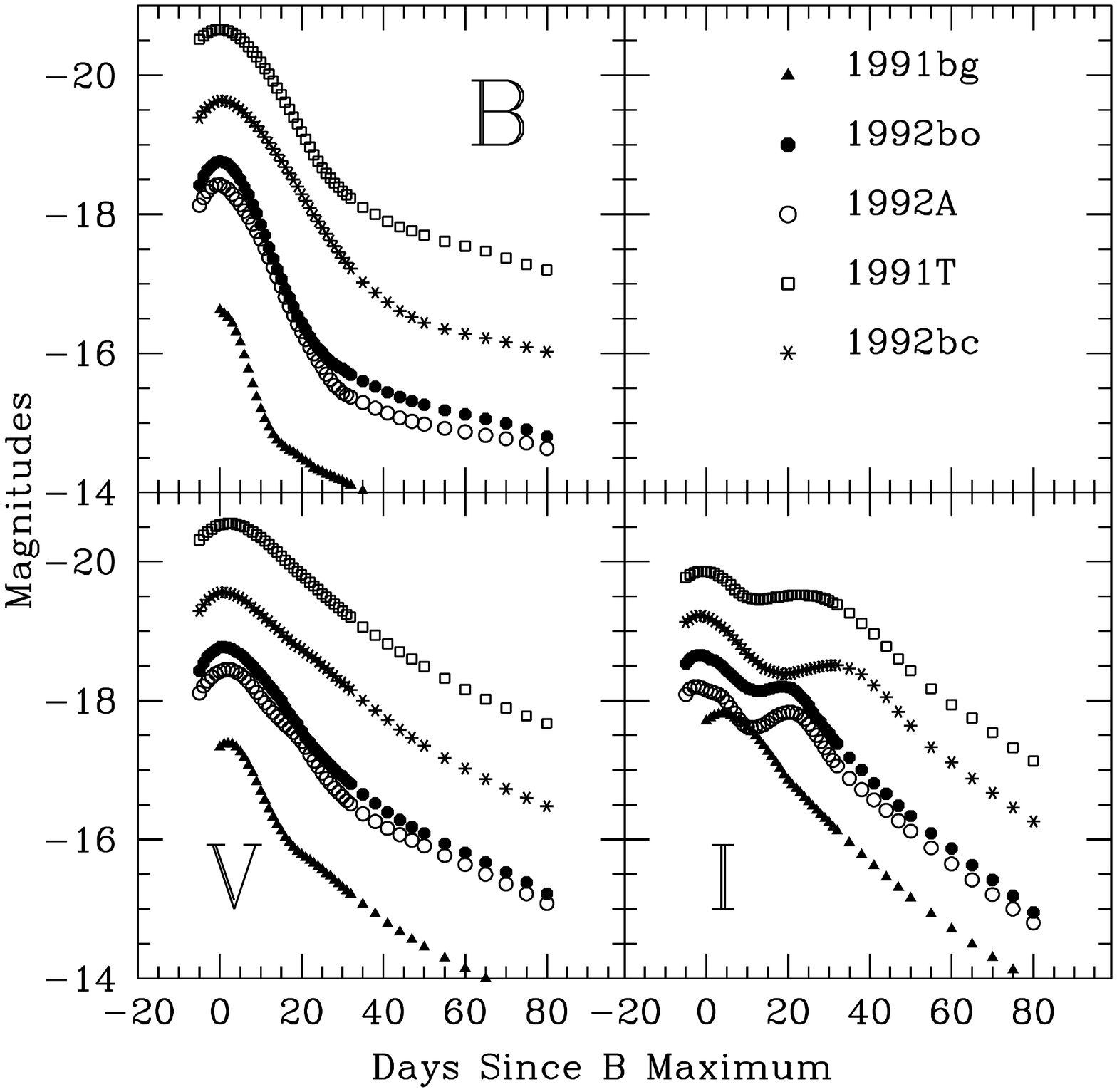}
\caption{Ќабор зависимостей ${\cal M}(t)$ дл€ SN~Ia в фильтрах {\it BVI}.
»сточник: шаблоны CTIO.
}
\label{ctioSNIa}
\end{center}
\end{figure}

ƒл€ примера приводим на рис.~\ref{ctioSNIa} абсолютные кривые блеска ${\cal M}(t)$
дл€ нескольких SN~Ia в фильтрах {\it BVI} --- синем (blue),
зелЄном (visible) и ближнем инфракрасном (infrared).

—ейчас твЄрдо установлено, что SN~Ia €вл€ютс€ продуктами термо€дерного
взрыва вырожденных звЄзд.
ћоделирование надЄжно показывает, что свет, т.е. энтропи€ в SNe~Ia
порождаетс€ в радиоактивных распадах: \nifsx переходит в \cofsx,
а тот в \fefsx.
Ёто в частности объ€сн€ет, почему кривые блеска Ia фильтрах {\it B,V} по форме столь
похожи (Fig.~\ref{ctioSNIa}).
¬ последнее врем€ гамма-линии, сопровождающие распад   \cofsx $\to$ \fefsx,
были непосредственно обнаружены в наблюдени€х на спутнике »нтеграл
\cite{ChurazovEa2014}.

—верхновые типа Ia  видны в больших количествах до $ z \sim 1.7 $ \cite{Poznanski07}, а по спектрам
отождествлены
3 сверхновых типа IIn на $ z = 0.808, 2.013, 2.357  $ \cite{Cooke09}.
—ама€  далека€ известна€ SN~Ia имеет красное смещение z=1.914~\cite{Jones2013}.

Ѕлагодар€ сверхновым Ia было открыто
 \cite{Schmidt98,Riess98,Perlm99}, что в рамках фридмановских моделей
космологический л€мбда-член не равен нулю,  $\Omega_\Lambda > 0$, т.е. сейчас
¬селенна€ расшир€етс€ с ускорением.
¬ более общем классе моделей это можно интерпретировать, как наличие
в космосе ``тЄмной энергии''.
ќдной из важнейших задач фундаментальной физики в насто€щее врем€ €вл€етс€
установление реальности и свойств тЄмной энергии (и тЄмной материи).
¬ решении этой задачи сверхновые, видимые на космологических рассто€ни€х, будут
продолжать играть ключевую роль.  ак уже отмечалось, за это открытие присуждена Ќобелевска€ преми€ по физике 2011 года.

ќсновы использовани€ сверхновых дл€ измерени€ рассто€ний можно кратко
сформулировать так.

\begin{enumerate}
 \item “еоретическа€ зависимость фотометрического рассто€ни€
 (\ref{dph}) от красного смещени€ $z$,
$D_{\rm ph \; theor}(z)[\Omega_m, \Omega_{\rm DE}, w(z), \ldots ]$,
определ€етс€ космологическими модел€ми, например, как в выражени€х
(\ref{HzLamDM})--(\ref{vz}).
Ќо, конечно, теоретические модели могут быть гораздо богаче, чем использованные
в этих формулах, они могут основыватьс€ на неэйнштейновской гравитации,
на дополнительных измерени€х пространства и т.д.
“огда там будут совсем другие параметры.
 \item —равнение предсказанной зависимости
 $D_{\rm ph \; theor}(z)$  с ``наблюдаемой'' $D_{\rm ph \; obs}(z)$
даЄт параметры наилучшим образом
(best-fitting) воспроизвод€щие наблюдени€, такие как $\Omega_m, \Omega_{\rm DE}, w(z)$, и т.д.
\end{enumerate}

ѕредложено несколько методов использовани€ сверхновых и их газовых остатков в космографии.
Ёти методы можно разделить на две группы.

¬ первой группе примен€ют идею стандартной свечи, которую мы подробно разберЄм в
разделе~\ref{SNsecondary}.
—тандартна€ свеча требует калибровки, она опираетс€ на 
лестницу космологических рассто€ний.
«десь сверхновые выступают как \textit{вторичные индикаторы рассто€ний}.

¬о второй группе методов сверхновые €вл€ютс€ \textit{первичными индикаторами рассто€ний}.
Ёти методы  рассмотрены в разделе  \ref{SNprimary}.

\subsection{–азнообразие кривых блеска SN~Ia и их использование дл€ космографии}
\label{SNsecondary}

 ак видно из рис.~\ref{ctioSNIa}, несмотр€ на похожие формы кривых
блеска SNe~Ia, их светимости в максимуме сильно различаютс€,
поэтому они не €вл€ютс€ стандартными свечами!

Ёто разнообразие не было столь очевидно многим исследовател€м три-четыре
дес€тилети€ назад.
“огда почти все считали, что сверхновые Ia все одинаковы,  т.е. SN~Ia €вл€ютс€ стандартными свечами в том смысле,
что максимумы абсолютной светимости (т.е. световой мощности) в разных сверхновых одинаковы.
ѕозже вы€снилось \cite{Pskovskij77,Pskovskii1984}, что это не так, но были предложены процедуры, позвол€ющие
найти абсолютную светимость, т.е. произвести стандартизацию свечи \cite{Rust,Phillips93}.

Ѕолее детально истори€ развити€ применимости сверхновых дл€ космологии может быть описана
так.
Ќекотора€ неоднородность максимальной светимости SN~I была известна давно, однако разброс
казалс€  намного меньшим, чем
у других быстро-переменных  астрофизических объектов, например, новых звезд.
¬ 1938 году Ѕааде \cite{Baade1938} заметил, что дисперси€ абсолютной звЄздной величины в
максимуме дл€ сверхновых (по его данным 1.1$^m$) намного  меньше, чем  дл€ новых.
ќн предложил использовать их в качестве индикаторов рассто€ний, использу€ одну и ту же
абсолютную  звездную величину дл€ всех SN.
«аметим, что в его выборке было всего 18 объектов и разделени€ SN~I на подтипы тогда еще не было.
»де€ использовани€ сверхновых дл€ космографии была тут же поддержана  \cite{Wilson1939,Zwicky1939}.
ѕараметры, характеризующие кривые блеска, стал вводить ѕсковский \cite{Pskovskii1967}
(его знаменитое $\beta$ дл€ наклона ``хвоста'' кривой блеска).
¬ 1967 году он измерил $\beta$ дл€ сверхновых I типа и показал, что все они имеют очень схожие
значени€ $\beta$, тем самым подтвердив вывод о пригодности SN~I в качестве индикаторов
рассто€ний.
» только в 1977, использу€ большие выборки SN, ѕсковский заметил св€зь между наклоном кривой
блеска ($\beta$) и абсолютной звездной величиной SN \cite{Pskovskij77}.
“огда же он выписал формулу дл€ этой коррел€ционной зависимости.

»так, ё.ѕ.ѕсковский  
первым св€зал форму кривых блеска сверхновых со светимостью в максимуме.
»стори€ этого открыти€ описана в статье \cite{Phillips05}. 
ѕсковский одним из первых установил не только разнообразие Ia, но и нашЄл
важную коррел€цию светимости в пике
$L_{\max}$ и темпа спада $L(t)$ \cite{Pskovskij77,Pskovskij77,Pskovskii1984}.
»менно такого рода коррел€ции используютс€ сейчас дл€ нахождени€
абсолютной светимости, т.е. дл€ \textit{стандартизации свечи}.
ѕодробности получени€ разных подходов к стандартизации свечи описаны в обзоре
\cite{Pruzh2015,Pruzh2016}.
»сследование \cite{Pruzh2015,Pruzh2016} установило, что
большую роль в открытии ѕсковского сыграла переписка с американским астрофизиком
–астом (B.W.~Rust), который, однако, не публиковал своих работ в доступных журналах.
“ем не менее, диссертаци€ –аста \cite{Rust} была замечена классиками исследовани€
расшир€ющейс€ ¬селенной \cite{Branch1992,Vaucouleurs1976} и сыграла свою роль
в установлении коррел€ции, котора€ позвол€ет стандартизовать свечу SN~Ia.

\begin{figure}[H]
\begin{center}
\includegraphics[width=0.6\linewidth]{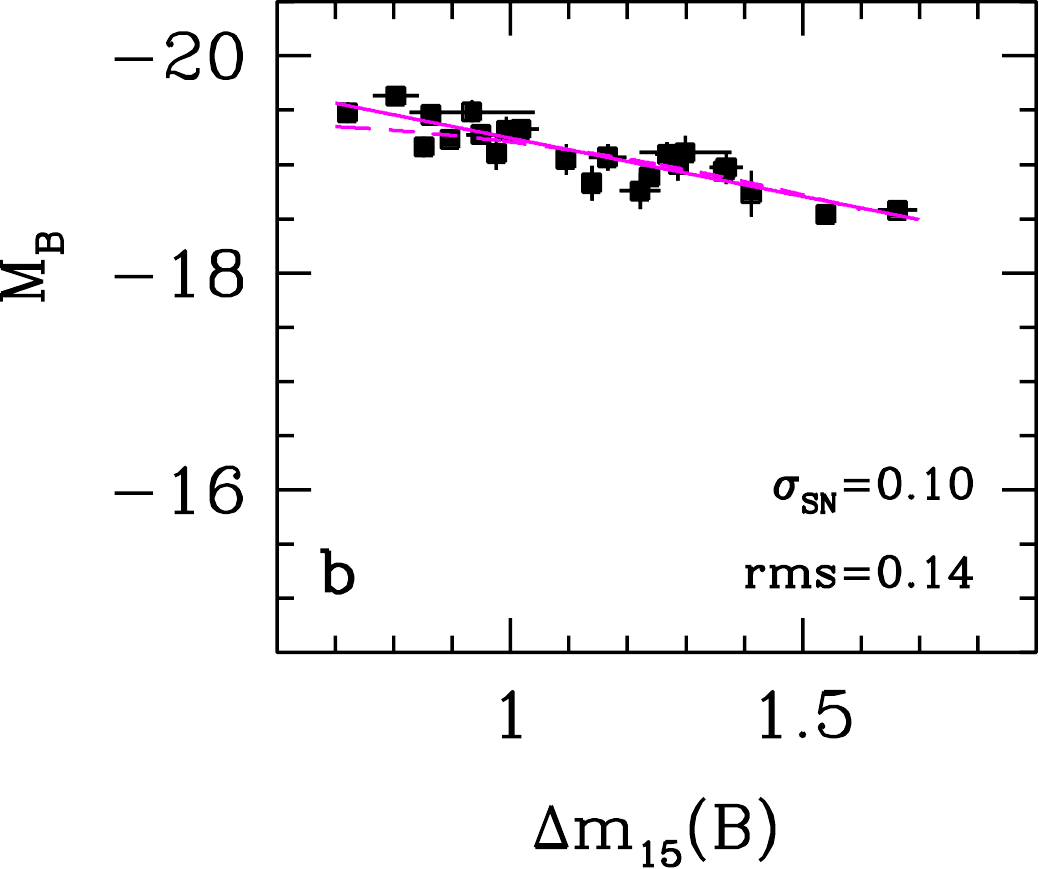}
\caption{—оотношение ѕсковского-‘иллипса по данным проекта Carnegi \cite{Folatelli2010}.
ѕо оси абсцисс отложено значение падени€ потока через 15 сут после максимума $\Delta m_{15}$,
по оси ординат -- абсолютна€ звЄздна€ величина в максимуме блеска.}
\label{PPGaston}
\end{center}
\end{figure}

’арактер этой коррел€ции состоит в том, что ``Ѕолее €ркие SNe~Ia €вл€ютс€ более
медленными''.
«десь ``€ркие'' строго значит ``имеющие высокую светимость'', а ``медленные''
относитс€ к темпу спада потока после максимума.
 ачественно это можно увидеть уже на рис.~\ref{ctioSNIa}, но эти данные не были
известны 40 лет назад.
ћ.‘иллипс \cite{Phillips93}  
после ѕсковского нашЄл аналогичную коррел€цию.
—ейчас такие коррел€ции называют
\emph{соотношением ѕсковского-‘иллипса, Pskovskij--Phillips, или PP},
а также WLR (Width--Luminosity--Relation),  и BDR (Brightness--Decline--Rate).
ѕример современных данных дл€ PP коррел€ции $B-\Delta m_{15}$ мы приводим на
рис.~\ref{PPGaston} из работы \cite{Folatelli2010} с разрешени€ авторов.
«десь $\Delta m_{15}$ обозначает изменение светимости в звЄздных величинах
через 15 сут после максимума в {\it B}-фильтре.
Ѕольшее $\Delta m_{15}$ означает более быстрый спад.
“огда, в самом деле, сверхнова€ слабее в максимуме.

≈сли измерен темп спада наблюдаемого потока $F(t)$ дл€ далЄкой сверхновой,
то из соотношений типа изображЄнного на рис.~\ref{PPGaston}, с учЄтом
фактора замедлени€ времени
$(1+z)$, можно получить $L$, и фотометрическое рассто€ние $D_{\rm ph}$.
»менно таким путЄм стандартизуютс€ нестандартные свечи.

«атем берЄтс€ формула типа (\ref{dL}) и методом наименьших
квадратов наход€тс€ космологические параметры, наилучшим образом воспроизвод€щие
наблюдени€.
Ќа практике и стандартизаци€, и нахождение космологических параметров
проводитс€ одним глобальным фитом.
Ёто позвол€ет уменьшить ошибки. “о есть $\chi^2$
содержит не только параметры, св€занные с космологией, но и параметры, св€занные с кривой блеска.
»менно таким путЄм было получено ненулевое значение
$\Lambda$ в работах \cite{Schmidt98,Riess98,Perlm99}.

“ака€ работа продолжаетс€, точность полученных параметров повышаетс€,
однако нельз€ сказать, что она очень высока.
Ќапример, первые результаты группы SN Legacy Survey (SNLS) \cite{Astier2006} дали
по $D_{\rm ph}(z)$ наилучший фит $\Om = 0.263 \pm 0.042 \mbox{(stat.)} \pm 0.032
\mbox{(syst.)}$, отсюда $\Omega_\Lambda \approx 0.74$ дл€ плоской $\Lambda$-
космологии.
ƒл€ тЄмной энергии, когда используютс€ выражени€ (\ref{Hzw}),(\ref{vz}),
они получили
$w = -1.023 \pm 0.090 \mbox{(stat.)} \pm 0.054 \mbox{(syst.)}$
при условии, что $w$ посто€нно в уравнении состо€ни€ $P=w\rho$.
Ёти результаты получены при комбинировании данных по сверхновым с данными
Sloan Digital Sky Survey (SDSS) по барионным акустическим осцилл€ци€м (BAO) -- т.е.
по распределению галактик в пространстве (см. раздел~\ref{BAO}).
Ѕолее поздн€€ важна€ работа, выполненна€ той
же самой французской группой~\cite{Betoule2014}, представл€ет собой компил€цию
данных от коллабораций SNLS, SDSS, Nearby SNe, HST. 
 омбинированные данные по CMB, BAO и SN~Ia привод€т согласно работе~\cite{Betoule2014} к значению
параметра ’аббла
$H_0=$\hoall~км/с/ћпк, что несколько ниже, чем даЄт группа ј.–иса
$H_0=73.8\pm2.4$ в \cite{Riess2011ApJ...730..119R}, но в хорошем согласии с последними результатами  \emph{WMAP}.
ƒл€ плоской \LCDM космологии в работе \cite{Betoule2014} найдено
$\Omega_m =$\bfomegamlcdm (stat+sys), что согласуетс€ со значени€ми, найденными по
измерени€м флуктуаций реликтового излучени€ (CMB) в экспериментах \emph{Planck} и  \emph{WMAP}.
— учЄтом данных по CMB был получен параметр $w=$\bfw (stat+sys) дл€ плоской вселенной.

\begin{figure}[H]
\begin{center}
\includegraphics[width=0.6\linewidth]{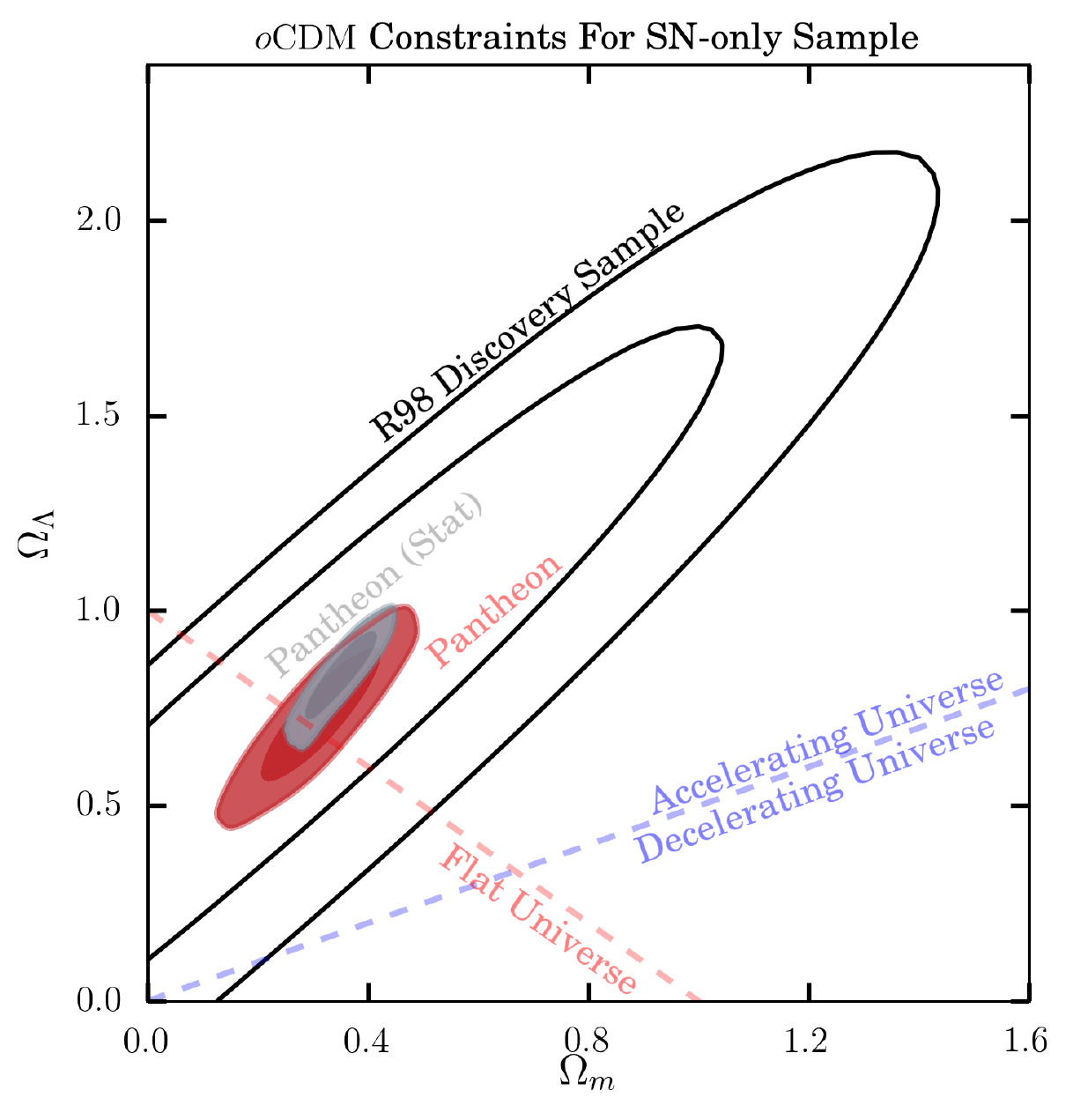}
\caption{—видетельства наличи€ тЄмной энергии на основе данных только о SN~Ia~\cite{Scolnic2017}.
ѕоказаны доверительные контуры при 68\% и 95\% дл€ космологических параметров $ \Omega_m $
и $ \Omega_\Lambda $
дл€ CDM~модели при фиксированном $ w = -1 $, но при переменном  $ \Omega_K $  как дл€
первого набора сверхновых  R98 \cite{Riess98}, так и дл€ набора Pantheon.
ќграничени€ Pantheon с учЄтом систематических неопределЄнностей показаны красным цветом, а чисто статистические неопределЄнности показаны серым.
}
\label{ScolnicPlot_omol}
\end{center}
\end{figure}

¬ следующем разделе \ref{BAO} мы кратко опишем возможности метода BAO дл€ космологических исследований.
Ётот метод €вл€етс€ важным дополнением к методам, опирающимс€ на свойства сверхновых.
«десь же на рисунке~\ref{ScolnicPlot_omol} приведЄм последние результаты, полученные на основе измерений только сверхновых  \cite{Scolnic2017}.
ќни собрали   276  SN~Ia ($0.03 < z < 0.65$) из наблюдений проекта Pan-STARRS1
с полезными оценками рассто€ний до SN~Ia из обзоров SDSS, SNLS, HST и др.,
что дало самый большой объединенный набор SN~Ia, общим числом 1049 сверхновых
в диапазоне $0.01 < z \lesssim 2$, который авторы назвали `Pantheon Sample'.


¬о всех упом€нутых выше результатах по космологическим параметрам, извлечЄнным из наблюдений
сверхновых использовалс€ обычный аппарат дл€ моделей FRW-космологии с учЄтом тЄмной энергии, когда
эволюци€ масштабного фактора описываетс€ через плотность энергии, усреднЄнной по очень большим масштабам
($> 100$~ћпк).
Ќа самом деле, усреднение нелинейных уравнений Ёйнштейна -- задача нетривиальна€.
ћогут возникать отклонени€ от решений, полученных, при обычном усреднении -- так называем эффект
backreaction.
Ќе услубл€€сь в споры вокруг этой проблемы дадим ссылки на недавние работы, где, в частности,
используютс€ данные по SN~Ia и обсуждаетс€ этот эффект \cite{Dam2017,Racz2017,Buchert2018}.

\subsection{Ѕарионные акустические колебани€, или осцилл€ции (BAO) }
\label{BAO}

Ѕарионными акустическими колебани€ми, или осцилл€ци€ми (далее BAO) называют первичные звуковые волны,
распростран€ющиес€ в гор€чей плазме фотонов и
барионов благодар€ давлению фотонов в ранней ¬селенной.
«десь мы остановимс€ на основных иде€х метода BAO дл€ определени€ космологических параметров (в том числе дл€ нахождени€ доли DE --- тЄмной энергии).
ѕодробности см. в хорошем обзоре~\cite{DWeinberg2013} --- там же много полезного
материала и о других методах исследовани€ DE, в том числе по сверхновым.
—пециально барионным акустическим колебани€м посв€щЄн обзор \cite{BassettHlozek2009}.
ќсновы, необходимые дл€ понимани€ теории BAO, можно найти и в старых
обзорах по развитию
возмущений во вселенной, например, полезен обзор \cite{KodamaSasaki1984}.


¬ ранней вселенной барионы и фотоны тесно св€заны, и возмущени€ в плазме
распростран€ютс€ в форме акустических колебаний. ѕосле рекомбинации водорода
при красном смещении $z_{r} \approx 10^3 $ фотоны и барионы практически перестают
взаимодействовать и давление света на барионы исчезает.  ак говор€т, происходит
расцепление барионов и фотонов.
¬скоре после этого, вследствие космологического расширени€ волны в барионах,
т.е. распростран€ющиес€ звуковые волны, останавливаютс€, оставл€€
след в распределении вещества на масштабе, соответствующем рассто€нию,
пройденному звуковыми волнами до этой эпохи (т.е. на масштабе акустического  горизонта).

«вуковые волны распростран€ютс€ до эпохи рекомбинации $t\approx 4\cdot 10^5$~лет,
со скоростью $\approx c/\sqrt{3}$ (чуть меньше перед самой рекомбинацией) и
застывают на масштабе, который в современную эпоху даЄт сопутствующий размер $l_{\rm BAO} \approx 150$~ћпк (т.е. $l_{\rm BAO} \approx 100 \hmpc$).
 оличественный вывод этого значени€ радиуса остановки  на основе упрощенных моделей
приведен в~\cite{Giovannini2005,Giovannini2008}, а также \cite{GorbunovRubakov2010} раздел 7.1.2. 

”точним эти утверждени€. ѕока барионы тесно св€заны с фотонами, 
имеем дл€ скорости звука (при $c\equiv 1$ )
\be
   u_s = \left(\frac{dP}{d\rho}\right)_S^{1/2} \approx \frac{1}{\sqrt{3}} .
   \label{soundRad}
\ee
«десь индекс $s$ в $u_s$ обозначает звук (sound), a заглавное  $S$ у скобки -- энтропию, так как
скорость звука -- адиабатическа€.
ѕоскольку $u_s = a(t) dr/dt$, где $r$ -- радиальна€ координата –обертсона-”окера из
(\ref{ds2-cosm}), имеем дл€ координаты акустического горизонта
\be
  r_s= \int_0^{t_r} \frac{u_s dt}{a(t)} ,
\ee
а дл€ его размера
\be
   l_s(t_r) = a(t_r) r_s= a(t_r) \int_0^{t_r} \frac{u_s dt}{a(t)} .
\ee
ѕри произвольном красном смещении $z$ после рекомбинации в коррел€ционной функции
материи должна быть особенность на длине
\be
   l_{\rm BAO}(z) = \frac{1+z_r}{1+z} l_s(t_r) .
\ee

≈сли приближенно положить до рекомбинации $u_s \approx  1/\sqrt{3}$ (\ref{soundRad})
и $a(t) \propto t^{1/2}$
(что на самом деле верно только в радиационно-доминированную эпоху), то получим
оценку
\be
   l_s(t_r) = t_r^{1/2} \int_0^{t_r} \frac{dt}{\sqrt{3} t^{1/2}} = \frac{2}{\sqrt{3}}t_r \approx 135 \; \mbox{кпк},
\ee
если дл€ рекомбинации $t_r \approx 380$~тыс€ч лет.
ƒл€  нерел€тивистского закона  $a(t) \propto t^{2/3}$ -- друга€ оценка $l_s = \sqrt{3} t_r \approx 200$~кпк.
Ќа самом деле надо учесть, что закон расширени€ лежит между этими двум€
предельными случа€ми.
 роме того,  скорость звука (\ref{soundRad}) отличаетс€ от $1/\sqrt{3}$ из-за вклада барионов в давление:
\be
 u_s^2= \frac{1}{3} \frac{1}{1+R_B} ,
\ee
где
\be
R_B = \frac{3\rho_b}{4\rho_\gamma} \approx
 \frac{3\cdot 10^4 \Omega_b h^2}{1+z} \left(\frac{T_0}{2.725 \, \mbox{K} }\right)^4 ,
\label{Rbaryon}
\ee
см. ниже выражение (\ref{Omega-b-h2}) дл€ $ \Omega_b h^2 $.

¬ результате получаетс€ $l_s (t_r) \approx 150$~кпк на эпоху рекомбинации, угловой размер п€тен
реликтового излучени€ именно этого размера и определ€ет основной пик в спектре мощности
возмущений температуры по сферическим гармоникам.
Ётот же масштаб даЄт
современный размер  коррел€ционной длины акустических осцилл€ций
$ l_{\rm BAO}(0) \approx 150$~Mпк, так как $z_r \approx 10^3$.

ƒанные  5 лет наблюдений реликтового фона WMAP дали \cite{Komatsu2009}:
\be
 l_{\rm BAO}(0) = 153.3 \pm 2.0 \; \mbox{ћпк}, \quad z_d = 1020.5 \pm 1.6 ,
\label{sWMAP}
\ee
т.е. это значение дл€ современного масштаба BAO получено из акустического
горизонта дл€ момента окончани€ так называемой ``эпохи увлечени€'' фотонов
барионами ---  drag epoch по-английски.
¬ выражении (\ref{sWMAP}) стоит обозначение $z_d$ вместо $z_r$, так как
в литературе различают дл€ эпохи рекомбинации два определени€ (дл€ пор€дковых оценок
это различие не существенно, но важно в точных расчЄтах), а именно,
отличают момент последнего рассе€ни€ фотонов от конца эпохи увлечени€ барионов фотонами.
ѕоследнее рассе€ние --- это момент,  когда оптическа€ глубина по комптоновскому рассе€нию фотонов до сегодн€шнего наблюдател€ падает ниже 1
--- без учЄта последующей реонизации.

ћомент $t_d$ конца эпохи увлечени€ определ€етс€ как врем€, когда барионы
выход€т из тесной св€зи с фотонами.
Ќа самом деле барионы зацеплены за электроны электрическими пол€ми,
а те зацеплены за фотоны комптоновским рассе€нием.
—ечение процесса то же самое, томсоновское $\sigma_{\rm T} $, что и при последнем рассе€нии фотона, но число фотонов
гораздо больше, чем число барионов (и электронов) в гор€чей (по энтропии)
вселенной, поэтому пробег фотонов $(n_e\sigma_{\rm T})^{-1}$ на много
пор€дков больше, чем пробег барионов $(n_\gamma\sigma_{\rm T})^{-1}$ в
море фотонов (поскольку $n_\gamma/n_b \sim 10^9 $ и до рекомбинации $n_e \sim n_b$).


Ќа самом деле дл€ правильного анализа выхода фотонов и барионов из режима тесной св€зи нужно сравнивать не пробеги частиц, а эффективность их обмена
импульсами.
—делаем это, следу€ работе \cite{HuSugiyama1996},
где показано, что в гидродинамическом приближении комптоновское рассе€ние приводит к обмену импульсами между жидкостью барионов и
излучени€ и стремитс€ уравн€ть средние скорости жидкостей $ u_b $ и $ u_\gamma $ соответственно.
ќднако плотности импульса
$(\rho_\gamma + p_\gamma)
u_\gamma = { 4 \over 3} \rho_\gamma u_\gamma$
и $(\rho_b + p_b)u_b \approx \rho_b u_b$ не равны при равенстве скоростей.
—охранение импульса (т.е. уравнение Ёйлера, точнее Ќавье-—токса) требует, чтобы ускорение барионной жидкости из-за комптоновского
увлечени€ (\textit{англ.} drag) было умножено на
$R_B^{-1}= {4 \over 3}
\rho_\gamma / \rho_b$ (ср. выражение \ref{Rbaryon})
в сравнении с той же величной дл€ фотонной жидкости.
“аким образом, в динамику процесса входит не огромное число $n_\gamma/n_b $,
а число $R_B$, близкое к единице при $z \approx 10^3$.
ѕрекрасное педагогическое изложение этого вопроса см. в книге~\cite{Dodelson2003}.
“ам показано, что аналитическа€ теори€~\cite{HuSugiyama1996} полностью подтверждаетс€ численным моделированием
кинетических уравнений дл€ фотонов и барионов в эпоху рекомбинации ¬селенной.


ѕоскольку гидродинамическое приближение дл€ фотонов становитс€ неприменимым после рекомбинации, различие между моментом
последнего рассе€ни€ и концом эпохи увлечени€ можно описать, перефразировав  ¬айнберга~\cite{Weinberg2008} стр. 441 русского издани€:
``—трого говор€, момент конца эпохи увлечени€ $t_d$
здесь следует выбрать на той стадии рекомбинации, когда типичный электрон перестаЄт обмениватьс€
ощутимым импульсом с фотонами, вместо несколько более раннего момента времени последнего рассе€ни€ $t_{ls}$, когда
типичный фотон перестает обмениватьс€ импульсом с электронами.
“ак как $R_B$ при последнем рассе€нии несколько
отличаетс€ от единицы, между этими временами есть небольша€ разница.''
Ќасколько конкретно конец эпохи увлечени€ $t_d$
отличаетс€ от момента последнего рассе€ни€, зависит от истории ионизации,
аппроксимирующие формулы
см. Appendix E статьи~\cite{HuSugiyama1996}.

—ледует иметь в виду, что  $t_d > t_{ls}$ только при $ \Omega_b \ll 1 $.
≈сли бы было $\Omega_b h^2 > 0.03$, то последнее рассе€ние было бы позже конца эпохи увлечени€ барионов.

“аким образом, характерный масштаб BAO представл€ет собой масштаб звукового горизонта в конце
эпохи тесной св€зи, когда фотонное давление больше не может
предотвращать гравитационную неустойчивость в барионах (что происходит
несколько позже последнего рассе€ни€ фотонов, потому что параметр
$ \Omega_b h^2 = 0.022 $ мал).

%
%
%
%

%

„асто говор€т, что BAO оставл€ют чЄткий  отпечаток на характерном масштабе скучивани€
(кластеризации) вещества, но следует иметь в виду, что этот след
про€вл€етс€ только статистически -- в качестве локального повышени€ в коррел€ционной функции
при $r \approx 150$~ћпк.

ѕо€сним, как используют BAO дл€ измерени€ космологических рассто€ний.
ѕровод€тс€ обзоры изображений галактик или скоплений галактик в определЄнных достаточно
широких по телесному углу област€х неба.
ѕри этом по показател€м цвета галактик определ€ют их так называемое фотометрическое
красное смещение.
“акие обзоры могут обнаружить отпечаток BAO
в кластеризации галактик по углу в достаточно узких интервалах по фотометрическому
красному смещению. ѕусть такой отпечаток имеет угловой поперечник
$\alpha$. ѕоскольку линейный поперечник $r\approx 150$~ћпк задаЄт нам стандартную линейку,
тем самым это позвол€ет измерить рассто€ние углового размера $D_A(z) = R/\alpha $ согласно формуле (\ref{dA}).
—пектроскопическое исследование галактик в том же объЄме позвол€ет измерить красные смещени€
гораздо более аккуратно, а также позвол€ет увидеть BAO вдоль луча зрени€, а не только по углу.
“ем самым оно даЄт более точное измерение $D_A(z)$.  роме того, измерение разницы скоростей
галактик на масштабе BAO позвол€ет напр€мую определить зависимость параметра ’аббла от красного смещени€ $ H(z) $.
ƒругие индикаторы распределени€ материи
также можно использовать дл€ измерени€ BAO. ѕоскольку масштаб BAO известен
в абсолютных единицах (на основе простого физического расчЄта
и значений параметров, хорошо измеренных по реликтовому фону), метод BAO
измер€ет $D(z)$ в абсолютных единицах --- ћпк, a  не в $\hmpc$ --- так что
BAO и измерени€ рассто€ний до сверхновых на том же красном смещении дают
различную информацию.
Ќедавние результаты по космологическим параметрам из подробного обзора
Baryon Oscillation Spectroscopic Survey (BOSS) см. в работе~\cite{Alam2016}.
Ќа основе данных Planck и BOSS получено $\Omega_K=0.0003\pm0.0026$
и $w=-1.01\pm0.06$, в близком согласии с плоской $\Lambda$CDM моделью.
ƒобавление данных о сверхновых Ia улучшает ограничение на параметр в уравнении
состо€ни€ тЄмной энергии до $w=-1.01\pm0.04$ и приводит к значению посто€нной
’аббла $H_0=67.9\pm0.9$~км/с/ћпк.

\begin{figure}[H]
\begin{center}
\includegraphics[width=0.4\linewidth,angle=-90]{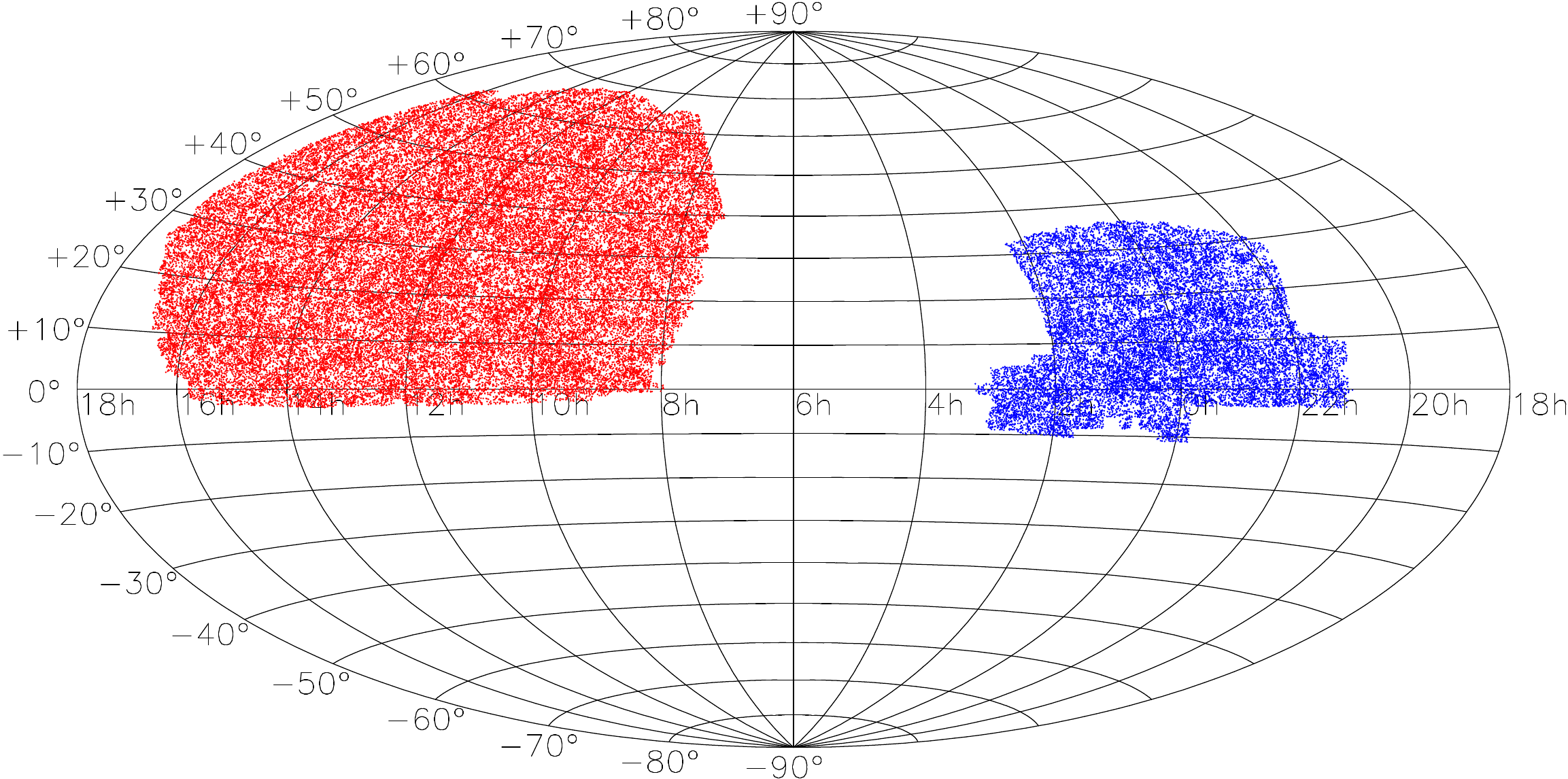}
\caption{Ќабор скоплений галактик \cite{Hong2015} в двух област€х неба.
–аспределени€ скоплений представл€ютс€ равномерными (ср. с рис.~\ref{rand1e4} справа), если не проводить подробный анализ.
}
\label{clustersH}
\end{center}
\end{figure}

\begin{figure}[H]
\begin{center}
\includegraphics[width=0.5\linewidth,angle=-90]{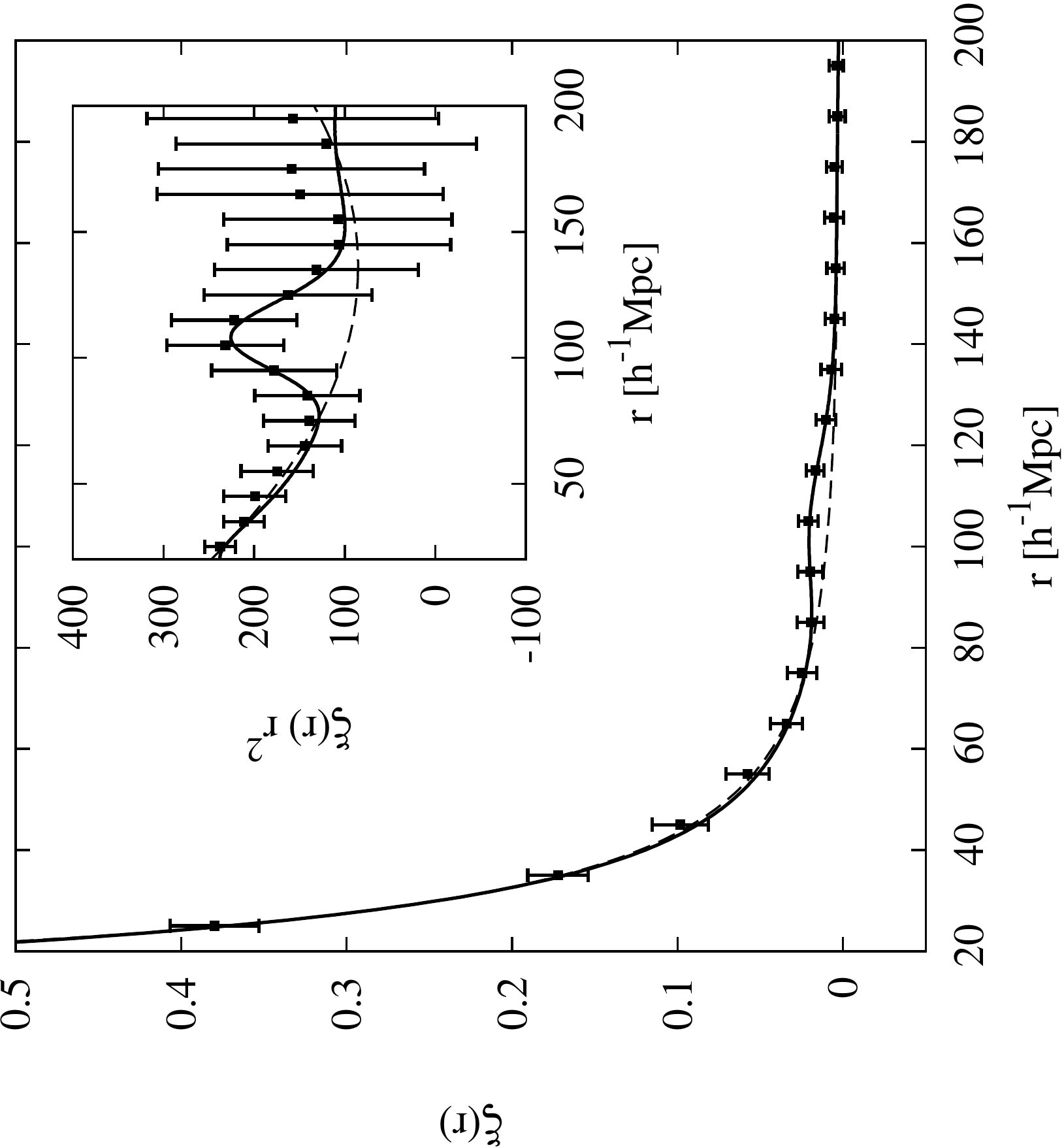}
\caption{ оррел€ционна€ функци€ дл€ скоплений галактик \cite{Hong2015} с €вным максимумом из-за BAO.}
\label{corrH}
\end{center}
\end{figure}

Ќагл€дно BAO можно увидеть пр€мо из наблюдательных данных по скоплени€м галактик,
например, по обработке в работе \cite{Hong2015}, см. рис.~\ref{clustersH} и \ref{corrH}.
Ќа рис.~\ref{clustersH} показаны две области на небесной сфере, точки отображают положени€ скоплений
галактик.
 ажетс€, что распределение точек случайное и совершенно равномерное.

—лед BAO в распределении материи про€вл€етс€ как в распределении скоплений галактик, так и
в распределении галактик пол€ \cite{Seo2016}. ≈го наблюдени€ оказались мощным и надежным зондом тЄмной
энергии  (смотри, например, \cite{DR11}). —равнение наблюдаемого масштаба BAO из обзоров галактик  и истинного физического
масштаба звукового горизонта, который можно  независимо оценить из исходных данных космического микроволнового фона, позвол€ет
установить соответствие  между наблюдательными координатами и физическими координатами; это соответствие чувствительно к истории
расширени€ и, следовательно, к свойствам тЄмной энергии, см., например, \cite{Eisen1998}.
“очность, с которой коррел€ционна€ функци€ ¬јќ может
быть измерена, и следовательно, ограничени€ на тЄмную  энергию, полученные из измерений BAO, зависит от силы следа BAO, и от того,
как хорошо мы можем отделить след BAO от широкополосного шума в спектре мощности
возмущений.
»з-за  образовани€ крупномасштабных структур, а также благодар€ наблюдаемому
эффекту пространственных искажений из-за красного смещени€, след BAO, как ожидаетс€, постепенно  деградирует/затухает вместе с
эволюцией структур, так что этот след значительно слабее при низких красных смещени€х, где расширение ¬селенной в основном
управл€етс€ тЄмной энергией.

\begin{figure}
\begin{center}
\includegraphics[width=0.45\linewidth]{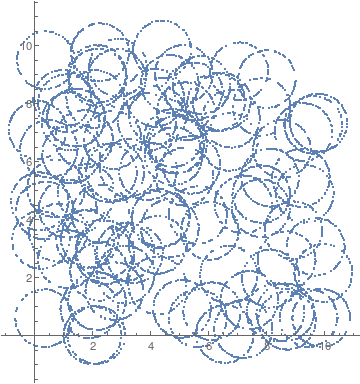} 
\hspace{7mm}
\includegraphics[width=0.45\linewidth]{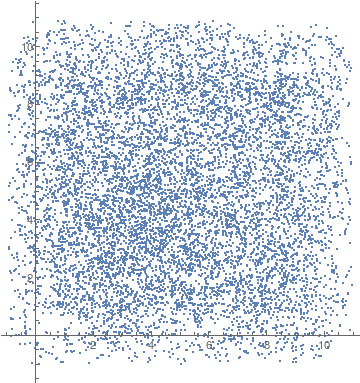} 
\caption{—лева: по 100 случайных точек на 100 случайно расположенных окружност€х.
—права: 1000 окружностей по 10 точек на каждой.
јлгоритм вз€т здесь:\\
{\footnotesize
{\tt http://mathematica.stackexchange.com/questions/57938/how-to-generate-random-points-in-a-region}}
}
\label{rand1e4}
\end{center}
\end{figure}

ѕриведЄм нагл€дные примеры замывани€ сигнала кольцевого распределени€ точек на двумерной плоскости:
когда много случайных точек расположены на небольшом числе колец, вы€вление сигнала не представл€ет
трудности. ≈сли то же число  точек разбросано по большому числу
колец (на  каждое кольцо приходитс€ мало точек), то требуетс€ специальна€ процедура
обработки и поиска коррел€ции, ---
см. рис.~\ref{rand1e4}.


Ќо главна€ причина трудности восстановлени€ (реконструкции) распределени€ плотности не в вы€влении сигнала из
случайной картины распределени€ точек, а в том, что пик плотности соответствует размеру 150 ћпк в насто€щую эпоху,
поэтому приходитс€ обследовать большой объЄм пространства, чтобы найти пик с приемлемой точностью.
ѕроблема реконструкции подробно освещена в обзорах~\cite{Eisen2007,Padm2009}.


»змерение характерного масштаба барионных акустических колебаний (BAO)
в коррел€ционной функции различных индикаторов распределени€ вещества --
это эффективный инструмент дл€ исследовани€ космического расширени€ и убедительный метод получени€ космологические параметров.
ѕик BAO в коррел€ционной функции при красном смещении $ z $
по€вл€етс€ на угловом удалении объектов $\Delta \theta = l_d/(1 + z)D_A(z)$,
где $ D_A = D_L / (1 + z)^2 $ - рассто€ние по угловому диаметру, а $ l_d = l_s (z_d) $ - звуковой горизонт
при красном смещении расцеплени€ $z_d$ (drag), то есть в эпоху, когда барионы отдел€ютс€ от фотонов.
 оррел€ционна€ функци€ BAO
также про€вл€етс€ в красном смещении $ \Delta z = l_d / D_H $, где
$ D_H \equiv c/H(z) $.
ѕоэтому измерение положени€ пика BAO при некотором $ z $ ограничивает
комбинации космологических параметров, которые определ€ют $ D_H / l_d $ и $ D_A / l_d $ на этом
красном смещении \cite{DHT}.

\subsection{BAO в коррел€ционной функции галактик}
\label{sec:bao1}

ѕик BAO наблюдалс€ в основном в коррел€ционной функции пар галактик,
полученных в обзорах красного смещени€.
Ќебольша€ статистическа€ значимость предыдущих исследований
давала ограничени€ только на $D_{\rm V}/l_d$, где
\begin{equation}
D_{\rm V}(z)\equiv\left\{z (1+z)^2 D_H D^2_{\rm A}\right\}^{1/3} .
\end{equation}

 ак физика, так и данные BAO завис€т от содержани€ вещества во вселенной.
—ледовательно, они априори завис€т от выбранной динамической структуры (см., например, обзор ~\cite{BAO}).
јнализ, проведЄнный в работе \cite{Tegmark:2006az}, показал, что
отношение $D_{\rm V}(z)/D_{\rm V}(z_0)$ слабо зависит от динамической структуры,
поэтому можно получать надЄжные ограничени€ на космологические параметры на основе
таких отношений.

–езультаты коллаборации Planck \cite{Ade:2013zuv} и работы \cite{Betoule2014},
по BAO-измерени€м величины  $D_{\rm V}/l_d$ при $z = 0.106$, 0.35, и 0.57 на
основе \cite{Beutler,Padmanabhan,Anderson} соответственно, а
также статей~\cite{Xia}
и \cite{Percival:2009xn} дл€ $z=0.35$  и $z=0.2$ дают:
\begin{equation}
\frac{D_{\rm V}(z)}{D_{\rm
V}(z=0.35)}=(0.335\pm 0.016,\; 0.576\pm 0.022,\; 1.539\pm 0.039),
\label{BetouleAde}
\end{equation}
при $z = (0.106,\, 0.2,\, 0.57)$  соответственно.

Ќа основе такого подхода к BAO и с использованием общедоступных данных по
сверхновым типа Ia, в работе~\cite{DHT} были получены
наблюдательные ограничени€ на класс моделей модифицированной гравитации,
которые при низких красных смещени€х привод€т к степенной космологии (т.е.
к модел€м со степенным ростом масштабного фактора в зависимости от времени).
Ѕыло показано, что при режиме расширени€
$a (t) \propto t^\beta$ с $\beta$, близким к 3/2, пространственно плоска€ вселенна€ хорошо подходит дл€ описани€ данных BAO и SN~Ia.

\subsection{ раткое резюме результатов по сверхновым совместно с BAO}
\begin{figure}[H]
\begin{center}
\includegraphics[width=0.6\linewidth]{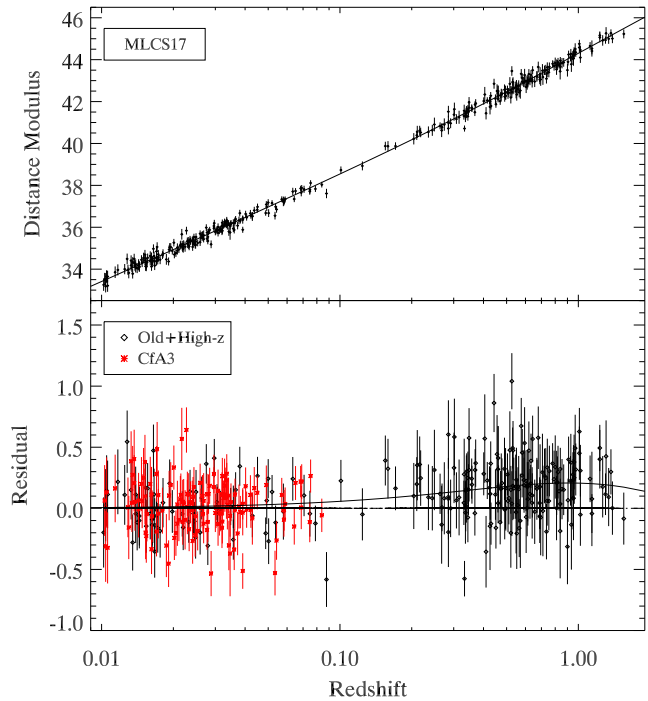}
\caption{
ƒиаграмма ’аббла по SN~Ia и еЄ нев€зки из каталога CfA3 \cite{Hicken2009}.
Ќовые  данные показаны красным, а старые - черными.
Ќев€зки относ€тс€ к вселенной без тЄмной
энергии, $ \Om = 0.27 $ и $ \Omega_\Lambda = 0 $. Ќаилучший космологический фит
показан в панели нев€зок.
Ѕолее современную версию диаграммы ’аббла по SN~Ia см. на рис.~8 из
работы~\cite{Betoule2014}
}
\label{Hdiag}
\end{center}
\end{figure}

\begin{figure}[H]
\begin{center}
\includegraphics[width=0.6\linewidth]{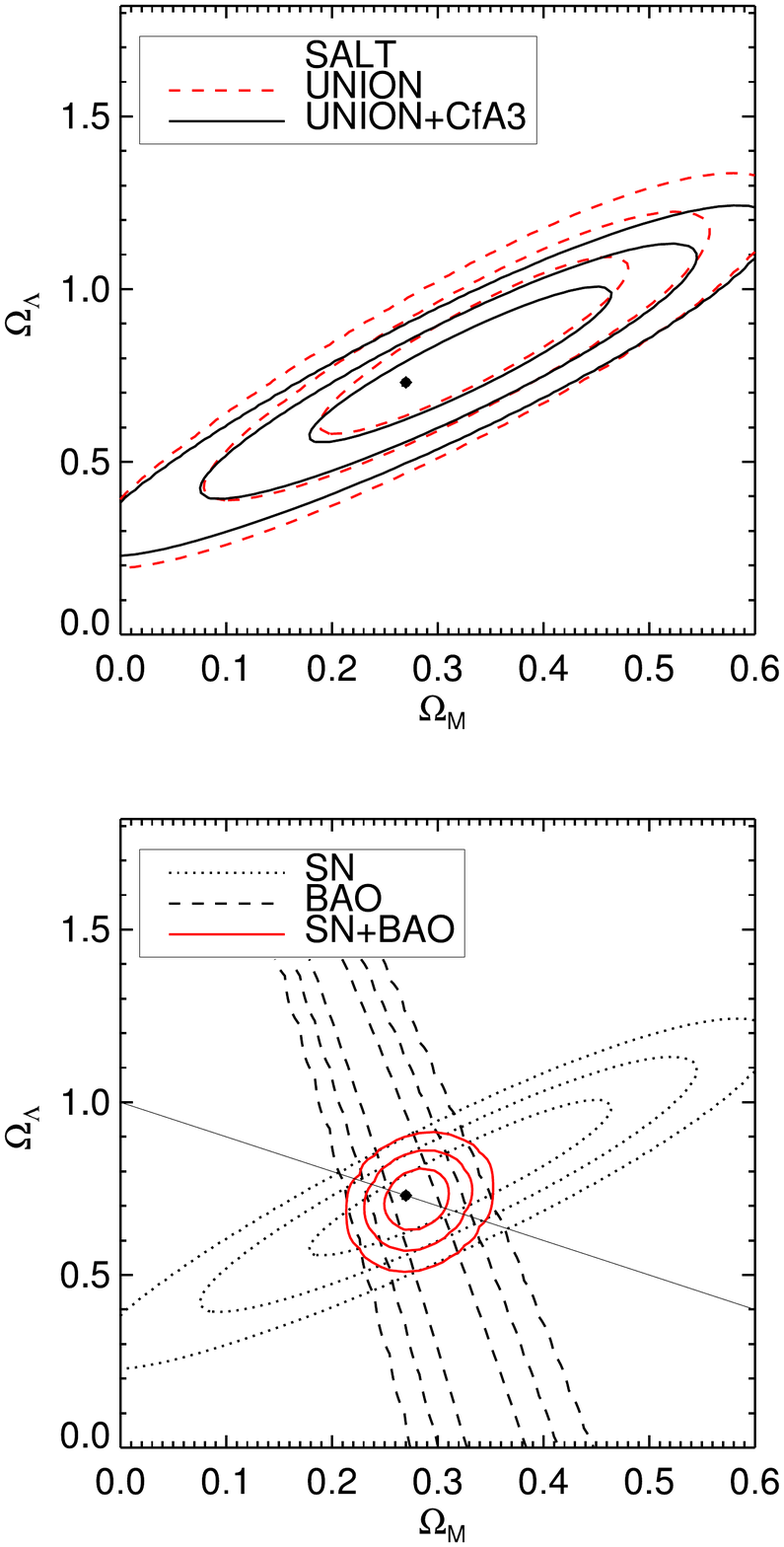}
\caption{ ѕараметры $\Omega$  из обзора CfA3 \cite{Hicken2009}.
 онтуры допустимых значений \OL и \Om~при $1+w=0$ без предположени€ о плоском пространстве.
—огласованна€ космологи€ (concordance cosmology) \OL$=0.73$, \Om$=0.27$
представлена точкой.
¬ерхний рисунок показывает, как добавление набора CfA3 значительно сужает
контуры вдоль оси~\OL. Ќижний рисунок показывает
сочетание контуров SN с ограничени€ми из BAO, а
пр€ма€ лини€ описывает плоскую вселенную \OL+\Om=1.
}
\label{OmOl}
\end{center}
\end{figure}

Ѕолее новые результаты той же группы SNLS \cite{Conley2011} дают на основе
более богатой статистики наблюдений дл€ двух разных процедур подгонки
кривых блеска к наблюдени€м (тут надо учесть, что наблюдаемые точки на кривой
блеска, как правило, расположены редко, нужны специальные процедуры вы€влени€
максимума, значений  $\Delta m_{15}$  и т.п.):
либо \Om$= 0.173^{+0.095}_{-0.098}$ и $w = -0.85^{+0.14}_{-0.20}$, либо \Om$=
0.214^{+0.072}_{-0.097}$
и $w = -0.95^{+0.17}_{-0.19}$ (все интервалы ошибок здесь чисто статистические).
ƒл€ параметра $w$ в уравнении состо€ни€ тЄмной энергии (который считаетс€ посто€нным
по крайней мере до $z = 1.4$) в плоской вселенной, они получили
$w = -0.91^{+0.16}_{-0.20}\mbox{(stat.)} ^{+0.07}_{-0.14} \mbox{(syst.)}$
на основе данных только по сверхновым.
Ёти значени€ $w$ согласуютс€ с космологической посто€нной,
котора€ требует $w \equiv -1$.
–езультаты обзора Center for Astrophysics CfA3 \cite{Hicken2009}, см.
рис.~\ref{Hdiag} и \ref{OmOl}, совместно с данными BAO, используемыми в качестве
априорного распределени€ веро€тностей, дают
$1+w=0.013^{+0.066}_{-0.068} (0.11 \rm~syst.)$, что тоже согласуетс€ с
космологической посто€нной.
ƒанные BAO совместно с их набором сверхновых дают также
\Om$=0.281^{+0.037}_{-0.016}$ и  \OL$=0.718^{+0.062}_{-0.056}$.
’от€ результаты разных групп формально согласуютс€, всЄ же видно, что различи€ между
ними больше, чем можно было бы ожидать по их собственным оценкам ошибок.

 осмологические параметры, извлекаемые из наблюдений сверхновых и BAO,
непрерывно уточн€ютс€.
Ќапример, стать€ \cite{Betoule2014} €вл€етс€ в каком-то смысле продолжением описанной выше работы \cite{Conley2011}.
ѕомимо этого там исправлены некоторые неточности в предыдущем анализе.
¬ \cite{Betoule2014} с учЄтом данных по BAO получено $w=$\bfwall дл€ параметра в уравнении
состо€ни€ тЄмной энергии, что согласуетс€ с их значением при учЄте CMB $w=$\bfw (stat+sys).

\begin{figure}[H]
\begin{center}
\includegraphics[width=0.6\linewidth]{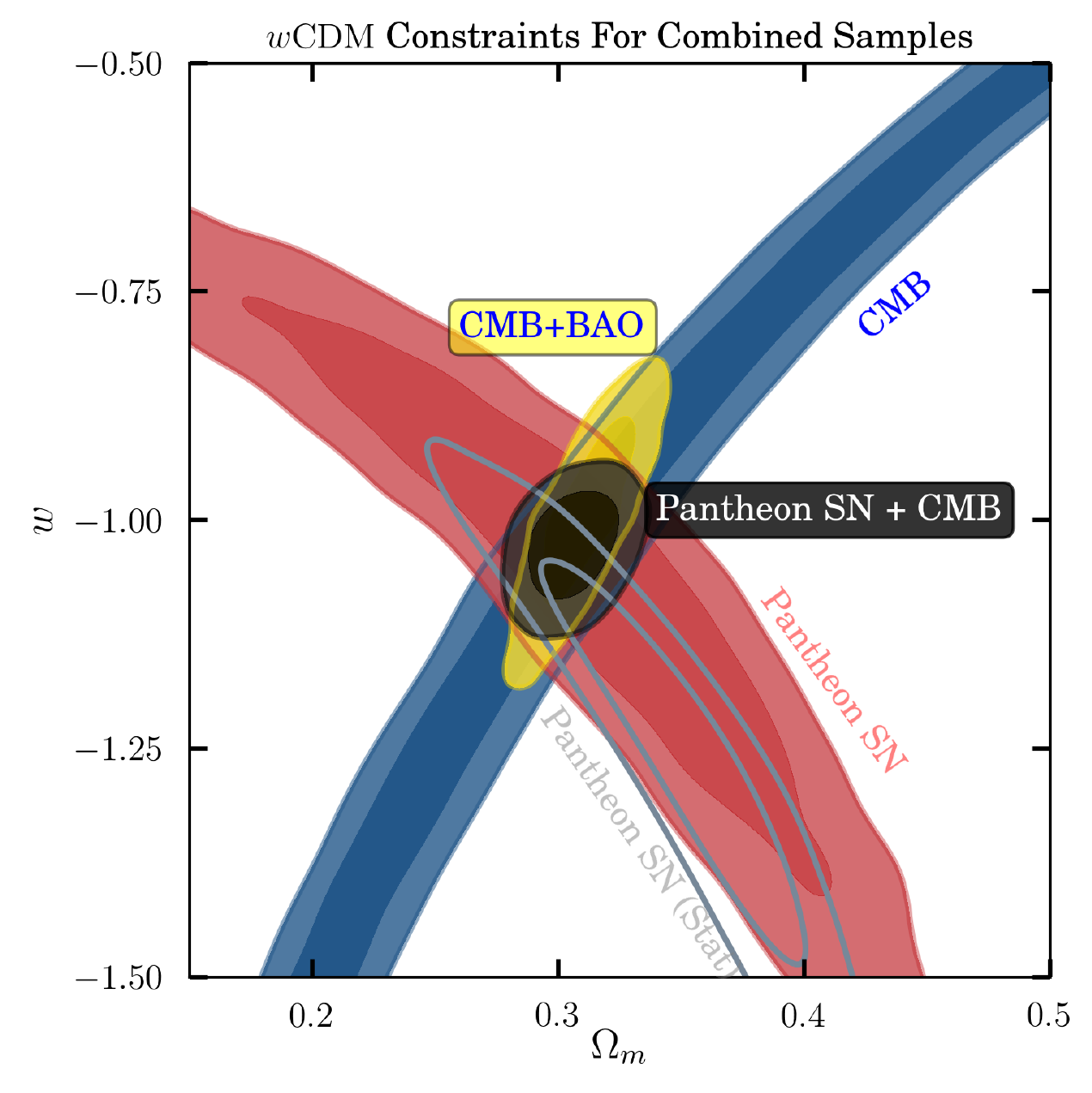}
\caption{ƒоверительные контуры 68\% и 95\% дл€ $\Omega_m$ и $w$
в CDM~модели по результатам~\cite{Scolnic2017}.
ѕоказаны ограничени€ по реликтовому фону CMB (оттенки синего цвета), SN -- с учЄтом
систематических неопределЄнностей (красный цвет), SN -- с учЄтом
только статистических неопределЄнностей (серые линии), и SN+CMB (фиолетовые).}
\label{ScolnicPlot_omw}
\end{center}
\end{figure}

 омбинированные результаты, представленные на рис.~\ref{ScolnicPlot_omw},
заимствованы из работы~\cite{Scolnic2017}, содержащей самый полный набор
SN~Ia на конец 2017~г.

ƒанные по BAO \cite{Kazin2014} в сочетании с измерени€ми флуктуаций
температуры реликтового излучени€ миссией
Planck, с измерени€ми пол€ризации WMAP (Wilkinson Microwave Anisotropy Probe) 9,
и 6dF обзором BAO, не вы€вил отклонений от плоской модели {$\Lambda$}CDM.
¬ этой модели получаетс€ параметр ’аббла
$H_{0}$ = 67.15 {$\pm$} 0.98 км с$^{-1}$ћпк$^{-1}$.
ƒл€ переменного уравнени€ состо€ни€ тЄмной энергии, получаетс€
$w_{\rm DE}$ = -1.080
{$\pm$} 0.135. ¬ неплоской {$\Lambda$}CDM модели получена кривизна
 {$\Omega$}$_{K}$ = -0.0043 {$\pm$} 0.0047, согласующа€с€ с нулЄм.

\subsection{—истематика, зависимость от $\boldsymbol z$}

ѕри переходе от описани€ соотношени€ ѕсковского-‘иллипса к получению диаграммы
’аббла нужно учесть важную добавку. ƒавно установлено, что светимость SN~Ia зависит не
только от формы кривой блеска (соотношение ѕсковского-‘иллипса), но и от цвета
сверхновой. » зависимость от цвета очень существенна. ѕоправка за цвет
вводитс€ во
все современные космологические анализы. ¬ наиболее попул€рной модели
стандартизации SALT2 \cite{Guy2007} абсолютна€ звЄздна€ величина SN~Ia определ€етс€
следущим образом: $M = M_{0} + \alpha X_1 - \beta C$, где $X_1$ -- это аналог
$\Delta m_{15}$ (отвечает за форму кривой блеска), а $C$ -- это цвет. ѕо-видимому,
первым параметр цвета ввел –оберт “рип \cite{Tripp}, хот€ и јдам –ис тоже как-то
его учитывал в своем методе MLCS.

  роме того, можно отметить (но эта поправка еще не внесена в космологический
 анализ), что достоверно установлено, что абсолютна€ звЄздна€ величина SN~Ia зависит от
 родительской галактики (host galaxy). ¬ качестве физической величины, характеризующей хост,
 используют деление по подтипам (спиральные, эллиптические \cite{Sullivan2003,
 Henne2017}), массу звездной составл€ющей галактики
 \cite{Sullivan2010,Johansson2013,Betoule2014}, глобальный и локальный темп звездообразовани€
 \cite{Sullivan2006,Neill2009}  (исследуют непосредственное окружение сверхновой, \cite{Rigault2015}).
 ѕока   что поправки за свойства родительской галактики внос€т в космологический
 анализ буквально руками. Ќапример, см. формулу (5) в~\cite{Betoule2014}: дл€ сверхновых,
 родительские галактики которых имеют звездную массу более $10^{10} M_\odot$, к $M_{0}$ прибавл€ют малую
 добавку $\Delta_M$, то есть просто сдвигают одну часть диаграммы ’аббла (с массивными родительскими
 галактиками) относительно другой (с маломассивными).

»меетс€ много факторов, которые могут повли€ть на результаты, полученные дл€
космологии с помощью сверхновых типа Ia, прокалиброванных по соотношению
ѕсковского-‘иллипса, например,
поглощение и рассе€ние света в межгалактическом пространстве или в родительских галактиках
(см.  Folatelli et al. \cite{Folatelli2010, Folatelli2013} 
и ссылки там), изменение металличности прародителей сверхновых и относительной
роли различных предсверхновых с изменением возраста ¬селенной.

ƒл€ читател€-физика попытаемс€ в нескольких словах объ€снить роль этих факторов
в космологических приложени€х сверхновых.

ќсобенно трудно определить поглощение и рассе€ние света сверхновых по пути к наблюдателю.
–ентгеновска€ астрономи€ даЄт очень много информации о гор€чем межгалактическом газе внутри
скоплений галактик (так называемый ``intracluster gas''), но о барионном газе \textit{между}
скоплени€ми галактик известно очень мало (``inter-cluster gas'').
»звестна проблема ``нехватки'' барионов во ¬селенной -- missing baryon problem:
космологический нуклеосинтез предсказывает
около 4--5\% плотности энергии в барионах, то же самое число получаетс€ из анализа
реликтового излучени€
CMB.
ќднако полное число всех наблюдаемых барионов (в газе и звЄздах) значительно меньше.
Ёти скрытые барионы могут находитьс€ в состо€нии тЄплого газа между скоплени€ми галактик
(inter-cluster gas).
“емпература этого газа может быть слишком высока дл€ его детектировани€ в видимом или
ближнем ультрафиолетовом диапазоне, но низка дл€ заметной эмиссии в рентгене.
¬ принципе такой газ может содержать примесь ``серой'' пыли (поглощение на которой
не зависит от длины волны).
“огда мы увидим далЄкие галактики ослабленными, что может быть ложно истолковано
как эффект ускорени€ космологического расширени€.
“ут можно упомюнуть работу Ѕогомазова и “утукова~\cite{Bogomazov2011}. ќни писали
про серую пыль и ее вли€ние на диаграмму ’аббла. ¬ той же работе они также
рассмотрели эффект от эволюции массы сливающихс€ белых карликов с ’аббловским
временем.
ќднако, этот эффект не должен присутствовать при красных смещени€х
больше $z \sim 1$: столь далЄкие сверхновые не показывают эффекта ускоренного
расширени€.

“акое построение с серой пылью может показатьс€ совершенно искусственным, тем не менее, его
нельз€ стопроцентно исключить, пока не решена проблема скрытых барионов.

ќчень мало известно также о так называемом
``покраснении'' в родительских галактиках сверхновых.
Ёто тот же эффект поглощени€/рассе€ни€ света, но теперь в среде, окружающей взрывающуюс€
звезду.
—войства этой среды могут сильно мен€тьс€ по пространству.
ƒл€ близких галактик эти свойства можно исследовать, но это очень трудно дл€ далЄких
объектов.

ћеталличность, т.е. содержание элементов т€желее гели€, в предсверхновых
может вли€ть на поведение кривых блеска сверхновых, т.е. на соотношени€ типа ѕсковского-‘иллипса
и тем самым на стандартизацию свечей.

ѕредложено несколько допустимых сценариев рождени€ сверхновых типа Ia в очень
сложной и запутанной эволюции двойных звЄзд (одиночные вырожденные звЄзды -- белые карлики
взрыватьс€ не могут).
ќтносительную роль этих сценариев в близкой вселенной трудно оценить (в литературе
имеютс€ противоречивые данные об этом).
“ем более трудно оценить это в далЄкой, молодой вселенной.

¬се эти факторы -- поглощение света, металличность, относительна€ роль сценариев эволюции --
могут измен€тьс€ с возрастом вселенной: звЄзды производ€т всЄ больше и больше металлов,
загр€зн€ющих межзвЄздную среду.
“аким образом по€вл€ютс€ систематические ошибки в определении рассто€ний и космологических
параметров с помощью сверхновых.

\begin{figure}[h!]
\begin{center}
\includegraphics[width=0.6\linewidth]{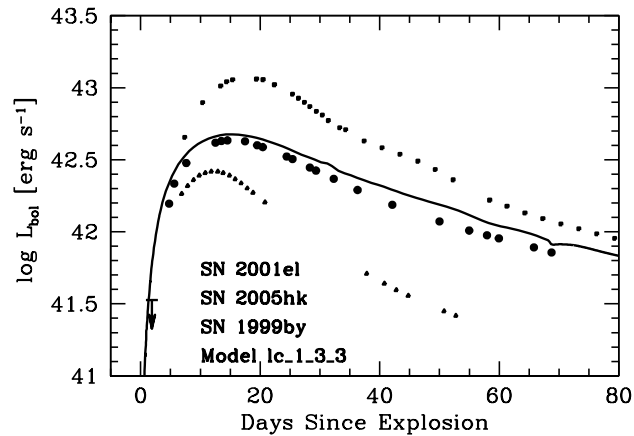}
\caption{SN~2005hk в сравнении с двум€ ``нормальнми'' SN~Ia.}
\label{sn05hk}
\end{center}
\end{figure}

ƒругим возможным источником ошибок €вл€етс€ неправильна€ классификаци€  и
примесь необычных событий типа SN~Ia.
Ќапример, был открыт своеобразный
подкласс SN~Ia,  подтип  SN~2002cx сверхновых  \cite{Li2003}.
ќни слабые, но медленные, см., например, рис.~\ref{sn05hk}, вз€тый из работы
\cite{Phillips2007},
т.е. ведут себ€ противоположно соотношению  ѕсковского-‘иллипса
(PP), которое используетс€ дл€ космологии и согласно которому
медленно спадающие SN~Ia €вл€ютс€ согласно PP самыми €ркими.
“еперь представим себе, что число событий  подтипа  SN~2002cx растет с
космологическим красным смещением $z$.
“огда, опира€сь на соотношение PP, которое установлено дл€ близких
SN~Ia, т.е. при $z=0$, мы получим, что сверхновые Ia  при больших $z$
в среднем кажутс€ \emph{более слабыми}, следовательно, фотометрическое
рассто€ние до них больше, чем при истинном значении $\Omega_\Lambda$.
“аким образом, будет получен ложный вклад в DE.

ћожно процитировать работу  онлей и др. 
\cite{Conley2011}:
``Evolution in the absolute magnitude of SNe Ia with redshift is not constrainable
without a detailed physical
model because it can in principle mimic any cosmology''.
\emph{``Ёволюцию  абсолютной величины сверхновых Ia с красным смещением
нельз€ получить без детальной физической модели, поскольку она в принципе может
имитировать любою космологию''}.
Ёто не означает, что сверхновые нельз€  использовать дл€ надежной космографии.
ѕросто нужно развивать новые подходы к этой задаче.
ќдин из таких новых подходов по SN~Ia  см. в работе~\cite{Fakhouri2015}.
«десь отмечаетс€:
``These results imply that at least 3/4 of the variance in Hubble residuals in current supernova
cosmology analyses is due to SN~Ia''.
\emph{``Ёти результаты означают, что по меньшей мере 3/4 дисперсии нев€зок параметра ’аббла в
текущем анализе космологии на основе сверхновых об€заны SN~Ia.''}
“акие результаты подтверждают наши опасени€  об использовании SN~Ia дл€ анализа DE, как делаетс€, например, в работах \cite{2007JCAP...02..011A,2009PhRvD..80j1301S}, 
так как
свойства звЄзд мен€ютс€ с возрастом вселенной.
јвторы работы \cite{Fakhouri2015} на основе наблюдений пытаютс€  оценить количественную меру
астрофизической неопределЄнности.
ќни аккуратно всЄ делают в локальной вселенной, улучша€ стандартизируемость SN~Ia,
но на больших рассто€ни€х всЄ равно неизвестна€ астрофизика  будет давать неизвестную систематику,
так как SN~Ia €вл€ютс€ вторичными индикаторами рассто€ний, и остаЄтс€ невы€сненным, как будет мен€тьс€
дл€ больших $z$ их улучшенна€ калибровка, полученна€ при  $z \sim 0$,
(их оценка ``standardizing high-redshift supernovae to within $0.06 \pm 0.07$ magnitudes'' остаЄтс€
принципиально статистической).

ѕри участии авторов данного обзора был предложен
новый пр€мой метод DSM дл€ измерени€ рассто€ний в космологии~\cite{SIB-h,SIB-h1,SIB-h2}, 
который обсуждаетс€ в следующем разделе.

\subsection{—верхновые как первичные индикаторы рассто€ний}
\label{SNprimary}

ƒл€ методов ``стандартной свечи'' требуетс€ знать рассто€ние до большого числа сверхновых,
измеренного совсем другим,
независимым методом с привлечением лестницы космологических рассто€ний \cite{Sandage1974}.
»наче, без набора большой статистики объектов с известным рассто€нием,
невозможно провести калибровку процедуры стандартизации свечи: см. обзоры, например,
\cite{Leibundgut01, Phillips05}.
“.е. в этой группе методов сверхновые используютс€ как вторичные индикаторы рассто€ний
(secondary distance indicators).

ƒруга€ группа методов позвол€ет использовать сверхновые как \emph{первичные индикаторы
рассто€ний (primary distance indicators)}.
Ќапример, сверхновые II~типа (SN~II) очень сильно различаютс€ по светимости
и по формам кривых блеска и не год€тс€, вообще говор€, дл€ метода стандартной свечи.
Ќо их большое преимущество --- это возможность пр€мых измерений рассто€ни€,
например, методом расшир€ющийс€ фотосферы (Expanding Photosphere Method, EPM) \cite{KK1974}.
ƒл€ этого  не нужна стандартизаци€ свечи и не нужна лестница космологических рассто€ний.

’орошие спектральные наблюдени€ сверхновой позвол€ют получить рассто€ние
методом SEAM (Spectral-fitting Expanding Atmosphere Method \cite{BaronSEAM}).
¬ отличие от EPM в этом методе нет необходимости делать предположение о чернотельном характере спектра сверхновой.

¬ насто€щем обзоре мы описываем развитие нового метода дл€ измерени€
рассто€ний в космологии,
частично основанного на EPM и SEAM и частично на ESM (Expanding Shock Front Method \cite{Bartel2007}).
¬ этом методе должны использоватьс€ сверхновые типа IIn наивысшей светимости,
которые в последние годы   стали не только активно открывать, но и детально исследовать \cite{Taddia2013}.

ѕредлагаетс€ называть этот метод DSM Ц Dense Shell Method,
так как светимость SN~IIn обусловлена распространением тонкого плотного сло€ в окружающей среде.
Ќа примере SN~2006gy и SN~2009ip показано, что этот метод €вл€етс€ работоспособным: получено рассто€ние до этих сверхновых,
независимо от обычных калибровок рассто€ний.
ƒл€ нового метода не требуетс€ приближени€ стандартной свечи, как дл€ сверхновых типа Ia,
и нет необходимости опиратьс€ на шкалу (так называемую лестницу) космологических рассто€ний.

¬ SN~II фотоны рождаютс€ в ударных волнах, распростран€ющимис€ в их оболочках
(за врем€ $< 10^4$ с в SN~1987A  и до $\sim10^7$ с и больше в SN~IIn).
¬ обычных SN~II ударна€ волна порождает не только кратковременную вспышку жЄсткого излучени€,
но и резервуар энтропии, обеспечивающий свечение на стадии ``плато'' в течение нескольких мес€цев.
” сверхновых SN~IIn, где ударна€ волна распростран€етс€ в окружающей среде несколько мес€цев,
она и €вл€етс€ источником свечени€ \cite{WooBliHeg2007,Grasberg1971,Grasberg1986,ChugaiEa04}.

 ривые блеска SN~II очень разнообразны по форме и амплитуде, что делает невозможным дл€ их описани€ введение стандартной свечи,
т.е. какой-то унифицированной кривой блеска.
 рива€ блеска сильно зависит от свойств оболочки, окружающей источник энергии сверхновой,
будь то коллапсирующее €дро или термо€дерное горение в €дре.
¬ то же врем€ оболочка делает SN~II гораздо менее зависимой от деталей взрыва.
¬ течение мес€цев в оболочке наблюдаетс€ реальна€ фотосфера, про€вл€юща€с€ на кривых блеска в виде классического плато.

јббревиатура EPM происходит от слов Expanding Photosphere Method.
»де€ этого метода восходит к работам Ѕааде \cite{Baade1926} и ¬есселинка \cite{Wesselink1946},
которые развивали его дл€ определени€ рассто€ний до пульсирующих звЄзд Ц цефеид.

ѕоскольку мы в состо€нии построить детальную модель SN~IIn,
на основе той же идеи возникает новый пр€мой метод Dense Shell Method (DSM),
позвол€ющий использовать €ркий свет SN~IIn дл€ космологии.

—уть метода в следующем.
≈сли известна скорость фотосферы $v$, то за врем€ измерений $dt$
ее радиус изменитс€ на $dr=v dt$,
и мы сразу узнаем изменение радиуса $dr$  безо вс€ких лестниц космологических рассто€ний.
»змер€емый поток излучени€ равен
\begin{equation}
 F = 4 \pi r^2\sigma T^4 / D^2,
\end{equation}
где $D$ - фотометрическое рассто€ние.
“емпература $T$ измерима, $dr$ и $dF$ измеримы, а $D$ не мен€етс€.
”добнее ввести $S = \sqrt{F}$
\begin{equation}
 S = 2\sqrt{\pi \sigma} r T^2 / D
\end{equation}.

ќтсюда, если $T$ не измен€етс€ значительно между двум€ измерени€ми, имеем
\begin{equation}
 dS = 2\sqrt{\pi \sigma} dr T^2 / D
 \label{eq:dS}
\end{equation}.

¬еличина $dr$ может быть измерена непосредственно в р€де случаев.
ј именно, $dr=v_{ph} dt$, если $v_{ph}$ есть скорость фотосферы.
“огда из измерени€ $dS$, $dr$ и $T$ пр€мым методом находим рассто€ние $D$.

јвторы \cite{KK1974} чЄтко пон€ли, что по спектральным лини€м мы видим
скорость $u$ вещества,
а сама фотосфера движетс€ относительно вещества (ведь при расширении падает коэффициент поглощени€ вещества).
ƒаже знаки $u$ и $v_{ph}$ могут быть противоположны, когда фотосфера сжимаетс€.
¬ этом состоит главна€ трудность методов EPM и SEAM:
чтобы эти методы работали, необходимо полагать,
что идЄт свободный разлЄт и скорость вещества составл€ет $u=r/t$.
“ак бывает, когда через какое-то врем€ после взрыва вокруг отсутствует плотное
вещество.
ј в SN~IIn как раз не так, вокруг много вещества и ударна€ волна мес€цами или даже
годами не прорываетс€ в разреженную среду.

«ато, как видно из графиков в работах \cite{ChugaiEa04} и \cite{WooBliHeg2007}, в
таких сверхновых всЄ вещество
за фронтом ударной волны сжалось в холодный плотный слой (Cold Dense Shell,CDS или DS).
‘отосфера приклеена к этому плотному слою, и как раз $u=v_{ph}$  Ц а это можно измерить. ¬сЄ описанное
соответствует идее Ѕааде \cite{Baade1926}, выдвинутой ещЄ в 1920-е годы.

»так, мы можем сформулировать новый метод DSM (Dense Shell Method) дл€ определени€ космологических рассто€ний
с помощью сверхновых типа SN~IIn.
Ётот метод состоит из следующих этапов.
\begin{itemize}
 \item Ќужно измерить \emph{узкие} компоненты спектральных линий дл€ оценки свойств (плотности, скорости)
 околозвЄздной оболочки.  «десь не требуетс€ высока€ точность измерений и моделировани€.
 \item »змерить \emph{широкие} эмиссионные компоненты линий и найти скорость на уровне фотосферы
 (с максимально возможной точностью).
 \item ’от€ закон  $u=r/t$ в этих сверхновых не применим, но $v_{ph}$ соответствует
 теперь \emph{истинной} скорости радиуса фотосферы (а не только
скорости течени€ вещества, как в SN~II-P).
 \item “еперь работает первоначальна€ иде€ Ѕааде \cite{Baade1926} дл€ измерени€ приращений радиуса $dr=v_{ph}dt$
 \emph{путЄм интегрировани€  по времени} (конечно, с необходимым учЄтом рассе€ни€, потемнени€/у€рчени€ к краю, и т.п.).
 ѕолучаемые значени€ изменений радиуса должны использоватьс€ при итераци€х оптимальной модели.
 \item Ќаблюдаемый поток теперь позвол€ет получить рассто€ние $D$ из выражени€
    \begin{equation}
	D = 2\sqrt{\pi \sigma} dr T^2 / dS \ .
	\label{eq:D}
    \end{equation}
\end{itemize}

ѕри значительном изменении $T$ такой простой подход не работает и нужно построить
модель, наилучшим образом воспроизвод€щую наблюдени€ широкополосной фотометрии и скорость $v_{ph}$.
котора€ контролируетс€ по наблюдени€м $dr(t)$.
“ака€ модель нужна дл€ вычислени€ эволюции $r$ и поправочного фактора $\zeta$,
а в реальности и дл€ детальных предсказаний теоретического потока $F_\nu$.

»зложим основной алгоритм метода на основе чернотельной модели с температурой $T$  с поправочным фактором $\zeta$.
Ѕудем считать, что наблюдени€ достаточно часты,
чтобы дл€ р€да точек можно было измерить изменени€ радиуса $dr=v_{ph}dt$.
ѕусть начальный (неизвестный нам) радиус есть $R_0$, а  $R_i=R_0+\Delta R_i$ дл€ $i=1,2,3,...$, где $\Delta R_i$
уже можно считать известными из суммировани€ измеренных $dr$.
“огда
\begin{equation}
 \zeta_i^2(R_0+\Delta R_i)^2 \pi B_\nu(T_{ci}) = 10^{-0.4 A_\nu} D^2 f_{\nu i} ,
\end{equation}
или, извлека€ корень,
\begin{equation}
 \zeta_i (R_0+\Delta R_i) \sqrt{\pi B_\nu(T_{ci})} = 10^{-0.2 A_\nu} D \sqrt{ f_{\nu i} } \ .
\end{equation}

’ороша€ модель даЄт нам набор $\zeta_i$, $T_{ci}$  дл€ всех точек наблюдений,
и по измеренным $f_{\nu i}$, $\Delta R_i$  мы находим методом наименьших квадратов неизвестные $R_0$
и комбинацию $10^{-0.4 A_\nu} D^2$.
„тобы получить рассто€ние $D$, мы должны либо знать поглощение $A_\nu$ из астрономических
наблюдений,
либо получить его из тех же уравнений, записанных дл€ различных спектральных фильтров.
¬ таком случае, уже зна€ $R_0$, мы получим р€д уравнений вида:
\begin{equation}
 10^{-0.4 A_\nu} D^2 = a_s \ ,
\end{equation}
где индекс $s$ обозначает один из фильтров UBVRI, а $a_s$ - некотора€ посто€нна€, завис€ща€ от
выбора фильтра.
Ёто дает нам разности $A_{s1} - A_{s2}$ и,
зна€ закон зависимости экстинкции от частоты,
можно получить $A_\nu$.

— помощью описанного здесь метода DSM в работах \cite{SIB-h,SIB-h1,SIB-h2}
были
получены рассто€ни€ до трЄх сверхновых типа IIn: SN~2006gy, SN~2009ip, SN~2010jl.
—равнение с рассто€ни€ми  до этих объектов,  измеренными  другими (косвенными) методами,
доказывает
работоспособность нового пр€мого метода, который позвол€ет избавитьс€ от
систематических ошибок,
вносимых лестницей космологических рассто€ний

\subsection{—ли€ние нейтронных звЄзд и метод стандартной сирены}

¬ 2015 году произошел прорыв в эру гравитационно-волновой астрономии:
впервые было объ€влено о наблюдении сли€ни€ пары массивных
чЄрных дыр.
–анее человечество наблюдало ¬селенную в электромагнитном и нейтринном каналах, теперь оно обрело новое ``зрение'':
абсолютно новый гравитационно-волновой канал дл€ исследовани€ объектов во ¬селенной,
изучени€ их свойств и проверки предсказаний теорий.
ѕервый сигнал гравитационных волн был получен 14 сент€бр€ 2015 года двум€ антеннами коллаборации LIGO~\cite{2016PhRvL.116f1102A}. 
јнализ показал, что источником сигнала были две массивные чЄрные дыры.
«атем последовало ещЄ несколько открытий сигналов сли€ни€ чЄрных дыр вплоть до 14 августа 2017 года (уже при участии
обновлЄнного детектора VIRGO).

“ри дн€ спуст€, 17 августа 2017 было сделано ещЄ одно важное открытие: был зарегистрирован сигнал гравитационных волн GW180717~\cite{2017PhRvL.119p1101A},
продолжительность и форма которого свидетельствовала не о сли€нии чЄрных дыр, а о сли€нии нейтронных звЄзд в тесной двойной системе.
ѕо временной задержке сигнала двух антенн LIGO с учЄтом данных VIRGO удалось определить область
нахождени€ источника --- примерно 30 квадратных градусов на небесной сфере.
 осмические обсерватории Fermi~\cite{2017ApJ...848L..13A} и »нтеграл~\cite{2017ApJ...848L..15S} увидели слабый короткий
гамма-всплеск GRB170817A в квадрате локализации источника гравитационных волн через 1.7 секунды после
потери сигнала LIGO.

ѕо амплитуде гравитационной волны и по картине изменени€ еЄ частоты можно оценить рассто€ние до источника.
ќно оказалось примерно 40~ћпк с ошибкой в несколько дес€тков процентов, но даже при большой ошибке на таком рассто€нии
в области локализации на небесной сфере совсем немного галактик,
поэтому наземные телескопы (в том числе робот ћј—“≈–~\cite{2017ApJ...850L...1L}) довольно быстро нашли вспышку слабой сверхновой (так называемой ``килоновой'') в галактике NGC 4993, котора€ находитс€ от нас на рассто€нии 40 мегапарсек.

¬ св€зи с этим выдающимс€ открытием мировой науки важно подчеркнуть, что первое предсказание такого €влени€ было сделано в 1980-х годах в работе одного из авторов насто€щего обзора с коллегами~\cite{1984PAZh...10..422B,1984SvAL...10..177B}.
«десь впервые было предсказано, что при сли€нии пары нейтронных звЄзд должен происходить выброс т€жЄлых элементов, электромагнитна€ вспышка, в частности гамма-всплеск, а не только излучение гравитационных волн.
ƒо этой работы предлагались модели сли€ни€ нейтронной звезды и чЄрной дыры, но с неэффективным гамма-всплеском внутри нашей
√алактики~\cite{1976ApJ...210..549L},
а Ѕлинников и др.~\cite{1984PAZh...10..422B,1984SvAL...10..177B} пр€мо писали про дес€тки мегапарсек и про энергию
взрыва пор€дка сверхновой.
–аньше этой работы была также замечательна€ стать€ \cite{1977ApJ...215..311C}, в которой было всЄ предсказано про гравитационно-волновой и нейтринный
сигнал при сли€нии нейтронных звЄзд в паре, но ничего не говорилось про электромагнитный сигнал, который в миллион раз слабее
первых двух, но гораздо легче регистрируетс€.
¬ дальнейшем сценарий работы~\cite{1984SvAL...10..177B} был подкреплЄн количественным расчЄтом~\cite{1990AZh....67.1181B,1990SvA....34..595B},
в которoм были получены все основные характеристики слабого
гамма-всплеска GRB170817A (полна€ энерги€  и жесткость фотонов в гамма-диапазоне, характерные скорости выброса т€жЄлых элементов).

‘отометрическое рассто€ние до источника гравитационной волны, измеренное чисто по гравитационно-волновому сигналу --- так называемый метод стандартной сирены, --- даЄт
возможность измерить посто€нную ’абла $H_0$, если известно красное смещение материнской галактики.
Ёто было впервые указано Ўутцем~\cite{1986Natur.323..310S} как раз после статьи~\cite{1984SvAL...10..177B},
когда стало €сно, что мощный сигнал в гравитационных волнах должен сопровождатьс€ слабым электромагнитным сигналом.
ќтождествление электромагнитного сигнала с конкретной галактикой гораздо надЄжней, чем по гравитационным волнам, а это повышает
точность измерени€ параметра ’аббла.
ќценка рассто€ни€ по гравитационно-волновому сигналу неизбежно содержит
неопределенности в основном из-за неопределенности угла наклона плоскости орбиты
(получено рассто€ние $43.8_{-6.9}^{+2.9}$~ћпк уже с учЄтом локализации килоновой \cite{2017Natur.551...85A}).
Ѕыл получен параметр ’аббла с использованием данных гравитационной волны дл€
GW170717 $H_0 = 70.0^{+12.0}_{-8.0}$~км/с/ћпк (на уровне достоверности 68.3\% \cite{2017Natur.551...85A}).
ќчевидно, что эта оценка совместима с измерени€ми по сверхновым $73.45 \pm 1.66$~км/с/ћпк~\cite{2016ApJ...826...56R,2018ApJ...861..126R} и по реликтовому фону по данным PLANCK: $H_0 = (67.8 \pm 0.9)$~\cite{PlanckCollaboration2016}, и более новое значение
$H_0 = (67.4 \pm 0.5)$~\cite{2018arXiv180706209P}.
 ак видно из оценок доверительных интервалов, между этими двум€ группами имеетс€ значимое расхождение (tension) в значении $H_0$.
 огда будет открыто большое количество сли€ний пар нейтронных звезд,
тогда точность измерени€ $H_0$ этим методом можно будет улучшить.
«аметим, что в насто€щее врем€ точность нашего метода DSM~\cite{SIB-h,SIB-h1,SIB-h2}, описанного выше, вполне конкуретноспособна с методом стандартной сирены, и его можно
развивать при меньших материальных затратах, чем последний, поскольку самы мощные оптические телескопы всЄ же дешевле, чем установки дл€ детектировани€ гравитационных волн.

\section{“Ємна€ энерги€ \label{s-DE} }

“Ємна€ энерги€ -- это общее название неизвестной субстанции, котора€, возможно,
вызывает космологическое ускорение. “Ємной энергией может быть люба€ форма
вещества, дл€ которой справедливо уравнение состо€ни€ $P= w \rho$ (\ref{P-w-rho})
с $w < -1/3$, см. уравнение (\ref{ddot-a}). Ќапомним, что согласно наблюдени€м
параметр $w$ весьма близок к (-1). —пециальный частный случай $w = -1$ отвечает
обсуждавшейс€ выше вакуумной энергии и принципиально отличаетс€ от вс€кого
другого.

 ак справедливо отмечено в работе \cite{Chernin2008}, из-за наличи€ вакуумной энергии с
таким уравнением состо€ни€ дл€ двух атомов
водорода в абсолютном вакууме нашей современной ¬селенной достаточно большим
рассто€нием, на котором их гравитационное прит€жение будет компенсироватьс€
гравитационным отталкиванием вакуумной энергии, будет уже один метр. —трого
говор€,  отталкивание на этом рассто€нии будет незаметным на фоне
наведенного электрического дипольного взаимодействи€ атомов. Ётот мысленный
эксперимент справедлив дл€ идеально нейтральных частиц с массой в 1~√э¬.
¬ реальном мире такое утверждение верно дл€ двух  галактик с массой ћлечного ѕути на рассто€нии в пару мегапарсек.

јльтернативным механизмом ускоренного космологического расширени€ могла
бы быть модификаци€ гравитации на больших (космологических) рассто€ни€х,
рассматриваема€ в следующем разделе.

—уществование тЄмной энергии было предсказано в работе~\cite{ad-nuff}, еще до
открыти€ ускоренного ускорени€. “ам и в  нескольких последующих
работах~\cite{ad-vector,ad-tensor} было сделано утверждение, что компенсационный
механизм должен иметь недокомпенсированный остаток с плотностью энергии,
близкой к космологической и нестандартным уравнением состо€ни€. ¬ частности,
в работе~\cite{ad-tensor} отмечалось, что может возникнуть уравнение состо€ни€
с $w<-1$, которое за конечное врем€ приведет к космологической сингул€рности,
впоследствии получившей название фантомной~\cite{phantom}.

ƒругие ранние работы, в которых отмечалось, что скал€рное поле может вести себ€
при некоторых потенциалах, как завис€ща€ от времени вакуумо-подобна€ энерги€,
приведены в стать€х~\cite{earlier-papers,earlier-papers1,earlier-papers2,earlier-papers3}.
¬ажным продвижением в описании тЄмной
энергии было обнаружение потенциалов, которые привод€т к так называемым
след€щим (tracking) решени€м, впервые найденным в работах~\cite{ratra-peebles}
и в несколько более поздних стать€х~\cite{wetter}.  “акие след€щие решени€
интересны тем, что за счет выбора потенциала скал€рного пол€
можно сделать так, чтобы
плотность энергии этого пол€ довольно естественно оказывалась близкой к
плотности энергии обычного вещества в насто€щее
врем€, тем самым реша€ проблему близости $\rho_m$ к $\rho^{\rm vac}$ в
современной вселенной.
Ёта тематика быстро стала очень попул€рной~\cite{quintessence-0,quintessence-1,quintessence-2,quintessence-3,quintessence-4,quintessence-5,quintessence-6,quintessence-7,quintessence-8,quintessence-9,quintessence-10,quintessence-11,quintessence-12,quintessence-13,quintessence-14,quintessence-15,quintessence-16,quintessence-17}, см.
также обзоры~\cite{rev-quint,rev-quint1,rev-quint2}.

¬озможные формы моделей тЄмной энергии, основанных на скал€рном поле, можно разделить
на несколько классов. ¬ частности,
это могло бы быть поле с обычным кинетическим членом, см.
уравнение (\ref{action-scalar}) в приложении B, однако потенциал такого пол€ $U(\phi)$,
как правило, выбираетс€ довольно необычным и, как минимум, приводит к неперенормируемой,
сильно нелинейной теории. Ёто поле было названо ``квинтэссенци€'', согласно предложению
 олдуэлла (Caldwell) и др.~\cite{quintessence-3}.

ƒругие варианты включают так называемую  -эссенцию
( -essence) - теорию, в которой кинетический член имеет нестандартную форму:
\be
A_{K} = \int d^4 x \sqrt {-g} \, P(\phi, X),
\label{A-K}
\ee
$X = (1/2) g^{\mu\nu} \partial_\mu \phi \partial_\nu \phi $ - обычный кинетический член,
но функци€ $P$ нелинейно зависит от него. „асто $P$ выбираетс€ в факторизованном виде
$ P(\phi,X) = f_1(\phi)\,f_2 (X)$. —вобода тут очень больша€, чем и объ€сн€етс€ поток работ
на эту тему.

≈ще один способ описани€ тЄмной энергии опираетс€ на скал€рное тахионное поле,
примером действи€ дл€ которого может служить выражение:
\be
A_T = \int d^4 x\, V(\phi) \  \sqrt{-\det [ g_{\mu\nu} - \partial_\mu \phi\, \partial_\nu \phi ]},
\label{A-T}
\ee
где $V(\phi)$ - некотора€ потенциальна€  функци€, подбираема€ {\it ad hoc}.  ќбратим
внимание на необычную размерность пол€ $\phi$, обратную первой степени энергии.

“акое действие напоминает действие Ѕорна-»нфельда~\cite{born-infeld}, где вместо
$\partial_\mu \phi\, \partial_\nu \phi $ стоит тензор ћаксвелла, $F_{\mu\nu}$ или же
модификацию гравитации~\cite{det-gravity}, где в действие под знаком корн€
подставл€етс€ $R_{\mu\nu}$. ¬се эти теории имеют р€д похожих свойств.

Ќаконец, стоит упом€нуть  ``фантoмное'' поле со знаком действи€, который противоположен
нормальному. “акое поле приводит к $w <-1$, как видно из выражени€ дл€ плотности
энергии и давлени€  (\ref{rho-P}) при изменени€ знака перед производными пол€ $\phi$.
–азумеетс€, неправильный знак кинетического члена приводит к неустойчивости
высокочастотных мод, что представл€ет серьезную проблему дл€ таких теорий уже
на классическом уровне, не говор€ уже о проблемах квантовой теории.

Ѕолее подробное обсуждение этих вариантов феноменологического описани€ (эффективного,
в духе эффективных лагранжианов) тЄмной энергии и обширный список литературы
можно найти в обзорах~\cite{rev-quint,rev-quint1,rev-quint2}. ¬озникает впечатление, что
всЄ, что считалось патологией ранее в нормальных теори€х, используетс€ дл€
``создани€''
ускоренного расширени€. ћы не можем, однако, исключить, что решение проблемы лежит
именно на этом пути.

ѕростейшим кандидатом на роль носител€ тЄмной энергии €вл€етс€ скал€рное поле
$\phi$ с каноническим кинетическим членом и с очень
малой (или даже нулевой) массой, $m <H_0$,  где $H_0$ -- параметр ’аббла в современную
эпоху. Ёто поле удовлетвор€ет уравнению  лейна-√ордона:
\be
\ddot \phi + 3 H\dot \phi - \frac{\nabla^2 \phi }{a^2(t)} + U'{\phi} =0\,,
\label{ddot-phi}
 \ee
где $U(\phi)$ --  потенциал этого пол€, а $U' = dU/d\phi$.

≈сли параметр ’аббла велик (ниже будут приведены конкретные услови€ необходимой
величины), то решение этого уравнени€ окажетс€ приближенно посто€нным.
¬о-первых,
пространственные неоднородности $\phi$ замоютс€ из-за космологического
расширени€, вызываемого фактором $1/a^2$ при пространственных производных.
ј втора€ производна€ по времени
окажетс€ пренебрежимой в силу большого хаббловского трени€. ќтметим, что
дл€ пространственно однородного пол€ $\phi = \phi (t)$ уравнение (\ref{ddot-phi})
совпадает с уравнением движени€ в ньютоновской механике, где роль координаты
играет $\phi (t)$. ѕоэтому его решение интуитивно легко
представить, зна€ форму потенциала $U(\phi)$. „лен
$3 H\dot \phi$ ведет себ€, как жидкое трени€ в механике, так что при большом
трении движение напоминает плавание в глицерине, когда скорость под воздействием
посто€нной силы стремитс€
к посто€нной величине.

»так, предположим, что
вторыми производными в уравнении (\ref{dot-phi}) можно пренебречь, найдем решение
и проверим, удовлетвор€ет ли найденное решение  исходным предположени€м о медленности изменени€ $\phi$. ¬ указанных предположени€х $\phi$ удовлетвор€ет
уравнению первого пор€дка:
\be
\dot\phi = -U'/3H.
\label{dot-phi}
\ee
Ёто так называемое приближение медленного скатывани€, широко используемое в
инфл€ционных модел€х.

≈сли основной вклад в космологическую плотность энергии $\rho$, дает $\phi$, то
в пределе медленного изменени€ этого пол€ $\rho$ определ€етс€ потенциалом
$U(\phi)$ и согласно уравнению (\ref{H2}) параметр ’аббла равен
\be
H^2 = \frac{8\pi U}{3 m_{Pl}^2}\,.
\label{H2-of-phi}
\ee

ѕриближение медленного скатывани€ будет справедливым, если
$ \ddot \phi \ll 3 H\dot \phi $ и  $\dot\phi^2 \ll 2 U(\phi)$. ƒл€ реализации этих
условий необходимо:
\be
 U''/ U  \ll \frac{8\pi}{3 m_{Pl}^2} \,,
\label{U-2prime}
\ee
что требует очень большой величины пол€ $\phi$.
¬ частности, дл€ массивного невзаимодействующего пол€, т.е. пол€ с гармоническим
потенциалом $U = m^2\phi^2/2$ указанные выше услови€ будут выполнены, если
\be
\phi^2 > (4\pi/3)\, m^2_{Pl}\,.
\label{phi2}
\ee

≈сли потребовать, чтобы плотность энергии пол€ $\phi$, а именно
$\rho_\phi = m^2_\phi \phi^2$
была пор€дка современной космологической плотности энергии, то его масса должна
быть ограничена очень низким значением: $m_\phi < 1/t_U \approx  10^{-42}$ э¬.

Ќекоторые сведени€ о теории скал€рного пол€ приведены в ѕриложении ¬.
“ам дан вывод тензора энергии-импульса скал€рного пол€, который имеет вид (\ref{T-mu-nu-of-phi}).
ƒл€ квази-посто€нного и квази-однородного
пол€ он становитс€ пропорциональным метрическому тензору,
$T_{\mu\nu} \sim g_{\mu\nu}$, реализу€ вакуумо-подобное уравнение состо€ни€
(\ref{T-vac}), т.е. $w \approx -1$, что и приводит к  почти экспоненциальному
ускоренному расширению. —ущественно, однако, что поле $\phi$, хоть и медленно,
но падает. ѕредполагаетс€, что в точке равновеси€, где $dU/d\phi =0$, потенциал
также обращаетс€ в ноль, $U = 0$. (Ёто требование отсутстви€ вакуумной энергии
и его выполнение, вообще говор€, не об€зательно.) ѕростыми примерами таких
потенциалов, стрем€щихс€ к нулю на бесконечности, €вл€ютс€ степенной
$U \sim 1/\phi^q $ или экспоненциальный $U \sim \exp (-\phi/\mu)$, где
$\mu$-постo€нный параметр размерности массы~\cite{exp-U}. Ёти потенциалы
были специально введены дл€ феноменологического описани€ ускоренного
расширени€, но фундаментальные причины дл€ их возникновени€ довольно шаткие.

ƒвижение $\phi(t)$ в таких потенциалах принципиально отлично от движени€
в потенциалах, имеющих минимум при конечном $\phi$, например,
$U(\phi) = m^2_\phi \phi^2 /2 $ или $U(\phi) = \lambda \phi^4/4$ и также нулевое
значение вакуумной энергии. ќтметим, что эти два потенциала естественны в
квантовой теории пол€, так как отвечают перенормируемой теории.
¬ таких потенциалах произойдет существенное изменение
режима расширени€, когда $\phi$ упадет до такого значени€, при котором
$H^2$  сравн€етс€ с $m_\phi^2$ или с $\lambda \phi^2$, т.е. $\phi$ упадет
заметно ниже
планковского значени€. ¬ этот момент квазиэкспоненциальный режим ускоренного
расширени€ перейдет в обычный замедл€ющийс€ фридмановский: либо
 нерел€тивистский ($w=0$) дл€
квадратичного потенциала, либо в рел€тивистский ($w=1/3$) дл€ квартичного.
ѕри этом $\phi$ начнет осциллировать вокруг минимума, порожда€ безмассовые
частицы, фотоны или гравитоны.

≈сли же потенциал не имеет минимума при конечном $\phi$, то поле монотонно
стремитс€ к нулю и режим расширени€ вечно будет ускоренным и в режиме
медленного скатывани€ (slow roll) параметр $w$ будет посто€нным и отрицательным,
обеспечива€ ускоренное расширение.  –ассмотрим дл€ примера экспоненциальный
потенциал~\cite{exp-U,exp-U2}:
\be
U(\phi) = U_0 \exp \left( -\frac{\phi}{\mu} \right).
\label{U-exp}
\ee
¬ предположении, что $\dot \phi^2 \ll U(\phi)$ уравнение движени€ $\phi$ сводитс€
к (\ref{dot-phi}), которое легко интегрируетс€, если в плотности энергии доминирует
потенциальный член и параметр ’аббла даетс€ выражением (\ref{H2-of-phi}). ѕоле
$\phi$ логарифмически растет со временем:
 \be
\phi (t) = 2 \mu\,\ln \left[ \sqrt{\frac{U_0}{96\pi \mu^4} } \, m_{Pl} (t- t_0) +
\exp \left( \frac{\phi_0}{2\mu} \right) \right],
\label{phi-for-exp}
\ee
где $\phi_0$-величина пол€ в начальный момент  $t_0$.

—оответственно при больших $t$:
\be
U(\phi) \approx \frac {96 \pi \mu^4}{(m_{Pl} t)^2},\,\,\,\,
(\dot \phi)^2 \approx \frac{4 \mu^2}{t^2}
\label{U-large-t}
\ee
и условое медленного скатывани€ выполн€етс€ при $m_{Pl} \lesssim \mu$.
ѕараметр ’аббла при этом падает обратно пропорционально времени:
\be
H \approx \frac{16 \pi \mu}{m_{pl} t},
\label{H-exp}
\ee
т.е. расширение будет степенным:
\be
a \sim t^{16\pi \mu/{m_{Pl}}}
\label{a-of-t-exp}
\ee
ѕри $16\pi \mu > m_{Pl} $ процессс расширени€ будет идти с ускорением.
ѕри $m_{Pl} /\mu \rightarrow 0$  закон расширени€ будет стремитьс€ к
экспоненциальному, а $ w \rar -1$.

\section{ ћодифицированна€ гравитаци€ \label{s-grav-mdf}}

 онкурирующей гипотезой феноменологического описани€  механизма
космологического ускорени€ €вл€етс€ предположение о модификаци€
гравитации при малой кривизне. ќбычно это осуществл€етс€ добавлением
некоторой функции скал€ра кривизны $R$ к действию √ильберта-Ёйнштейна:
\be
A = \frac{m_{Pl}^2}{16\pi} \int d^4 x \sqrt{-g}  \left[R+F(R)\right]+A_m,
\label{A1}
\ee
где  $m_{Pl}= 1. 2 2\cdot 10^{19}$ GeV - масса ѕланка,
$R$ - скал€р кривизны и $A_m$ - действие полей материи.
ƒополнительный член $F(R)$ мен€ет гравитацию на больших рассто€ни€х,
поэтому говор€т об инфракрасной модификации гравитации. Ќелинейна€ функци€
$F(R)$ подбираетс€ таком образом, чтобы уравнени€ движени€ имели решени€
$R= \mbox{const}$ в отсутствии материи. ќтметим, что, в принципе, можно бы
рассматривать более сложные варианты модификации гравитации с какими-то
функци€ми не только $R$, но и инвариантных комбинаций, построенных из
$R_{\mu\nu}R^{\mu\nu}$ или тензора –имана. ¬ насто€щее врем€, однако,
рассматриваютс€ лишь теории с $F(R)$. Ёто диктуетс€ не только простотой, но
и проблемами с духами и тахионами в более сложных теори€х.

”равнени€ Ёйнштейна дл€ $F(R)$-гравитации модифицируютс€ следующим образом:
\be
\left( 1 + F'\right) R_{\mu\nu} -\frac{1}{2}\left( R + F\right)g_{\mu\nu}
+ \left( g_{\mu\nu} D_\alpha D^\alpha - D_\mu D_\nu \right) F'  =
\frac{8\pi T^{(m)}_{\mu\nu}}{m_{Pl}^2}\,,
\label{eq-of-mot}
\ee
где $F' = dF/dR $, $D_\mu$ ковариантна€ производна€ в метрике
‘ридмана-–обертсона-”окера и $T^{(m)}_{\mu\nu}$ - тензор энергии-импульса
вещества. ¬з€в след по индексам этого уравнени€,
$g_{\mu\nu} \delta A/\delta g_{\mu\nu} $, получим:
\be
\label{eq:trace}
3 D^2 F'_R-R+RF'_R-2F=  8\pi T^\mu_\mu/m_{Pl}^2,
\ee
где $D^2\equiv D_\mu D^\mu$ - ковариантный оператор ƒ'јламбера. ѕри рассмотрении
$F(R)$-теории изучение решений этого уравнени€ часто оказываетс€ вполне достаточным.

—ледует, однако, заметить, что добавление нелинейного по кривизне слагаемого в действие
может привезти к серьезным патологи€м в теории, в частности, к нарушению унитарности и
по€влению духов и/или тахионов. ќбычно эти проблемы игнорируютс€ при кривизнах близких
к планковским --  кто знает, что там происходит -- но заметные отклонени€ от обычной гравитации
при малых кривизнах, т.е.  в пределе слабого пол€, могут привести к противоречию с хорошо
установленными наблюдаемыми фактами. ћы, однако, не будем останавливатьс€ на всем круге
этих проблем, а ограничимс€ лишь возможными отклонени€ми
от стандартной ќ“ќ на уровне решений классических уравнений движени€.

¬ пионерских стать€х по описанию космологического ускорени€ за счет модификации
гравитации~\cite{grav-mdf,grav-mdf1} функци€  $F(R)$ была выбрана в виде $F(R)= -\mu^4/R$,
где $\mu$ - малый параметр размерности массы. ќднако, как было показано в
работе~\cite{DolgKaw},  такой выбор функции $F(R)$ приводит к очень сильной
экспоненциальной неустойчивости теории при наличии материи, так что
обычна€ теори€ гравитации будет сильно искажена. ƒействительно рассмотрим
уравнение, описывающее эволюцию кривизны как функцию времени (\ref{eq:trace}),
дл€ теории с $F(R)= -\mu^4/R$:
\be
D^2 R - 3 \,\frac{(D_\alpha R)\,(D^\alpha R)}{R} =
\frac{R^2}{2}  - \frac{R^4}{6\mu^4}  - \frac {\tilde T\,R^3 }{6\mu^4}\,.
\label{D2-R}
\ee
 «десь ${\tilde T=8\pi T_\nu^\nu/m_{Pl}^2>0}$.

Ёто уравнение в отсутствии вещества имеет очевидное решение
$ R^2 = 3\mu^4$, описывающее вселенную де —иттера с посто€нным скал€ром
кривизны.

ѕредположим теперь, что источником гравитационного
пол€ €вл€етс€ обычное небесное тело, например, «емл€ или —олнце,
так что создаваемое гравитационное поле оказываетс€ слабым
и фоновое пространство €вл€етс€ €вл€етс€ четырехмерно плоским с метрикой ћинковского. Ѕудем искать решение уравнени€ (\ref{D2-R}) по теории
возмущений, предполага€ малое отклонение от стандартной ќ“ќ. ¬ низшем пор€дке
кривизна алгебраически выражаетс€ через след тензора  энергии-импульса:
$R_0 =-\tilde T$.  ќбычное решение в пустоте $R = 0$ теперь оказываетс€
 приближенным, так как функци€ $F(R)$, как уже отмечалось, подбираетс€ в виде, чтобы
уравнение (\ref{eq:trace}) в отсутствие материи имело бы решение $R = R_c = \mbox{const}$,
 где $R_c $ равн€етс€ наблюдаемой космологической кривизне и пренебрежимо мало
с кривизной внутри любых материальных тел, в отношении космологической
плотности энергии к плотности энергии этих тел. ћожно проверить также, что
стационарное решение  вне гравитирующего тела в модифицированной теории
быстро убывает при удалении от этого тела.
Ќа этом основании можно  сделать вывод, что стационарные решени€ в
модифицированной гравитации хорошо согласуютс€ с ньютоновским пределом
стандартной ќ“ќ при достаточно малом $\mu$.

Ќа первый взгл€д введение дополнительного члена в действие $\mu^4/R$ при
достаточно малом $\mu$ не приводит к значительным отклонени€м от обычной
теории гравитации, но это совсем не так. ƒело в том, что в модифицированной теории
скал€р $R$ становитс€ динамической величиной, эволюци€ которой определ€етс€
уравнением второго пор€дка с малым коэффициентом  перед старшей производной
по времени. ¬ силу этого возникает очень сильна€ неустойчивость решени€ в
присутствии материальных тел~\cite{DolgKaw}.

ѕрименим уравнение (\ref{D2-R}) к пертурбативному вычислению гравитационного
пол€, точнее скал€ра кривизны, внутри какого-либо небесного тела, например,
—олнца, «емли или просто облака газа в галактике с завис€щей от времени
плотностью энергии. Ѕудем искать решение в виде
 $R = R_0 + R_1$, где $R_0$ -  обычное решение ќ“ќ, т.е.
${ R_0 = -\tilde T }$. ѕолага€, как отмечалось выше, фоновую метрику плоской,
найдем, что отклонение от ќ“ќ, $R_1$, удовлетвор€ет уравнению
\be
\ddot R_1 -\Delta R_1 - \frac{6\dot {\tilde T}}{\tilde T}\, \dot R_1
+\frac{6\partial_j \tilde T}{\tilde T}\, \partial_j R_1
{ {  +R_1 \left[ \tilde T + 3\,\frac{ (\partial_\alpha \tilde T)^2}{\tilde T^2}
 {-\frac{\tilde T^3}{6\mu^4} }\right] } }
 = \Delta \tilde T + \frac{\tilde T^2}{2} - \frac{3 (\partial_\alpha \tilde T)^2}{\tilde T}\,,
\label{ddotr1}
\ee
где $(\partial_\alpha \tilde T)^2 =  \dot{\tilde T}^2 - (\partial_j \tilde T)^2 $.

ѕоследнее слагаемое в квадратных скобках в левой части приводит к
экспоненциальной неустойчивости малых флуктуаций, а также и к неустойчивости
гравитационного пол€, создаваемого регул€рно мен€ющейс€ со временем плотностью
массы/энергии рассматриваемого тела. ’арактерное врем€ неустойчивости
оказываетс€ очень коротким:
\be
\tau_{\rm instab}=\frac{\sqrt{6}\mu^2} {T^{3/2} }
\sim 10^{-26}\, \mbox{сек}\,
\left(\frac{\rho_m}{ \mbox{ г/см}^{3} }\right)^{-3/2}\, ,
\label{t-eff}
\ee
где ${\rho_m}$ - плотность массы/энергии рассматриваемого тела, а
$ \mu^{-1} \sim t_u \approx 3\cdot 10^{17} $ сек. ¬ силу малости $\mu$
 слагаемое, содержащее $\mu^4$ в знаменателе в уравнении~(\ref{ddotr1}), намного превосходит все
другие члены.

ќбычно пространственные производные ликвидируют или подавл€ют неустойчивость.
Ќапример, джинсовска€ неустойчивость устран€етс€ противодействующим гравитации
давлением. ќднако в рассматриваемом случае слагаемое, содержащие лапласиан,
$\Delta R_1$, которое обратно пропорционально квадрату размера системы, намного
меньше, чем член, вызывающий неустойчивость. Ћегко убедитьс€, что неустойчивость
может быть подавлена на масштабах менее, чем комптоновска€ длина волны протона,
т.е. короче, чем $10^{-14}$ см.

„тобы решить проблему неустойчивости, была предложена дальнейша€ модификаци€
модифицированной гравитации. ћы рассмотрим здесь лишь класс моделей,
предложенных в работах~\cite{mdf-mdf,mdf-mdf1,mdf-mdf2}. Ќекоторые другие формы модификации
гравитации рассмотрены в обзоре~\cite{odin-rev}.
–азличные формы действи€, изучаемые в работах~\cite{mdf-mdf,mdf-mdf1,mdf-mdf2},
имеют вид:
\be\label{eq:cr1}
F_{\rm HS}(R) &=&  - {R_{\rm vac} \over 2} {c \left({R \over R_{\rm vac}}\right)^{2n} \over 1+
c\left({R \over R_{\rm vac}}\right)^{2n}}\,,  \label{eq:cr201} \\
F_{\rm AB}(R) &=&  {\epsilon \over 2}\,
{\rm log} \left[ {\cosh\left({R \over \epsilon}-b\right) \over \cosh b} \right]
 - \frac{R}{2}\,, \\
F(R)_S &=& \lambda R_0 \,\left[ {\left(1+ \frac{R^2}{R_0^2}\right)^{-n}} - 1 \right]\,.
\label{F-AAS}
\ee
Ќесмотр€ на внешнее различие, все эти выражени€ привод€т практически к
совпадающим следстви€м. Ќиже мы будем следовать анализу, проведенному в
работе~\cite{appl-bat-star}, а в конкретных примерах мы будем использовать
выражение (\ref{F-AAS}).

¬вед€ обозначение ${f = R +F(R)}$, перепишем полевые уравнени€ в виде:
\begin{equation}
f' R_{\mu}^{\nu} - \frac{f}{2}\delta_{\mu}^{\nu}+
(\delta_{\mu}^{\nu}\Box - D_{\mu} D^{\nu})f'= \frac{ 8\pi\,T_{\mu}^{\nu}}{M_{Pl}^{2}}\,.
\label{FReq}
\end{equation}
«десь и всюду ниже штрих у $f$ или $F$ означает производную по $R$.

¬з€в след по индексам $\mu$ и $\nu$, придем  к замкнутому уравнению дл€ $R$:
\begin{equation}
3\Box f'(R) + Rf'(R) - 2f(R)= 8 \pi M_{\rm Pl}^{-2}T^\mu_\mu \,.
\label{trace-eq}
\end{equation}

“ака€ модифицированна€ теори€ приведет к ускоренному космологическому
расширению при условии, что уравнение
\be
R f'(R) - 2 f(R) = 0
\ee
имеет решение ${R=R_1>0}$, с (приближенно) посто€нным $R_1$

ƒл€ того, чтобы избежать возможных патологий в теории, необходимо потребовать
выполнение следующих условий:\\
1. ”стойчивость космологических решений в будущем:
\begin{equation}
F'(R_1)/F''(R_1)>R_1\,.
\end{equation}
{2.  лассическа€ и квантова€ устойчивость (гравитационное прит€жение и отсутствие
духов):
\begin{equation}
F'(R)>0\,,
\label{stabil}
\end{equation}
{3. ќтсутствие неустойчивости в материальных телах, продемонстрированное выше:}
\begin{equation}
F''(R)>0\,.
\end{equation}

Ќесмотр€ на существенное улучшение эти дважды модифицированные версии теории
по-прежнему привод€т к серьезным проблемам. ¬о-первых, в космологической
ситуации, когда плотность энергии падает со временем, в недалеком прошлом должна
была существовать сингул€рность, когда кривизна имела бесконечно большую
величину хот€ плотности энергии оставалась конечной.

Ѕолее того, астрономические системы с растущей плотностью энергии либо должны
уже прийти к сингул€рному состо€нию, либо прийдут к нему  в близком
будущем~\cite{frolov,ea-ad}. —леду€ работе~\cite{ea-ad}, рассмотрим версию
(\ref{F-AAS}) модифицированной гравитации в пределе
${R \gg R_0} $,  когда можно приближенно прин€ть
\be
F(R) \approx -\lambda R_0 \left[ 1 -\left(\frac{R_0}{R}\right)^{2n} \right]
\label{F-large-R}
\ee
и проанализируем эволюцию ${R}$  в массивных астрономических объектах
с растущей со временем плотностью массы/энергии ${\rho \gg \rho_{cosm}}$.

ѕредположим, как и выше, что гравитационное поле изучаемых объектов €вл€етс€ слабым,
так что ковариантные производные могут быть заменены на обычные. ¬ этом приближении
уравнение (\ref{trace-eq}) принимает вид:
\be
(\partial^2_t - \Delta) R -(2n+2) \frac{\dot R^2 - (\nabla R)^2}{R} +
 {{\frac{R^2}{3n(2n+1)} \left[\frac{R^{2n}}{R_0^{2n}} -(n+1) \right] }} \nonumber \\
 -\frac{R^{2n+2}}{6n(2n+1)\lambda R_0^{2n+1}} (R + \tilde T)=0\,.
\label{eq-for-R}
\ee
”равнение кардинально упроститс€, если вместо $R$ ввести другую неизвестную
функцию ${w \equiv F' = - 2n\lambda \left({R_0}/{R}\right)^{2n+1} }$ котора€ подчин€етс€
уравнению:
\be
(\partial^2_t - \Delta) w  + dU(w)/dw = 0\,,
\label{eq-for-w}
\ee
где потенциал ${U(w)}$ равен:
\be
U(w) = \frac{1}{3}\left( \tilde T - 2\lambda R_0\right) w +
{{\frac{R_0}{3} \left[ \frac{q^\nu}{2n\nu} w^{2n\nu}+ \left(q^\nu
    +\frac{2\lambda}{q^{2n\nu} } \right) \,\frac{w^{1+2n\nu}}{1+2n\nu}\right]\,,
}}
\label{U-of-w}
\ee
с параметрами:  ${\nu = 1/(2n+1)}$ и ${q= 2n\lambda}$. «десь мы следуем обозначени€м
работы~\cite{ea-ad} и надеемс€, что совпадени€ обозначени€ $w$ дл€ новой функции
кривизны и космологического параметра, св€щывающего давление и плотоность
энергии, не приведет к путанице.

«аметим, что обсуждаемой сингул€рности $R = \infty$ отвечает $w = 0$. Ёто
обсто€тельство при анализе сингул€рности позвол€ет спокойно пренебречь
пространственными производными в этом уравнении.
ƒействительно, обычно пространственные производные не дают функции расти. Ќаоборот,
они ``тащат'' еЄ в ноль. ≈сли сделать преобразование ‘урье по координате, то
$-\Delta w = k^2 w$, т.е. это слагаемое ведет себ€, как потенциал гармонического
осцилл€тора с минимумом в нуле. “аким образом, если сингул€рность возникает в
пренебрежении членом $\Delta w$, то она тем более возникнет и при учете этого члена.

≈сли ограничитьс€ лишь главными по малому отношению $R_0/R$ слагаемыми
и пренебречь
пространственными производными, то уравнение (\ref{eq-for-w}) превращаетс€ в обычное
осцилл€торное уравнение:
\be
\ddot w + \frac{\tilde T}{3} - \frac{q^\nu (-R_0)}{3w^\nu}=0
\label{eq-w-simple}
\ee
с потенциалом, который в этом приближении принимает вид
\be
U(w) =\frac{\tilde T w}{3} - \frac{ q^\nu (-R_0) w^{1-\nu}}{ 3(1-\nu)}.
\label{U-of-w2}
\ee
ѕотенциал {$U$}} будет зависеть от времени, если плотность массы/энергии рассматриваемого
небесного тела мен€етс€ со временем. ¬озьмем дл€ примера
объект, состо€щий из
нерел€тивистского вещества с растущей по линейному закону плотностью массы:
\be
\tilde T\equiv\tilde T(t) =\tilde T_0 (1 +t/t_{\rm contr})\, ,
\ee
где $t_{\rm contr}$-характерное врем€ сжати€ системы.  –азумеетс€, линейное приближение
когда-то должно стать несправедливым, но результаты качественно не мен€ютс€.

ƒл€ дальнейшего удобно ввести безразмерные  переменные:
${t = \gamma \tau}$ and ${w = \beta \zeta}$, где
\be
\gamma^2 = \frac{3q}{(-R_0)} \left(-\frac{R_0}{\tilde T_0}\right)^{2(n+1)}\,,\nonumber\\
\beta = \gamma^2 \tilde T_0/3 = q \left(-\frac{R_0}{\tilde T_0}\right)^{2n+1}.
\label{gamma-beta}
\ee
¬ их терминах уравнение (\ref{eq-w-simple})  становитс€ очень простым по форме:
\be
z'' - z^{-\nu} + (1+\kappa \tau) = 0\,.
\label{eq-for-z}
\ee
«десь штрих означает производную по безразмерному времени $\tau$.

Ћегко видеть, что минимум потенциала движетс€ к $\zeta=0$,  а глубина потенциала в минимуме
уменьшаетс€.  ажетс€ совершенно очевидным, что даже если в исходном состо€нии $\zeta$
неподвижно сидело в минимуме потенциала, то при росте плотности массы возникнут
осцилл€ции $\zeta$ вокруг ползущего минимума, и в процессе раскачки $z$ достигнет нул€.
„исленные расчеты, проведенные в работе~\cite{ea-ad}, показывают точно такую картину.

ƒл€ качественного анализа поведени€ осцилл€ций удобно использовать
интегральный закон эволюции
энергии, который дл€ осцилл€ционого уравнени€ общего вида:
\be
\zeta'' + \partial U/ \partial\zeta = 0
\label{d2-z-genrl}
\ee
с потенциалом, который может €вно зависеть от времени, имеет вид:
\be
\frac{(\zeta')^2}{2} + U (\zeta,\tau) - \int d\tau \frac{\partial U}{\partial\tau}  = \mbox{const}.
\label{evolv-enrgy}
\ee
—оотношение (\ref{evolv-enrgy}) также позвол€ет провести общий анализ возникновени€
сингул€рности $R$, как было сделано в работе~\cite{lr-sing}.

ћожно избежать бесконечных значений $R$, если добавить к действию
слагаемое пропорциональное $R^2$:
\be
F(R) \rar F(R) - R^2/6m^2 .
\label{F-of-R-R2}
\ee
 осмологические модели с квадратичным по кривизне действием
были впервые рассмотрены в работах~\cite{aas-R2,aas-R2a,aas-R2b}.
„лены высшего пор€дка по кривизне могут возникнуть в результате радиационных поправок
к обычному действию √ильберта-Ёйнштейна
при рассмотрении вакуумного среднего тензора энергии-импульса в искривленном
пространстве. «аметим, что радиационные поправки генерируют не только $R^2$, но и
$R_{\mu\nu} R_{\mu]nu}$, которые не столь безобидны, как  предыдущие, так как привод€т
к патологии в теории в виде тахионов и духов.

»нтeресным свойством действи€ с квадратичными по кривизне членами
€вл€етс€ ранн€€ инфл€ционна€ стади€,
как это имеет место в модели —таробинского~\cite{aas-infl}. ¬ этой модели инфл€ци€
естественно заканчиваетс€ за счет рождени€ частиц новой скал€рной гравитационной
степенью
свободы - скал€ра кривизны, который становитс€ динамической переменной  за счет
$R^2$-поправок. ѕроцесс разогрева ¬селенной, знаменующий окончание инфл€ции,
за счет гравитационного рождени€ частиц в $R^2$-теории рассматривалс€ в ранних
работах~\cite{pp-R2,pp-R2a,pp-R2b,pp-R2c,pp-R2d} и в недавних работах~\cite{GP,ea-ad-lr}.

ѕроцесс рождени€ частиц в поздней ¬селенной
за счет высокочастотных осцилл€ций скал€ра кривизны
в теории с действием (\ref{F-AAS})  рассматривалс€ в работах~\cite{ea-ad-lr-2,ea-ad-lr-2a}, где
был сделан вывод, что рожденные частицы могут давать заметный
вклад в поток космических лучей высоких энергий. Ётот результат критиковалс€
в работах~\cite{GT,GTa}, в которых ставилась под сомнени€ высока€ эффективность
рождени€ частиц. ќднако, как было показано в работе \cite{adr-reply}, эта критика
€вл€етс€ необоснованной.

”равнение движени€, определ€ющее эволюцию $R(t)$, при наличии $R^2$-члена, не
удаетс€ €вно
привести к осцилл€торному   уравнению типа~(\ref{d2-z-genrl}), оно принимает более
сложный вид~\cite{ea-ad}:
\be
\left[ 1-\frac{R^{2n+2}}{6\lambda n(2n+1) R_0^{2n+1} m^2 }\right]\,\ddot R
- (2n+2) \,\frac{\dot R^2}{R} -
\frac{R^{2n+2} (R+T)}{6\lambda n (2n+1) R_0^{2n+1}} = 0 \,.
\label{eq-for-R-mdf}
\ee
“ем не менее можно  видеть, что наличие второго слагаемого в коэффициенте перед $\ddot R$
не дает кривизне добратьс€ до бесконечности. »нтересно, что наивна€ оценка обрезани€ $R$
на величине, когда это второе слагаемое будет по пор€дку величины близко к единице,
оказываетс€ сильно заниженной. –ост $R$ прекращаетс€ при гораздо более высоком значении,
как это показано в работах~\cite{ea-ad-lr-2}, где был проведен  более подробный анализ эволюции
 $R(t)$ в инфракрасно модифицированных теори€х с действием
типа (\ref{eq:cr1} - \ref{F-AAS}) с добавленным слагаемым $R^2/6m^2$ дл€ астрономических систем
с растущей плотностью энергии. Ѕыли получены
аналитическое, с использованием соотношени€ (\ref{evolv-enrgy}), а также численное решени€,
наход€щиес€ в хорошем согласии между собой.

ѕри больших частотах численное решение неустойчиво и ``взрываетс€'' при довольно малых временах.
»з-за этого не удаетс€ надежно продвинутьс€ к асимптотическому режиму. ќднако при достаточно
малых частотах вычислени€ вполне надежны и хорошо согласуютс€ с аналитическими. — другой
стороны, аналитические результаты станов€тс€ более точными с ростом частоты. Ѕлагодар€ этому
удаетс€ покрыть весь существенный диапазон частот.

ƒл€ анализа уравнение движени€ (\ref{eq-for-R-mdf}) было преобразовано к
форме~(\ref{d2-z-genrl}), однако €вного выражени€ дл€ потенциала в этом случае получить не
удаетс€. “ем не менее полученные приближени€ адекватно описывают решени€.
ƒл€ рассматриваемого случа€ астрономических систем с растущей плотностью
справедливы соотношени€: $|R_c|\ll|R|\ll m^2$ и тогда $F$ примерно равна
\be\label{eq:model_approx}
F(R)\simeq -R_c\left[1-\left(\frac{R_c}{R}\right)^{2n}\right]-\frac{R^2}{6m^2}\,.
\ee
–ассмотрим приближенно однородное распределение материи с растущей плотностью массы,
котора€ остаетс€ достаточно малой (например, облако газа в процессе образовани€ галактики или
звезды). ¬ этом случае фоновую метрику можно полагать метрикой ћинковского.
¬ этом приближении уравнение (\ref{eq:trace}) принимает вид:
\be\label{eq:trace_approx}
3\partial_t^2F' -R-T = 0\,.
\ee
¬ведем, как и выше, безразмерные величины
\be\label{eq:definitions}
\begin{gathered}
 z\equiv \frac{T(t)}{T(t_{in})}\equiv \frac{T}{T_0}= \frac{\rho_m(t)}{\rho_{m0}}\,,
 \qquad y\equiv -\frac{R}{T_0}\,, \\
g\equiv \frac{T_0^{2n+2}}{6 n(-R_c)^{2n+1}m^2}= \frac{1}{6 n  (m t_U)^2} \,\left( \frac{\rho_{m0}}{\rho_c}\right)^{2n+2}\,,
\qquad \tau\equiv m\sqrt g\,t\,,
\end{gathered}
\ee
где $\rho_c \approx 10^{-29}~$г/см$^3$ -- космологическа€ плотность энергии в современной
вселенной, $\rho_{m0}$ -- начальное значение плотности энергии в рассматриваемой системе,
и $T_0 = 8\pi \rho_{m0}/m_{Pl}^2$.

ƒалее введем новую неизвестную функцию:
\be\label{eq:xi_definition}
\xi\equiv  \frac{1}{2 n}\left(\frac{T_0}{R_c}\right)^{2n+1}F_{,R}  = \frac{1}{y^{2n+1}}-gy\,,
\ee
дл€ которой уравнение движени€~(\ref{eq:trace_approx}) может быть переписано в простой
осцилл€торной форме:
\be\label{eq:xi_evol}
\xi''+z-y=0\,,
\ee
где штрих у $\xi$ означает производную по безразмерному времени $\tau$.
ѕодстановка (\ref{eq:xi_definition})
аналогична, но не тождественна той, что была сделана в работе \cite{ea-ad},
даже при отсутствии $R^2$-члена,
¬ рассматриваемом случае это несколько более удобно технически.

ѕотенциал этого осцилл€тора, очевидно, определ€етс€ условием:
\be
\frac{\partial U}{\partial \xi}= z - y(\xi).
\label{U-prime}
\ee
“еперь
$y$ не выражаетс€  через $\xi$ аналитически и €вное выражение дл€ потенциала отсутствует.
¬се же  удаетс€ получить достаточно аккуратные
приближенные выражени€ дл€ $U(\xi)$, особенно при
малых $g$, дл€ положительных и отрицательных $\xi$ по отдельности.
ƒетали можно найти в работах~\cite{ea-ad-lr-2}.

\begin{figure}[!t]
\centering
 \includegraphics[width=.4\textwidth]{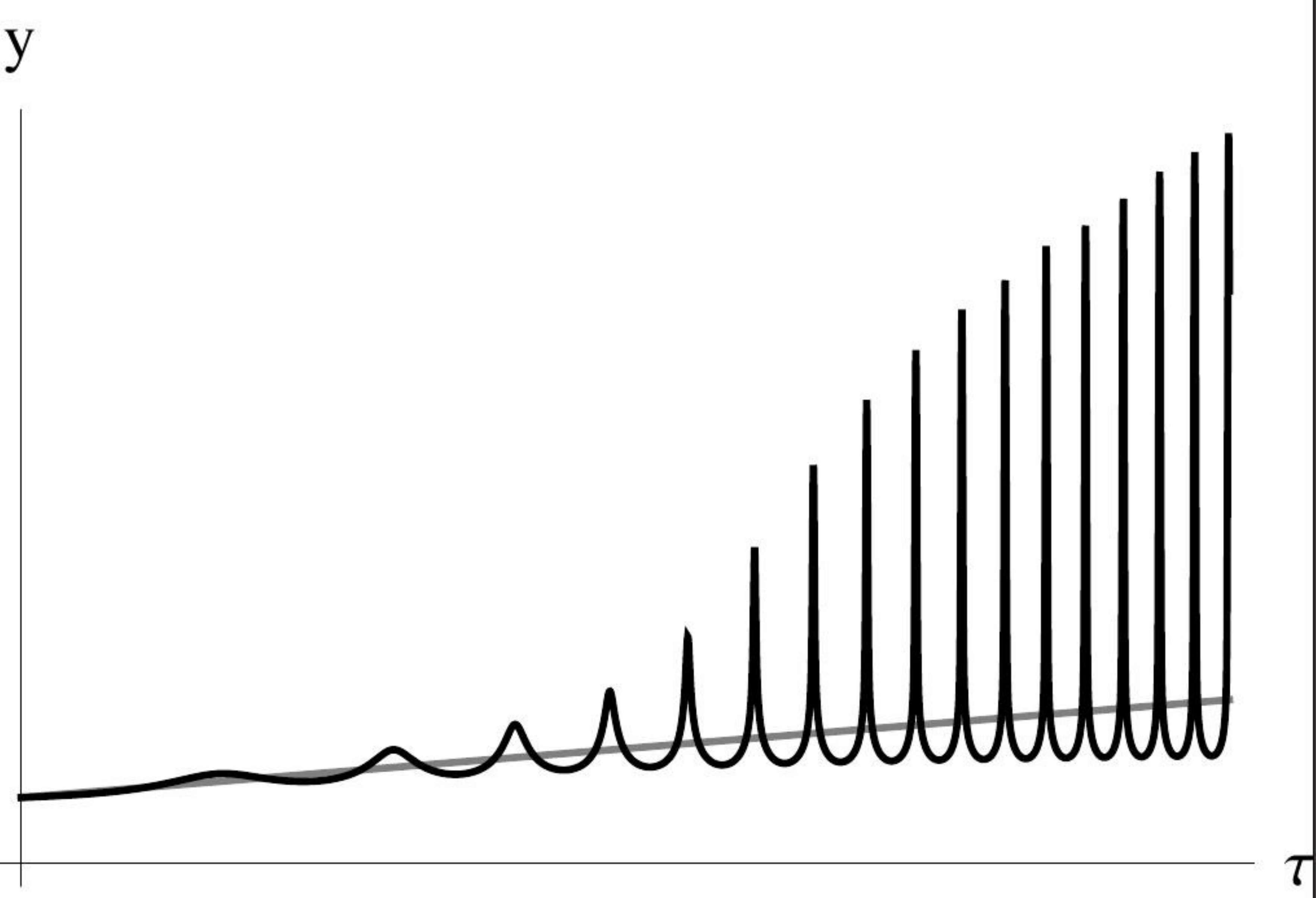}
\includegraphics[width=.4\textwidth]{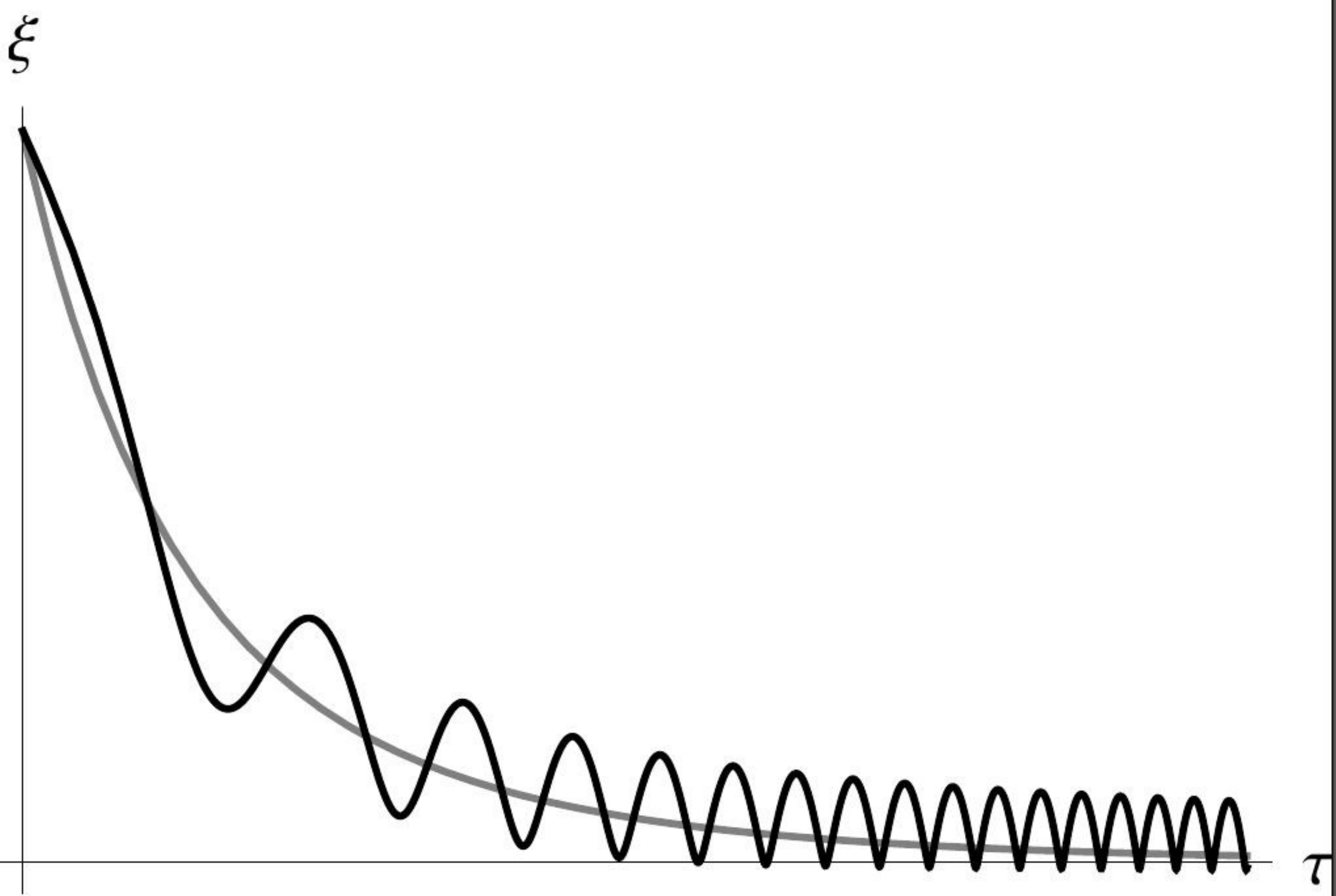}
\caption{ѕики в осцилл€ци€х кривизны. –езультаты представлены дл€
$n=2$, $g=0.001$, $\kappa=0.04$ и $y'_0=\kappa/2$.}
\label{fig:spikes}
\end{figure}

ќсцилл€ции $\xi(t)$ довольно близки к гармоническим, но физическое поле $R(t)$ колеблетс€
весьма далеким от гармонической формы образом, демонстриру€ резкие короткие пики с
большой амплитудой. в которых $y\gg 1$, т.е. решение далеко уходит от стандартного значени€ в
ќ“ќ, $R_{GR} = -T_0$. Ёти пики отвечают нереализованным сингул€рным точкам, когда достижение
кривизной бесконечной величины останавливаетс€ $R^2$-членом.
ѕоведение кривизны в зависимости от времени изображено на рис.~\ref{fig:spikes}.
—ильна€ ангармоничность колебаний $y$ или, что то же, $R$ приводит к эффективной
перекачке энергии низкочастотных мод, которые возбуждаютс€ при медленном сжатии
системы, в высокочастотные.

¬ысокочастотные осцилл€ции кривизны с большой амплитудой приведут к рождению космических
лучей высоких энергий в период образовани€ крупномасштабной структуры вселенной. ѕотоки
таких частиц могут быть наблюдаемы дл€ довольно широкого интервала значений параметров
теории.

ћы довольно подробно остановились на этом необычном поведении кривизны еще и из-за
того, что на этих решени€х нарушаетс€ теорема …ебсена-Ѕиркгофа (Jebsen-Birkhoff)
и в частности возникает поразительное €вление гравитационного отталкивани€, антигравитации  дл€ систем конечного размера~\cite{adr-anti}. ¬озникновение
гравитационного отталкивани€ дл€ бесконечных систем, например, ускоренного
космологического расширени€, не противоречит ќ“ќ.
ќбсуждение нарушени€
теоремы …ебсена-Ѕиркгофа в $F(R)$-модифицированных теори€х можно
найти в работах~\cite{JB-theorem,JB-theorem1}.

ћы продемонстрируем возникновение гравитационного отталковани€ на примере
сферически симметричного распределени€ материи с метрикой, определ€емой
обычным выражением:
\be
ds^2 =  A (r,t) dt^2 - B(r,t) dr^2 - r^2 (d\theta^2 + \sin^2 \theta\,d\phi^2) .
\label{ds2-repuls}
\ee
ћетрика такого типа дл€ $F(R)$ теорий анализировалась в работах~\cite{sch-dombriz,cembranos}, но
там не были прин€ты во внимание осцилл€ции кривизны, на которых и сидит эффект антигравитации.

ѕредположим, что метрические коэффициенты $A$ и $B$ близки к единице, т.е. что метрика близка к
плоской метрике ћинковского. »менно в этом предположении и были получены количественные
результаты об осцилл€ци€х $R$.
Ќенулевые компоненты тензора –иччи, отвечающие метрике~(\ref{ds2-repuls}), равны:
\be
R_{00}&=&\frac{A''-\ddot B}{2B}+\frac{(\dot B)^2-A'B'}{4B^2}+\frac{\dot A \dot B - (A')^2}{4AB}+\frac{A'}{rB}\, ,
\label{R00}\\
R_{rr}&=&\frac{\ddot B -A''}{2A}+\frac{(A')^2-\dot A \dot B}{4A^2}+\frac{ A' B' - (\dot B)^2}{4AB}+\frac{B'}{rB}\, ,
\label{R11}\\
R_{\theta \theta }&=&-\frac{1}{B}+\frac{rB'}{2B^2}-\frac{rA'}{2AB}+1\, ,
\label{R22} \\
R_{\varphi \varphi }&=&\left(-\frac{1}{B}+\frac{rB'}{2B^2}-\frac{rA'}{2AB}+1\right )\sin^2\theta  = R_{\theta \theta} \sin^2\theta \, ,
\label{R33} \\
R_{0r}&=&\frac{\dot B}{rB}\, .
\label{R01}
\ee
«десь штрих и точка над символом функции означают производные по $r$ и $t$, соответственно.
—оответствующий скал€р кривизны, $R = g^{\mu\nu} R_{\mu\nu}$, равен:
\be
R&=&\frac{1}{A}R_{00}-\frac{1}{B}R_{rr}-\frac{1}{r^2}R_{\theta \theta}-\frac{1}{r^2\sin^2\theta }R_{\varphi \varphi}  \nonumber \\
&=&\frac{A''-\ddot B}{AB}+\frac{(\dot B)^2-A'B'}{2AB^2}+\frac{\dot A \dot B - (A')^2}{2A^2B}+\frac{2A'}{rAB}-\frac{2B'}{rB^2}+\frac{2}{r^2B}-\frac{2}{r^2}
\label{Rscalar}\\
&=& \frac{2}{A}R_{00}-\frac{2B'}{rB^2}+\frac{2}{r^2B}-\frac{2}{r^2} \, .
\nonumber
\ee
ѕолага€, как сказано выше, слабое отличие метрики от плоской, т.е. что
\be
A_1 = A - 1 \ll 1\,\,\,{\rm and} \,\,\, B_1 = B-1 \ll 1
\label{A1-B1}
\ee
проверим, самосогласованность этого предположени€ дл€ осцилл€ционных решений, найденных в
работах~\cite{ea-ad-lr-2}, дл€ которых $R$ заметно превосходит свое значение в ќ“ќ. ƒл€ этого
удобно использовать уравнени€ (\ref{eq-of-mot}) в форме:
\be
R_{00}  - R/2 &=& \frac{ \tilde T_{00} + \Delta F_{,R} + F/2 - RF_{,R}/2}{1+F_{,R}},
\label{R-00} \\
R_{rr}  + R/2 &=&  \frac{ \tilde T_{rr} + (\partial_t^2 +\partial_r^2 - \Delta) F_{,R} - F/2 + RF_{,R}/2}{1+F_{,R}} ,
\label{R-rr}
\ee
так как их лева€ часть не содержит вторых производных кривизны. ¬ пределе слабого гравитационного пол€
можно пренебречь квадратами производных $A(r,t)$ и $B(r,t)$, и в результате получим дл€
$R_{00}$ и $R_{rr}$ компонент тензора –иччи и дл€ скал€ра кривизны $R$ следующие выражени€:
\be
R_{00}&\approx&\frac{A''-\ddot B}{2}+\frac{A'}{r}\, ,
\label{R00-weak}\\
R_{rr}&\approx&\frac{\ddot B -A''}{2}+\frac{B'}{r}\, ,
\label{R11-weak}\\
R&\approx&A''-\ddot B+\frac{2A'}{r}-\frac{2B'}{r}+\frac{2(1-B)}{r^2}\, .
\label{R-weak}
\ee

≈сли плотность энергии материи внутри облака, т.е. дл€ $ r<r_m$ намного больше, чем средн€€ космологическа€
плотность энергии, то будут верны следующие соотношени€:
\be
F_{,R} \ll 1\,\,\,\,\, \mbox{и} \,\,\,\,\,\, F \ll R
\label{bounds-on-F}
\ee

ƒл€ статических решений эффекты модификации гравитации в этом пределе невелики и решение окажетс€
близким к обычному шварцшильдовскому решению в согласии с имеющейс€ литературой.  ак было отмечено выше, в
работах~\cite{ea-ad-lr-2} показано, что в системах с растущей плотностью энергии возбуждаютс€
высокочастотные осцилл€ции кривизны. ƒл€ таких решений можно пренебречь пространственными
производными  $F'_R$ по сравнению с временными, так как характерное врем€ осцилл€ций микроскопически
мало, а пространственные вариации макроскопически велики. “ак что из уравнени€ (\ref{eq:trace}) следует
что $(\partial_t^2 - \Delta ) F_{,R} = (\tilde T + R)/3 $ и мы получим
\be
B_1' + \frac{B_1}{r} & = & r \tilde T_{00},
\label{B1-prime} \\
A''_1-\frac{A_1'}{r} &  = & -\frac{3B_1}{r^2} +\ddot B_1 +\tilde T_{00}-2\tilde T_{rr} +
\frac{\tilde T_{\theta\theta}}{r^2} +
 \frac{\tilde T_{\varphi\varphi}}{r^2\sin^2\theta}
\equiv S_A\, .
\label{A1-two-prime}
\ee

ѕредполага€ малое отклонение от метрики ћинковского, мы пренебрегли поправками к
$T_{\mu \nu}$, вызванными кривизной пространства-времени. Ќиже мы проверим, в каком случае
справедливо это предположение.

”равнение~(\ref{B1-prime}) имеет решение:
\be
B_1(r,t) = \frac{C_B(t)}{r}+\frac{1}{r}\,\int_0^r dr' r'^2 \tilde T_{00} (r',t)\, .
\label{B1-new}
\ee
„тобы избежать сингул€рности при $r=0$, необходимо положить $C_{B}(t) \equiv 0$.
“огда выражение дл€ $B_1$ формально совпадает с обычным шварцшильдовским решением, а
решение дл€ $A_1$ обладает дополнительной свободой:
\be
A_1(r,t)=C_{1A}(t)r^2+C_{2A}(t)+\int^{r_m}_r dr_1\,r_1 \int^{r_m}_{r_1} \frac{dr_2}{r_2}\, S_A(r_2,t)\, .
\label{A1-common}
\ee
ѕределы интегрировани€ выбраны таким образом, чтобы не возникло сингул€рности при $r_2=0$.

»спользу€ уравнение (\ref{B1-new}) с $C_{B}=0$, перепишем $S_A$ в следующем виде:
\be
S_A = -\frac{3}{r^3}\,\int_0^r dr' r'^2 \tilde T_{00} (r',t)
+\frac{1}{r}\,\int_0^r dr' r'^2 \ddot {\tilde {T}}_{00} (r',t) \, +
\tilde T_{00}-2\tilde T_{rr} + \frac{\tilde T_{\theta\theta}}{r^2} +
 \frac{\tilde T_{\varphi\varphi}}{r^2\sin^2\theta}
\label{S-A}
\ee
и в итоге найдем:
\be \nonumber
 A_1(r,t) = C_{1A}(t)r^2+C_{2A}(t)
+\int_{r}^{r_m} dr_1\,r_1\int_{r_1}^{r_m} \frac{dr_2}{r_2}\, \left(\tilde T_{00} (r_2,t)-2\tilde T_{rr}(r_2,t) +  \frac{\tilde T_{\theta\theta}(r_2,t)}{r^2} +
\right.\\  
 \left.  \frac{\tilde T_{\varphi\varphi}(r_2,t)}{r^2\sin^2\theta}\right)  
-\int_{r}^{r_m} dr_1\,r_1\int_{r_1}^{r_m} \frac{dr_2}{r_2} \, \left(\frac{3}{r_2^3}\,\int_0^{r_2} dr' r'^2 \tilde T_{00} (r',t) -
\frac{1}{r_2}\,\int_0^{r_2} dr' r'^2 \ddot {\tilde {T}}_{00} (r',t)\right).
\label{A-1}
\ee

Ќайдем сначала посто€нные интегрировани€ в случае, когда отсутствует осциллирующее слагаемое в кривизне.
ќпределим массу внутри радиуса $r$ обычным образом:
\be
M (r,t) =\int_0^{r} d^3r\,  T_{00} (r,t)=4\pi \int_0^{r} dr\,  r^2\,  T_{00} (r,t)
\label{M}
\ee
≈сли все вещество сосредоточено внутри радиуса $r_m$, то полна€ масса системы
оказываетс€ равной $M \equiv M(r_m)$ и, в соответствии с классическим решением
Ўварцшильда, не зависит от времени. ѕоскольку
$\tilde T_{00}=8\pi T_{00}/m_{Pl}^2$, то дл€ $r>r_m$, получаем, как и ожидалось, $B_1=r_g/r$, где $r_g=2M/m_{Pl}^2$ - обычный внешний радиус Ўварцшильда.

¬ычислим теперь $A_1$ (\ref{A-1}). ќчевидно, что при $r>r_m$ первый интеграл обращаетс€ в нуль, так как
$r_2 >r_m$, и по предположению там $T_{\mu \nu}=0$. »нтеграл, содержащий
$\ddot {\tilde {T}}_{00} $,
тоже занул€етс€ в силу посто€нства внешней массы.
ќстающийс€ интеграл легко беретс€:
\be
\int_{r}^{r_m} dr_1\,r_1\int_{r_1}^{r_m} \frac{dr_2}{r_2} \, \frac{3}{r_2^3}\,\int_0^{r_2} dr' r'^2 \tilde T_{00} (r',t) =
\frac{r_g}{r}+\frac{r_g\,r^2}{2r_m^3}-\frac{3r_g}{2r_m}\, .
\label{int-A}
\ee
“аким образом, метрический коэффициент вне источника равен:
\be
A_1 = -\frac{r_g}{r}+\left[C_{1A}(t)-\frac{r_g}{2r_m^3}\right] r^2 +\left[C_{2A}(t)+\frac{3r_g}{2r_m}\right] \, .
\label{A-1-Sch}
\ee
¬ыберем
\be
C_{1A}=r_g/(2r_m^3),\,\,\,\,\,\,\, C_{2A}=-3r_g/(2r_m).
\label{C1-C2-sch}
\ee
ѕервое условие необходимо, чтобы устранить слагаемое, пропорциональное $r^2$ на бесконечности, второе
же необ€зательно, но оно всегда может быть наложено переопределением времени.
«аметим, что такой выбор произвольных посто€нных справедлив, когда в решении нет быстро осциллирующих
членов.

¬ модифицированной гравитации внутреннее решение имеет ту же форму, что приведена выше:
(\ref{B1-new}) и (\ref{A-1}), однако коэффициент $C_{1A}$ может нетривиально зависеть от времени.
Ётот коэффициент можно найти из уравнени€ (\ref{R-weak}), если известна скал€рна€ кривизна $R(t)$.
»спользу€ выражени€ (\ref{B1-new}) и (\ref{A-1}) и сравнива€ их с уравнением (\ref{R-weak}),
мы можем заключить, что доминирующий вклад в кривизну даетс€ суммой
$A''+2A'/r$ и соответственно $C^{(\rm osc)}_{1A}(t)=R(t)/6$, со скал€ром кривизны,
вычисленным в работах~\cite{ea-ad-lr-2}:
\be
R(t) = R_{GR} (r) y (t),
\label{R-of-t}
\ee
где $R_{GR} = - 8\pi  T(r)/ m_{Pl}^2$ - решение в общей теории относительности, а быстро осциллирующа€
$y(t)$ функци€ может быть много больше единицы. ¬ соответствии с работой~\cite{ea-ad-lr-2} максимальное
значение $y$ в пиковой области составл€ет:
\be
y(t) \sim 6 n (2n+1)  m t_u \left(\frac{t_u}{t_{\rm contr} }\right) \left[\frac{\rho_m (t)}{\rho_{m0}} \right]^{(n+1)/2}
\left(\frac{\rho_c}{\rho_{m0}} \right)^{2n+2} ,
\label{y-of-t}
\ee
где $t_u$ - возраст вселенной, $t_{\rm contr} $ - характерное врем€ сжати€ системы, так что плотность сжимающегос€
облака ведет себ€, как $\rho_m (t) = \rho_{m0} (1 + t/t_{\rm contr})$, а $\rho_{m0} $ и $\rho_c~=~10^{-29}$~g/cm$^3$
 - начальна€ плотность массы/энергии облака и современна€ космологическа€ плотность соответственно.
¬ соответствии с результатами работы~\cite{ea-ad-lr}, параметр $m$, вход€щий в выражение~(\ref{F-of-R-R2}),
должен превосходить $10^5$ √э¬, чтобы не возникло противоречи€ с первичным нуклеосинтезом.
ѕосему множитель $m t_u $ принимает гигантское значение: $ m t_u \geq 10^{47}$ и $y$ может быть много
больше единицы, если только не будет подавлен малой величиной отношени€
$( \rho_c/\rho_{m0})^{2n+2}$ при большом показателе $n$.

ќтметим, что имеетс€ существенное различие между вакуумными решени€ми в обычной и модифицированной
гравитации. ¬ стандартном случае пропоциональное $r^2$ слагаемое по€вл€етс€ как вне ($r>r_m$), так и
внутри ($r<r_m$) облака с одним и тем же коэффициентом и поэтому оно должно обратитьс€ в нуль.
— другой стороны, в модифицированной гравитации такого услови€ нет и, следовательно, слагаемое
$C_{1A} r^2$ может присутствовать при $r<r_m$ и отсутствовать при $r \gg r_m$.

¬акуумное решение дл€ $R$, по-видимому, имеет вид $R\sim R_c$, где $R_c$ - маленька€ космологическа€
кривизна, плюс возможное осциллирующее слагаемое. “аким образом, метрические функции внутри
облака равны:
\be
B (r, t) & =& 1 + \frac{2M(r,t)}{m_{Pl}^2 r} \equiv  1+ B_1^{\rm (Sch)}\, ,
\label{B-of-r-t}\\
A(r,t)  &=& 1 + \frac{R(t )\,r^2}{6} + A_1^{\rm (Sch)} (r,t) \label{A-of-r-t}\, .
\ee
»ными словами, мы построили решение, предполага€, что оно состоит из двух слагаемых: швардшильдовского
и осциллирующего, причем быстро осциллирующа€ часть возникает в системах с медленно мен€ющейс€ со
временем плотностью энергии. ¬ыражение дл€ $A_1^{\rm (Sch)} (r,t) $ определ€етс€ интегралами (\ref{A-1}) с
посто€нными $C_{A1}=r_g/2r_m^3$ и $C_{A2}=-3r_g/r_m$, как это следует из выражений (\ref{C1-C2-sch}).

„то касаетс€ интегралов в равенстве~(\ref{A-1}) во внутренней области, то мы вычислим их, полага€, что
вещество €вл€етс€ нерел€тивистским, так что пространственные компоненты $T_{\mu\nu}$ пренебрежимо малы по
сравнению с $T_{00}$. ѕредположим также дл€ простоты, что плотность массы/энергии $T_{00} \equiv \rho_m (t)$
посто€нна по пространству, но может зависеть от времени. ѕервые два интеграла в выражении (\ref{A-1})
взаимно сокращаютс€ и выживает лишь интеграл пропорциональный второй производной плотности массы. “ак
что в результате получим:
\be
A_1^{(Sch)} (r,t) = \frac{r_g r^2}{2r_m^3} -\frac{3r_g}{2r_m}+
 \frac{ \pi \ddot\rho_m}{3 m_{Pl}^2}\, ( r_m^2 - r^2)^2\, .
\label{A-1-2}
\ee

 ак мы уже отмечали, $R(t)$ в модифицированной теории, как правило, выше своего значени€ в ќ“ќ,
т.е. $|R_{GR}|=8 \pi \rho_m /m_{Pl}^2$, так что второе слагаемое в уравнении (\ref{A-of-r-t}), т.е. $R(t)r^2/6$,
дает доминирующий вклад в $A_1$ при достаточно больших  $r$. ƒействительно,
$r^2 R(t) \sim r^2 y R_{GR}$, причем $y > 1$, в то врем€, как канонический шварцшильдовский вклад пор€дка
$r_g/r_m \sim \rho_m r_m^2/m_{Pl}^2 \lesssim r_m^2 R_{GR} $.

¬ низшем пор€дке по гравитационному взаимодействию уравнение движени€ пробной
нерел€тивистской частицы (геодезическое уравнение) имеет вид:
\be
\ddot r = - \frac{A'}{2} = -\frac{1}{2}\left[ \frac{R(t) r}{3} + \frac{r_g r}{r_m^3} \right],
\label{ddot-r}
\ee
где  $A$ определено в  уравнении (\ref{A-of-r-t}).  ѕоскольку в рассмотренной здесь модификации
гравитации $R(t)$ отрицательно и велико по абсолютной величине, то внутри сжимающегос€ облака с
$\rho>\rho_c$ возникают эффекты гравитационного отталкивани€, которое окажетс€ сильнее гравитационного
прит€жени€, если
\be
\frac{|R|r_m^3}{3r_g} = \frac{|R|r_m^3m_{Pl}^2}{6 M} = \frac{|R|r_m^3m_{Pl}^2}{8\pi \rho\,r_m^3} =
\frac{|R|}{\tilde T_{00}} \equiv y > 1\,,
\ee
¬ результате могут образоватьс€ €чеистые структуры, в частности, космические пустоты (voids).

»нтересно, что в этой теории сила гравитационного отталкивани€ про€вл€етс€
короткими, но сильными ударами, заметно превышающими обычное гравитационное
прит€жение.

«аметим, что хот€ результат был получен в приближении плоского фонового
пространства, он представл€етс€ более общим и справедливым и в случае
искривленного пространства времени.
Ѕолее того, естественно ожидать, что антигравитаци€ возникает
и в других похожих верси€х $F(R)$ теорий.

\section{«аключение}

ѕосле того, как наш обзор был представлен в журнал, по€вилс€ еще р€д работ, имеющих
непосредственное отношение к рассматриваемым нами проблемам. Ќиже, по предложению
рецензента, мы кратно опишем  некоторые из новых работ. ”кажем, прежде всего, обзор~\cite{2017PhR...692....1N}, 
где рассмотрены различные типы модификации гравитации, а не только $F(R)$, как в нашем
обзоре. ¬ этом обзоре~\cite{2017PhR...692....1N} описана космологическа€ эволюци€, начина€ от инфл€цинонной
эпохи  до наших дней, а также  модели отскока от сингул€рности при схлопывании ¬селенной.

¬ работе~\cite{2018PhLB..779..425N} 
выведены ограничени€ на параметры скал€рно-тнезорных и $ F(R)$-теорий на основе
анализа наблюдений гравитационных волн от  сли€ни€ нейтронных звезд.

ƒетальное обсуждение космологической эволюции в $F(R)$-теори€х  приведенo в неопубликованной на момент написани€
нашего обзора работе~\cite{Capozziello2018}. 

јвторы обзора~\cite{2016ARNPS..66...95J} провод€т строгое различие между тЄмной энергией и модифицированной гравитацией
на основе сильного и слабого принципов эквивалентности.
–ассмотрены феноменологи€ и характерные наблюдаемые характеристики этих двух категорий моделей.

¬ обзоре~\cite{2018RPPh...81a6902B}  подчЄркнута квантова€ природа проблемы тЄмной энергии и необходимость такого решени€, которое
обходит no-go теорему ¬айнберга~\cite{rho-vac-rev}.

 ритический анализ широкой совокупности различного типа астрономических наблюденой проведен в
работе~\cite{2016ApJ...833L..30R}. 
ѕо утверждению авторов, факт космологического ускорени€ €вл€етс€ твердо установленным.

¬опрос о том, насколько твЄрдо  сверхновые SNIe могут считатьс€ стандартными свечами обсуждаетс€
в работе~\cite{WrightLi2017} 
с учетом того, что гравитаци€ может быть модифицирована. јвторы делают вывод, что
в теори€х, в которых сила гравитации зависит от времени, сверхновые SNIe не могут считатьс€
стандартными свечами и вывод об ускоренном расширении, вообще говор€, может быть не справедлив.
ќднако, остаЄтс€ открытым вопрос, насколько необходима€ зависимость гравитационной посто€нной
от времени остаетс€ совместимой с полным набором данных в пользу антигравитирующей тЄмной энергии.








Ќедавно возникло новое течение дл€ описани€ ускоренного расширени€ ¬селенной и тЄмной
энергии, под названием  ``заболоченна€ почва'' или просто ``болото'' (\textit{англ.} `swamp\-land')~\cite{Brennan2017,Obied2018}
как альтернатива струнному ландшафту.
ѕод последним понимаетс€ почти
бесконечное количество вакуумных состо€ний (более, чем $10^{100\,000}$)
при компактификации суперструнных моделей на четырехмерное пространство. —толь гигантское количество вакуумных состо€ний открывает путь антропному решению проблемы вакуумной энергии, см. выше раздел 3 примерно через страницу после уравнени€ (\ref{m-p}).

јвторы работ~\cite{Brennan2017,Obied2018}  
отмечают, что общеприн€то полагать, будто все самосогласованные эффективные квантовые полевые теории, со взаимодействием с гравитацией с большой веро€тностью
могут быть получены тем или иным образом как результат компактификации струнной теории, что
делает струнные теории не слишком полезными с точки зрени€ низкоэнергетической феноменологии.
—огласно точке зрени€, высказанной в работах~\cite{Brennan2017,Obied2018}, 
это не так, и на самом деле могут существовать эффективные
теории пол€, не привод€щие к самосогласованной теории гравитации при ультрафиолетовом замыкании.
јвторы предлагают называть набор таких теорий, которые не привод€т к струнной теории при
ультрафиолетовом замыкании,  болотом по контрасту с ландшафтом.

 осмологические эффекты, к которым могла бы привести ``заболoченна€'' картина,
рассмотрены в работе~\cite{2018PhLB..784..271A}. 
 ритический анализ таких моделей и их следствий приведен в работе~\cite{Akrami2018}. 

¬ насто€щее врем€ количество работ по УболотнойФ тематике составл€ет уже несколько дес€тков,
и их анализ требует отдельного обзора.

%
%
%
%
%


\textbf{Ѕлагодарности}

ћы благодарим одного из рецензентов за конструктивные замечани€, а также √астона ‘олателли (Gaston Folatelli) за обсуждени€ вопросов о систематических ошибках в наблюдаемых
параметрах SN~Ia и ћарию ѕружинскую за внимательное прочтение текста и многочисленные поправки по истории и по космологическим
приложени€м сверхновых.

—.Ѕ. благодарен Shun Saito и Tomomi Sunayama за ценные указани€ на литературу по BAO и критические замечани€ и гранту –‘‘» 19-52-50014 за поддержку работ по моделированию сверхновых.

ј.ƒ. благодарен за поддержку гранту –Ќ‘ 16-12-10037.

\section{ѕриложени€} 
\subsection{ѕриложениe ј.  ¬ывод уравнений ‘ридмана \label{A-derive-Friedman}}
\subsubsection{Ёлементарный вывод из ньютоновской теории}

–ассмотрим пространство, заполненное однородно и изотропно  распределенным
веществом, и выделим в нем небольшой шар радиуса $ r a(t)$. «десь $r$ -- это
фиксированна€, так называема€ сопутствующа€, координата,
а возможное изменение радиуса определ€етс€
функцией $a(t)$. “акие обозначени€ выбраны дл€ соответстви€ с
используемыми в тексте космологическими обозначени€ми.

–ассмотрим поведение материальной точки на границе этого шара.  ак
известно, в сферически симметричном случае внешность шара не оказывает
никакого гравитационного действи€ на пробную частицу. —умма кинетической
и потенциальной энергий этой частицы должна сохран€тьс€:
\be
\frac{r^2\, \dot a^2}{2} - \frac{ 4\pi}{3}\, \frac{\rho r^2 a^2}{m_{Pl}^2} = \mbox{const}
\label{en-conserv}
\ee
ќчевидно, что после делени€ на $r^2 a^2/2$ это уравнение совпадает с
уравнением (\ref{H2}).

–ассмотрим далее баланс плотности  энергии среды,
${dE = -P\,dV}$, где ${ E =\rho V}$ и, следовательно,  ${ dE = V d\rho + 3 (da/a) V\rho  }$.
ќтсюда, очевидно, следует уравнение (\ref{dot-rho}).

ƒифференциру€ уравнение (\ref{H2}) и использу€ уравнение (\ref{dot-rho}), получим
уравнение (\ref{ddot-a}).

Ќа этом этапе может возникнуть законный вопрос, каким образом на основании
нерел€тивистской механики Ќьютона удалось получить рел€тивистскую теорию
гравитации Ёйнштейна, где гравитирует не только масса, но и давление? ќтвет
состоит в том, что мы предположили, что источником гравитации  €вл€етс€, по
крайней мере частично, не плотность массы, а плотность энергии. ¬ этом и
состоит критическое отличие от нерел€тивистской теории Ќьютона.

\subsubsection{¬ывод из вариационного принципа}

ќбычно в учебниках по ќ“ќ основные уравнени€ космологии, т.е. уравнение
‘ридмана, вывод€т из уравнений Ёйнштейна в предположении однородной и
изотропной метрики (\ref{ds2-friedman}) или (\ref{ds2-cosm}).
¬ самом деле, уравнение (\ref{Lambda}) в смешанных компонентах даст при $\Lambda=0$ дл€
тензора Ёйнштейна $G_\mu^\nu$:
\be
G_\mu^\nu \equiv R_\mu^\nu - \frac{1}{2} g_\mu^\nu R  =
\frac{8\pi}{m_{Pl}^2}\, T_\mu^\nu\, ,
\label{EinsMixed}
\ee
откуда дл€ метрики FRW (\ref{ds2-cosm}) имеем
\be
G_0^0 = R_0^0 - \frac{1}{2} R  =
\frac{8\pi}{m_{Pl}^2}\, T_0^0\,.
\label{EinsM00}
\ee
¬ычисление тензоров –иччи и Ёйнштейна даЄт
\begin{equation}
G_0^0={{3\,k+3\,\left(\dot a\right)^2 }\over{a^2}} .
\label{G00}
\end{equation}
ѕоскольку $ T_0^0 = \rho$,  отсюда сразу получаем (\ref{H2}).


ѕокажем, что это уравнение можно вывести в рамках ќ“ќ, не пользу€сь
тензором энергии-импульса $T_i^k$ и
уравнени€ми Ёйнштейна, а пр€мо из действи€ $S$.
„тобы выводить из вариационного принципа нужно знать только скал€р кривизны $R$.
¬ этом выводе есть р€д поучительных моментов.

ƒействие $S$ состоит из суммы двух слагаемых: гравитационной части и действи€
материи, $S_m$, определ€емого лагранжианом ${\cal L}_m$:
\beq
  S=S_H+S_m=\varkappa \int R \sqrt{-g}d^4x
  + \int {\cal L}_m  \sqrt{-g}d^4x \ .
  \label{fullAction}
\enq

 онстанта $\varkappa$ в выражении (\ref{fullAction}) равна:
\beq
 \varkappa \equiv \frac{1}{16\pi G_N} \equiv \frac{m_{Pl}^2}{16\pi}  .
 \label{varkappa}
\enq

¬озьмем простой случай совершенной (идеальной) жидкости, когда плотность
лагранжиана $ {\cal L}_m$ есть просто энерги€ в единице объЄма, ${\rho} $.
Ёто подробно рассмотрено в книге ¬.ј.‘ока \cite{Fock}.


ѕодставл€€ выражени€ дл€ скал€ра кривизны, $R$, через масштабный фактор и его
производные и плотность энергии, ${\rho}$, умноженные на $\sqrt{-g}=a^3$,
получим функцию Ћагранжа дл€ величины $a(t)$:
\begin{equation}
 L[a(t)] =  -6\varkappa(a^2\ddot a+ a {\dot a}^2 +ka)
 + \rho a^3.
\end{equation}
ћы можем избавитьс€ от второй производной по времени,
$\ddot a$, проинтегрировав соответствующее слагаемое в действии по част€м:
\be
  \int_{t_{in}}^t a^2 \ddot a dt =   \int_{t_{in}}^t a^2 d \dot a =
 a^2 \dot a \biggr|_{t_{in}}^t -\int_{t_{in}}^t \dot a d a^2 =
\mbox{const} - 2 \int_{t_{in}}^t {\dot a}^2 a dt .
\label{int-by-parts}
\ee
“еперь обсудим пределы при интегрировании по времени.
Ќижний предел хотелось бы вз€ть за нуль $t_{in}=0$ (то что называют моментом
Ѕольшого ¬зрыва - Big Bang), но
не надо забывать, что решение в этом нуле в идеализированных модел€х окажетс€
сингул€рным.
—труктуру этой  сингул€рности в рамках ќ“ќ уже нельз€ изучать.
¬ерхний предел хотелось бы увести на бесконечность, но сделать это тоже нельз€,
так как наша модельна€ вселенна€ может оказатьс€ конечной во времени.

»збавившись от второй производной, мы получим лагранжиан в стандартном  виде
лишь с первыми производными,
но в отличие от принципа экстремума действи€ классической механики, когда мы
фиксируем только координаты частиц на концах траектории,
теперь мы должны фиксировать и скорость (здесь $\dot a$).
“олько при $\delta a = 0$ и $\delta \dot a = 0$ на границах области
интегрировани€ мы можем забыть о неинтегральном члене.
¬ этом случае мы получим
лагранжиан дл€ масштабного фактора $a(t)$: 
\begin{equation}
 L = -6\varkappa (-{\dot a}^2 a +ka) + {\rho} a^3.
\end{equation}

ћы можем вз€ть любое баротропное уравнение состо€ни€, чтобы св€зать
$\rho$ и плотность числа частиц, $n$. ¬ частности, можно предположить
что $ \rho $ зависит только от $a$ (через $n$), но не от $\dot a$.
“огда мы имеем `гамильтониан'
$$
{\cal  H}= \frac{\partial L}{\partial \dot a}\dot a - L =
 6\varkappa ( a {\dot a}^2 + ka) -  {\rho }a^3.
$$
«десь нет никакой €вной зависимости от времени, следовательно, ''энерги€''
должна быть посто€нной:
\begin{equation}
  (a {\dot a} ^2 + ka) - \frac{ {\rho} a^3}{6\varkappa} = \mbox {const} .
\label{friedmkappa}
\end{equation}
ћы можем обратить константу в правой части в нуль, добавив к плотности энергии
плотность какой-то пылевидной материи: $\rho \to \rho + \Delta \rho$ с
$\Delta \rho = \rho_0 (a_0/a)^3$. »ными словами, произвольна€ посто€нна€ может
быть поглощена в определении плотности энергии за счет выбора $\rho_0 a_0^3$.
 азалось бы, этого произвола нет при выводе такого же уравнени€ обычным путЄм из
уравнений  Ёйнштейна, приведЄнном выше.



 Ќо произвол всЄ равно остаЄтс€, только константа определ€етс€ правильным выбором нулевого
отсчЄта энергии материи.

¬ыбор нул€ дл€ плотности энергии в (\ref{EinsMixed}) только кажетс€ естественным:
там где нет никакой кривизны, имеем нулевые тензоры –имана, –иччи и Ёйнштейна,
там же полагаем и $\rho=0$.
Ќо если истинные уравнени€ гравитации содержат Ћ€мбда-член, то никакой естественности
не получаетс€: ведь можно писать либо Ћ€мбда-член слева, либо ненулевую энергию вакуума
справа с уравнением состо€ни€ $P=-\rho^{\rm vac}$.
ј можно и произвольные доли Ћ€мбда слева, а вакуумной энергии справа, получа€
совершенно эквивалентную динамику вселенной.

»так, при занулении константы в  уравнении (\ref{friedmkappa}) получаем,
поделив его на $a^3$ и использу€ $\varkappa$, определ€емое  выражением 
(\ref{varkappa}):
\begin{equation}
  \left({{\dot a}\over a}\right)^2={{8\pi G}\over 3}{\rho}
  -{{k}\over a^2} ,
\label{friedm}
\end{equation}
что и даЄт известное первое уравнение ‘ридмана (\ref{H2}).


¬ернЄмс€ к вопросу о граничных членах, возникающих при интегрировании по част€м
в выводе уравнений движени€.
—в€зь вариационного принципа
 √ильберта с уравнени€ми Ёйнштейна вовсе не тривиальна из-за внеинтегральных членов.

«ануление вариаций производных на границе в учебниках обычно упоминают,
но уравнени€ из вариационного принципа получают, сразу забыв
об этом ограничении. Oдно известное нам исключение -- книга ”олда~\cite{Wald}.
Ѕолее свежа€ книга Eric Poisson~\cite{Poisson04}
 -- см. там формулу (4.7) и ниже.
»з книг по космологии это обсуждаетс€ в книге~\cite{KolbTurner}
(стр. 451-464 -- on Hamiltonian constraint).

¬ этих книжках показано, что к действию √ильберта надо добавить члены с внешней кривизной.
“акое действие обычно называют действием √иббонса-’окинга~\cite{GibH1977}, иногда добавл€ют …орка (GibbonsЦHawkingЦYork action),
хот€, по-видимому, …орк~\cite{York1972} был первым, кто в €вном в виде добавил внешнюю кривизну в действие.

Ќапример, одна из часто цитируемых статей \cite{HawkHor1996}
предлагает такое действие ``for a metric $g$ and generic matter fields $\phi$:
\beq
 I(g,\phi)= \int_M \left[{R \over 16\pi} + L_m(g,\phi)   \right] + {1 \over 8\pi} \oint_{\partial M} K
\label{Hawk}
\enq
where $R$ is the scalalr curvature of $g$, $L_m$ is the matter Lagragian, and $K$ is the trace of the
extrinsic curvature of the boundary.''

¬ формуле (\ref{Hawk}) выбраны единицы, в которых и $m_{Pl}=1$.
ћы не будем давать формальное определение  внешней кривизны $K_{ik}$.
≈го можно вз€ть, например, из книги \cite{MTW} стр.153 том 2
русского издани€ и далее, и убедитьс€, что в нашем примере с метрикой ‘ридмана (и
–обертсона-”окера) с нашей сигнатурой метрики мы имеем
\beq
 K_{ik}=0.5{\dot g_{ik}} ,
 \label{extCurv}
\enq
ср. с формулой (21.70) из \cite{MTW}.
«десь латинские индексы  --- это 1,2,3 ( не так, как у Ћандау-Ћифшица~\cite{LL}!).

—мысл такого выражени€ дл€ внешней кривизны нетрудно пон€ть: в нашем случае граница многообрази€
$\pd M$ -- это сечени€ при $t=\mbox{const}$, т.е. всЄ наше трЄхмерное пространство.
¬нешн€€ кривизна получаетс€ по \cite{MTW} из векторов, которые смотр€т в 4-е измерение, в нашем случае -- это $t$.
“ак и по€вл€етс€ производна€ по времени в (\ref{extCurv}).

ѕотом надо  вз€ть след
$$
 K \equiv K_i^i \, ,
$$
и мы увидим, что при при том выборе коэффициентов, что сделан в выражении
(\ref{Hawk}), наш внеинтегральный член в (\ref{int-by-parts}), а именно,
$a^2 \dot a \biggr|_{t_0}^t$, который
содержит в себе точно ту же производную по времени, как и (\ref{extCurv}),
исчезает. Ќе следует забывать и про другие коэффициенты, которые тут не выписаны.
“аким образом, наш простой пример делает пон€тным выбор добавки к действию √ильберта в (\ref{Hawk}).

\section{ѕриложение Ѕ.  осмологические параметры. \label{B-cosm-par}}

«десь мы приведем значени€ основных космологических параметров и
кратко опишем способы их измерени€. „тобы не загромождать список
литературы, мы сошлемс€ лишь на~\cite{pdg},  где имеютс€ ссылки
на оригинальные работы и краткие обзорные статьи.

¬ажной величиной в космологии €вл€етс€ так называема€ критическа€
плотность энергии:
\be
\rho_c = \frac{3 H^2 m_{Pl}^2}{8\pi}\,.
\label{rho-c}
\ee
¬ англо€зычной литературе ее называют ``critical''  или ``closure'' density.
ѕоследний термин
св€зан с тем, что  $\rho=\rho_c$ -- граничное значение, при котором происходит
переход от открытой к замкнутой вселенной. —огласно астрономическим данным
полна€ плотность энергии всех форм материи во вселенной, $\rho_{\rm tot}$, очень близка к критической, см. ниже,  уравнение (\ref{Omega-tot}).

 осмологическа€ плотность энергии той или иной формы материи характеризуетс€
безразмерным параметром:
\be
\Omega_a = \frac{\rho_a}{\rho_c}
\label{Omega-a}
\ee
”становлено, что
\be
\Omega_{\rm tot} = \rho_{\rm tot}/\rho_c =  1.006  \pm 0.006 .
\label{Omega-tot}
\ee
Ётот результат получен главным образом на основе анализа спектра угловых флуктуаций
микроволнового фона (CMB). ‘изическа€ длина волны, отвечающа€ первому максимуму
на момент  рекомбинации водорода ($z\approx  10^3$) известна, а тот угол, под которым
мы ее наблюдаем сегодн€, а значит и положение первого максимума, зависит от геометрии вселенной. Ёто наблюдаемое положение как раз отвечает плоской, эвклидовой геометрии.

 осмологическа€ плотность барионной материи определ€етс€ несколькими
независимыми способами: по наблюдаемым обили€м легких элементов, гели€, $^4$He
и дейтери€ D$\equiv ^2$H, рожденных во врем€ первичного нуклеосинтеза, по
отношению высот пиков в флуктуаци€х микроволнового фона и по масштабу, на
котором начинаетс€ диффузинонное, или  силковское затухание
(Silk damping). ¬се эти методы привод€т к близким значени€м:
\be
\Omega_B h^2 = 0.022,
\label{Omega-b-h2}
\ee
где $h$ - безразмерна€ посто€нна€ ’аббла, нормированна€ на 100:
 $h = H /100$ км/сек/ћпк. —огласно прин€тому значению~\cite{pdg}:
\be
h = 0.673 \pm 0.012
\label{h-cmb}
\ee
Ётот результат получен на основании анализа угловых флуктуаций микроволнового фона и
крупномасштабной структуры вселенной~\cite{planck-h}.
Oднако стоит указать на систематическое, и довольно значительное
расхождение этого результата с
измерени€ми $h$ традицинными астрономическими методами~\cite{h-astro}:
\be
h =0.738 \pm 0.024.
\label{h-astro}
\ee
ѕока не€сно, означает ли это, что значение параметра ’аббла зависит от рассто€ни€ необычным
образом и, следовательно, закон расширени€ не тот, что принимаетс€ в обычной
$\Lambda$CDM - космологии, см. например работу~\cite{DHT}, где проанализированы
возможные вариации закона расширени€, или же расхождение может быть
объ€снено наличием нестабильной, но долгоживущей компоненты тЄмной
материи~\cite{BDT,BDT1}. ¬прочем, не исключено и даже веро€тнее более
прозаическое  объ€снение,
что различие св€зано со систематическими ошибками
в определении астрономической лестницы рассто€ний. ¬ св€зи с последней
возможностью большой интерес приобретает метод пр€мого измерени€ $H$, предложенный в
работах~\cite{SIB-h}, который свободен от неоднозначности  обычных астрономических методов.

“ак или иначе, плотность барионной материи близка к
\be
\Omega_B = 0.05.
\label{Omega-B}
\ee
«аметим, что из этого 5\%-го вклада барионов в космологическую плотность, менее
половины наблюдаютс€ непосредственно. ќставша€с€ половина находитс€ неизвестно
где и неизвестно в чем.

—уммарна€ плотность барионов и тЄмной материи определ€етс€ по совокупности
целого р€да независимых космологических наблюдений. Ќекоторые, исторически
более ранние методы, не могут конкурировать с получившими развитие в недавнее
врем€, такими как анализ флуктуаций микроволнового фона или барионных
акустических осцилл€ций, но мы их упом€нем, чтобы показать, что анализ физически
различных €влений во вселенной приводит к близким значени€м плотностей
тЄмной и барионной материи и, совокупно определ€емой плотности тЄмной
энергии. ¬ качестве альтернативы тЄмной материи нередко предлагаетс€ модификаци€
динамики или гравитационного взаимодействи€, но им не под силу описать всю
совокупность имеющихс€ указаний на наличие именно тЄмной материи.

»так, сам факт наличи€ тЄмной материи и величина
космологической плотности нормально гравитирующей (барионной плюс невидимой, тЄмной
материи) определ€етс€ по: \\
1) по плоским кривым вращени€:  скорости частиц газа или малых галактик-спутников при
удалении от свет€щегос€ центра (большой) галактики не падают с рассто€нием а стрем€тс€ к
посто€нной величине. Ётот закон справедлив до рассто€ний пор€дка 10 галактических радиусов,
внутри которого находитс€ видимое свет€щеес€ вещество.\\
2) по  гравитационному линзированию отдаленных объектов, которое позвол€ет оценить полное
количество гравитирующего вещества вдоль луча зрени€.\\
3) по равновесию гор€чего газа в богатых галактических скоплени€х; дл€ удержани€ газа внутри
кластера требуетс€ примерно в п€ть раз больше гравитации, чем создаетс€ видимым веществом.\\
4) по эволюции образовани€ галактических скоплений как функции красного смещени€; в
пространственно плоской вселенной, т.е. во вселенной с $\Omega = 1$, в которой вс€ масса
``сидела''  бы в нормально гравитирующей материи, число галактических кластеров при $z \sim 1$
было бы раза в три меньше, чем наблюдаетс€. ≈сли же основна€ часть вещества находитс€ в виде
антигравитирующей тЄмной энергии, наблюдаема€ картина согласуетс€ с теорией.

¬се эти методы согласно дают дл€ полной плотности кластеризованной нормально
гравитирующей материи значение $\Omega_{\rm DM} + \Omega_B \approx 0.3$.
Ѕолее точные методы, основанные на анализе барионных акустических осцилл€ций и угловых
флуктуаци€х температуры микроволнового фона привод€т (после вычета вклада барионов) к:
\be
\Omega_{\rm DM} = 0.27.
\label{Omega-DM}
\ee
Ќедостающий до единицы ``кусок'' обеспечиваетс€ тЄмной энергией:
\be
\Omega_{\rm DE} = 0.68.
\label{Omega-GE}
\ee
 ак уже отмечалось, указание на антигравитационные свойства тЄмной энергии получены на
основании открыти€ того, что отдаленные сверхновые оказались более тусклыми, чем ожидалось
при стандартном замедл€ющимс€ космологическом расширении.  роме того, ускоренное
расширение необходимо дл€ разрешени€ кризиса возраста вселенной. Ѕез антигравитирующей
тЄмной энергии вселенна€ была бы раза в полтора моложе, чем следует из возраста старых
звездных скоплений и из €дерной хронологии.   такому же результату о необходимости
ускоренного расширени€ приводит и анализ крупномасштабной структуры вселенной. ”скоренное
расширение приводит к подавлению образовани€ структур на больших масштабах в согласии
с наблюдени€ми. ≈сли параметризовать равнение состо€ни€ тЄмной материи в форме
$P = w \rho$, то дл€ параметра $w$ имеем (см. \cite{pdg} и \cite{w-combine}):
\be
w =  -1.01 \pm 0.04 .
\label{w}
\ee
Ћюбопытно, что данные не только не исключают фантомное значение $w <-1$, но и €вно
указывают на него. ќстаетс€ возможным также вакуумное значение $w=-1$.  ¬ 
работе~\cite{w-of-time} на основе данных о барионных акустических осцилл€ци€х был сделан
вывод о возможной вариации тЄмной энергии при космологическом расширении.
  сожалению, имеющиес€ данные пока не позвол€ют определить, чем вызываетс€ ускоренное
расширение: тЄмной энергией или инфракрасной модификацией гравитации.

\section{ ¬. —кал€рное поле} \label{C-scalar-field}

¬ этом приложении мы приведем сводку необходимых формул о скал€рном поле в космологии.
ƒействие дл€ действительного скал€рного пол€ обычно выбираетс€ в виде:
\be
A [\phi ] = \int d^4 x \sqrt{-g} \left[ (1/2
) g^{\mu\nu}  \partial_\mu \phi  \partial_\mu \phi - U(\phi)
\right]\,
\label{action-scalar}
\ee
и приводит к уравнению движени€
\be
D^2 \phi + U'(\phi) = 0\,,
\label{D2-phi}
\ee
где $U'= dU/d\phi$,  $D^2 = g^{\mu\nu} D_\mu D_\nu$, а $D_\mu$-ковариантна€ производна€
во внешнем гравитационном поле. ¬ FRW-метрике (\ref{ds2-cosm})  с $k=0$ это уравнение принимает вид:
\be
\ddot \phi + 3H \dot \phi  - (1/a^2) \nabla^2 \phi + U' = 0
\label{ddot-phiH}
\ee
ќтвечающий этому действию тензор энергии-импульсa равен
\be
T_{\mu\nu} =2\, \frac{\delta A}{\delta g^{\mu\nu}}= \partial_\mu \phi \partial_\nu \phi -
g_{\mu\nu} \left[ \frac{1}{2}(\partial \phi)^2 - U(\phi)\right]\,.
\label{T-mu-nu-of-phi}
\ee
—оответственно плотность энергии и давлени€ равны:
\be
\rho = \frac{\dot \phi^2 + (\nabla \phi)^2/a^2 }{2} + U(\phi) , \,\,\,\,
P_{ij} = \delta_{ij}\left[\frac{\dot \phi^2 -(\nabla\phi)^2/a^2}{2} - U(\phi) \right] +
\frac{\partial_i\phi  \partial_i\phi}{a^2}\,.
\label{rho-P}
\ee
«аметим, что при $U>0$ среднее давление всегда меньше, чем плотность энергии,
$P_i^i /3  < \rho$. ќчевидно, что в случае медленно измен€ющегос€ пол€ $\phi$, как во времени
и в пространстве будет приближенно справедливо вакуумо-подобное уравнение состо€ни€:
$P \approx -\rho$. ¬ обратном предельном случае быстро измен€щегос€ $\phi(t)$, когда в $T_{\mu\nu}$
доминируeт производна€ по времени, а поле $\phi$ пространственно однородно,
имеет место максимально жесткое уравнение состо€ни€: $P=\rho$. ƒл€ такого уравнени€
состо€ни€ скорость звука равна скорости света.
Ёто уравнение могло бы реализоватьс€ при космологическом сжатии.

»ногда рассматриваютс€ так называемые тахионные уравнени€ состо€ни€, возникающие при $U < 0$,
например, при $U = -m^2 \phi^2/2$, т.е. теории с отрицательным квадратом массы. ѕрин€то считать,
что в таких теори€х должно быть сверхсветовое распросранение сигнала, но это неверно. ƒело в том,
что скорость сигнала - это скорость распространени€ его фронта, котора€ опередел€етс€ асимптотикой
коэеффициента преломлени€ при частоте/энергии, стрем€щейс€ к бесконечности. ¬ этом пределе
можно пренебречь потенциалом и мы приходим к нормальному распространению со скоростью света.
–азумеетс€, группова€ скорость оказываетс€ сверхсветовой, но это означает лишь деформацию
волны, что естественно так как вакуумное состо€ние неустойчиво при $m^2 < 0$.
 ак известно, в средах с аномальной дисперсией  имеет  место така€ же картина: группова€ скорость
оказываетс€ сверхсветовой, но волна деформируетс€, превраща€сь в ударную волну  с неаналитическим
фронтом.

«аметим, что уравнение состо€ни€ в форме  $P = P (\rho)$ не всегда существует. ’от€ всегда можно
определить параметр $w(t) \equiv P(t)/\rho(t)$, однако функциональной св€зи $P = P (\rho)$ может
и не быть. Ќапример, соотношение между $P$ и $\rho$ может быть нелокальным во времени.
“ем не менее, несмотр€ на отсутствие уравнени€  состо€ни€, набор космологических уравнений
окажетс€ полным так как недостающим уравнением будет уравнение движени€ пол€ (\ref{ddot-phiH}).

 вантование пол€ проводитс€ согласно обычной процедуре. ¬ однородном пространстве оператор
пол€ разлагаетс€ по его пространственным ‘урье-модам с коэффициентами, завис€щими от времени:
\be
\phi(t,{\bf x}) = \int \frac{d^3 k}{\sqrt{2E_k\,(2\pi)^3}} \left[ a_k e^{i \bf{kx}} f_k(t) +
a^\dagger_k e^{-i\bf{kx}} f^*_k(t) \right]\,,
\ee
где $E_k = \sqrt{m^2 + k^2 } $.
‘ункции $f_k (t)$ €вл€ютс€ решени€ми ‘урье преобразованного уравнени€ (\ref{D2-phi}) или
уравнени€ (\ref{ddot-phiH}), если речь идет о  FRW-метрике.
¬ частности, в плоском пространстве получим известный результат,
$f_k (t) = \exp (- iE_k t)$.
¬ космологическом случае дл€ FRW метрики
решение уравнени€ (\ref{ddot-phiH}) также известно аналитически. ћожно
показать, что оно выражаетс€ через функции Ѕессел€.

ќператоры $a_k$ и $a^\dagger_k$
€вл€ютс€  операторами уничтожени€ и рождени€ квантов пол€ $\phi$ (элементарными частицами).
ќни удовлетвор€ют соотношению коммутации:
\be
[a^\dagger_k, a_{k'}] = 2 E_k\,(2\pi)^2 \,\delta^3 ({\bf k} - \bf{ k'} )
\label{a-comm}
\ee
ќператор $a_k$, действу€ на вакуум, уничтожает его, $a_k |vac\rangle =0$, а оператор
$ a^\dagger_k$ создает одночастичное состо€ние с импульсом $k$: $a^\dagger_k |\mbox{vac}\rangle = |k\rangle$. ‘актор $2E$ в выражении (\ref{a-comm}) возникает лишь в пространстве ћинковского.
¬ общем случае искривленного пространства-времени необходимо писать вронскиан решений
уравнени€ движени€ дл€ $f_k(t)$.

‘ермионные пол€ квантуютс€ аналогично, только коммутатор операторов рождени€-уничтожени€
замен€етс€ на антикоммутатор. Ётот факт имеет важное значение дл€ взаимного уничтожени€
расход€щихс€ частей вакуумных энергий бозонов и фермионов, см. раздел (\ref{s-lambda}).

\newpage

\textbf{Cosmological acceleration}

S.I.Blinnikov and A.D.Dolgov

An overview is given of the current status of the theory and observations of the acceleration of expansion
the observable part of the universe.


\begin{thebibliography}{99}

\bibitem{ein-eq}
A. Einstein,
Sitzungsberichte der Preussischen Akademie der Wissenschaften zu Berlin: 844 (1915);
\bibitem{ein-eq1}
A. Einstein, Annalen der Physik, 49 (1916) 76,
``Die Grundllage der aillgeminen Relativit{\"a}tstheorie''.

\bibitem{Friedman1922}
A. Friedmann, Z. Phys. {\bf 10} (1922) 377.

\bibitem{Friedman1924} 
A. Friedmann, Z. Phys. {\bf 21} (1924) 326.

\bibitem{Robertson1933}
{H.P.} {{Robertson}},
  {\rmp} \textbf{{5}} (1933) {62}.

\bibitem{Walker1933}
{A.~G.} {{Walker}},
  {\mnras} \textbf{94} (1933) {159}

\bibitem{hubble}
E. Hubble, PNAS, \textbf{15} (1929) 168.

\bibitem{lemaitre}
G. Lemaitre, Annales de la Soc{\'e}t{\'e} scientifique de Bruxelles, S{\'e}rie A, 47 (1927), 49;
английский перевод  MNRAS, 91 (1931) 483;\\ 
см. также H. Nussbaumer, L. Bieri,  arXiv:1107.2281v2 [physics.hist-ph].

\bibitem{ein-lambda}
A. Einstein, Sitzgsber. Preuss. Acad. Wiss. {\bf 1} (1918) 142.

\bibitem{Chernin2008} A.D.Chernin, Physics Uspekhi, 178, 267 (2008)

\bibitem{lambda-OK}
W. de Sitter, Proc. Kon. Ned. Acad. Wet. 20 (1917) 229;
\bibitem{lambda-OK1}
G. Lema{\^i}tre, Annales de la Soci{\'e}t{\'e} Scientifique de Bruxelles 47 (1927) 49;
\bibitem{lambda-OK2}
G. Lema{\^i}tre, Nature 127 (1931) 706;
\bibitem{lambda-OK3}
A.S. Eddington, Monthly Notices of the Royal Astronomical Society, 90 (1930) 668.

\bibitem{prokopec}
J.F. Koksma, T. Prokopec; arXiv:1105.6296 [gr-qc].

\bibitem{PauliFQ} W. Pauli, ``Feldquantisierung'' 1950Ц1951, translation into English:
W. Pauli, Pauli Lectures on Physics: Vol 6, Selected Topics in Field Quantization,
MIT Press, 1971 (editor C.P. Enz)

\bibitem{zeld-vac}
я.Ѕ. «ельдович, ”спехи ‘изических Ќаук, 95 (1968) 209

\bibitem{Visser2016} Visser, M.\ 2016.\ Lorentz invariance and the zero-point stress-energy tensor.\ ArXiv e-prints arXiv:1610.07264

\bibitem{golfand}
ё.ј. √ольфанд, ≈.ѕ. Ћихтман. ѕисьма в ∆Ё“‘, 13 (1971) 452;\\
ƒ.¬. ¬олков, ¬.ѕ. јкулов, ѕисьма в ∆Ё“‘, 16 (1972) 621;\\
J. Wess, B. Zumino, Phys. Lett. {\bf 49} (1974)  52.

\bibitem{q-cond}
M. Gell Mann, R.J. Oakes, B. Renner, Phys. Rev. {175} (1968) 2195.

\bibitem{g-cond}
M.A. Shifman, A.I. Vainshtein, V.I. Zakharov, Nucl. Phys. { B147} (1978)  385.

\bibitem{rho-vac-rev}
S. Weinberg, Rev. Mod. Phys. {\bf 61} (1989) 1;
\bibitem{rho-vac-rev1}
A.D. Dolgov,{\it Proc. of the XXIVth Rencontre de Moriond; Series: Moriond
Astrophysics Meetings}, Les Arcs, France, p.227 (1989). Ed. J. Adouse and
J. Tran Thanh Van;
\bibitem{rho-vac-rev2}
A.D. Dolgov, {\it Fourth Paris Cosmology Coloquium}, World Scientific, 1998,
Singapore, 161. Ed. H.J. De Vega and N. Sanchez;
\bibitem{rho-vac-rev3}
P. Binetruy, Int. J. Theor. Phys. {\bf 39}, 1859 (2000) (lectures at Les
Houches
summer school ``The early Universe'' and Peyresq 4 meeting, July 1999);
\bibitem{rho-vac-rev4}
V. Sahni, A. Starobinsky, Int. J. Mod. Phys. {\bf D9}, 373 (2000);
\bibitem{rho-vac-rev5}
S. Weinberg, astro-ph/0005265; talk given at Dark Matter 2000,
February, 2000;
\bibitem{rho-vac-rev6}
Y. Fujii, Grav. Cosmol. {\bf 6}, 107 (2000);
\bibitem{rho-vac-rev7}
A. Vilenkin, hep-th/0106083,
talks given at ``The dark Universe'' (Space Telescope Institute)
and PASCOS-2001 in April 2001;
\bibitem{rho-vac-rev8}
V. Sahni, astro-ph/0202076, invited review at ``The Early Universe and
Cosmological Observations: a Critical Review'', UCT, Cape Town,
July 2001;
\bibitem{rho-vac-rev9}
N. Straumann, astro-ph/0203330,
Invited lecture at the first {\it S\'eminaire Poincar\'e}, Paris, March 2002;
\bibitem{rho-vac-rev10}
P.J.E. Peebles, B. Ratra, Rev. Mod. Phys. {\bf 75}, 559 (2003);
\bibitem{rho-vac-rev11}
J.E. Kim,  hep-ph/0402043, talk presented at COSPA-03, Taipei, Taiwan,
Nov. 13-15, 2003;
\bibitem{rho-vac-rev12}
C.P. Burgess, hep-th/0402200, contribution to the
proceedings of SUSY 2003, University of Arizona, Tucson AZ, June 2003;
\bibitem{rho-vac-rev13}
R. Bousso,  Gen. Rel. Grav. {\bf 40}, 607 (2008),  arXiv:0708.4231 [hep-th].

\bibitem{anthrop-early}
R.H. Dicke, Nature
{\bf 192} 440 (1961)
\bibitem{anthrop-early1}
B. Carter, in IAU Symposium 63: Confrontation of Cosmological Theories with Obser- vational Data, ed. by M. Longair (Reidel, 1974), page. 291;
\bibitem{anthrop-early2}
B.J. Carr, M.J. Rees, Nature
{\bf 278}, 605 (1979)
\bibitem{anthrop-early3}
I. L. Rozental', Usp. Fiz. Nauk {\bf 131} 239 (1980)
[Sov. Phys. Usp. 23,296 (1980).

\bibitem{book-anthrop}
».Ћ.  –озенталь, {\it Ёлементарные частицы и структура ¬селенной}, Ќаука 1984
\bibitem{book-anthrop1}
J. Barrow, F. Tipler, {\it The Anthropic Cosmological Principle}, Clarendon Press, 1986
\bibitem{book-anthrop2}
».Ћ.  –озенталь, {\it √еометри€, ƒинамика, ¬селенна€}, Ќаука 1987
\bibitem{book-anthrop3}
I.L. Rozental, {\it Big Bang, Big Bounce: How Particles and Fields Drive Cosmic Evolution},  Springer-Verlag (1988).


\bibitem{vilenkin-infl-83}
A. Vilenkin, Phys. Rev. D27, 2848 (1983).

\bibitem{linde-chaotic}
A.D. Linde,
Phys. Lett. B {\bf 175}, 395 (1986).

\bibitem{sakharov-lambda}
ј.ƒ. —ахаров,
∆Ё“‘, {\bf 87} 375 (1984).

\bibitem{kklt}
S. Kachru,  R. Kallosh, A. Linde, S.P. Trivedi
 Phys.\ Rev.\  D {\bf 68}, 046005 (2003).

\bibitem{douglas}
R.M. Douglas, M. R.,
 JHEP {\bf 0305}, 046 (2003).

\bibitem{suskind-landscape}
L. Susskind; arXiv:hep-th/0302219 .

\bibitem{carter-83}
B. Carter, W. H. McCrea, Phil. Trans. Roy.Soc. London A {\bf 310}, 347 (1983).

\bibitem{weinberg-anthrop}
S.  Weinberg,
Physical Review Letters {\bf 59} 2607 (1987).

\bibitem{linde-87}
A.~D.~Linde,
Inflation And Quantum Cosmology,
in: Three hundred years of gravitation. Cambridge Univ.
Press, Eds. Hawking, S.W. and Israel, W., 604-630 (1987).

\bibitem{tegmark-rees-98}
M. Tegmark, M.J. Rees, Ap. J. 499, 526 (1998).

\bibitem{vilenkin-2004}
A. Vilenkin,
in "Universe or Multiverse?", ed. by B.J. Carr (Cambridge University Press,
Cambridge 2007); arXiv:astro-ph/0407586.


\bibitem{Carriga2000}
J.Garriga, A, Vilenkin,
Phys.Rev. D61 083502, On likely values of the cosmological constant
(2000)

\bibitem{Carriga1999}
J. Garriga, M. Livio, A. Vilenkin, Phys. Rev. D 61, 023503,
The cosmological constant and the time of its dominance
(1999)

\bibitem{Mersini2008}
L.Mersini-Houghton, F.Adams,
Class.Quant.Grav.25:165002
Limitations of anthropic predictions for the cosmological constant: Cosmic Heat Death of Anthropic Observers (2008)

\bibitem{Hong2012}
Sungwook E. Hong, Ewan D. Stewart, Heeseung Zoe, Phys. Rev. D 85, 083510,
Anthropic Likelihood for the Cosmological Constant and the Primordial Density Perturbation Amplitude
(2012)

\bibitem{Hartle2013}
James Hartle, Thomas Hertog
Phys. Rev. D 88, 123516,
Anthropic Bounds on Lambda from the No-Boundary Quantum State (2013)

\bibitem{kane-anti-anthrop}
G.L.  Kane, M. J. Perry, A.N. Zytkow,
New Astronomy {\bf 7}, 45 (2002); arXiv:astro-ph/0001197.

\bibitem{KoyamaSakstein}   Koyama, K., Sakstein, J.,
  \prd, 91, 124066 (2015); arXiv eprint 1502.06872

\bibitem{Vainshtein} Vainshtein, A.~I.,
    To the problem of nonvanishing gravitation mass.
  Physics Letters B, 39, 393-394, (1972)

\bibitem{ad-nuff}
A.D. Dolgov, The very early universe, Proceedings of Nuffield Workshop, 1982, eds. B.W.
Gibbons and S.T. Siklos, Cambridge University Press.

\bibitem{ad-vector}
A.D. Dolgov, JETP Lett. 41 (1985) 345 [Pisma Zh.Eksp.Teor.Fiz. 41 (1985) 280].

\bibitem{ad-tensor}
A.D. Dolgov, Phys. Rev. D55 (1997) 5881.

\bibitem{mb}
M. Bronstein, Physikalische Zeitschrift der Sowjetunion, {\bf  3} (1933) 73.

\bibitem{notari}
T. Biswas, A. Notari,  Phys.Rev. D74 (2006) 043508.

\bibitem{Rothman1982} T. Rothman, R. Matzner, Astrophys. J. 257, 450 (1982)

\bibitem{Accetta1990}
 F. S. Accetta, L. M. Krauss, P. Romanelli, Phys. Lett. B248, 146 (1990)

\bibitem{Copi2004}
 C. J. Copi, A. N. Davis, L. M. Krauss, Phys. Rev. Lett. 92, 171301 (2004)

\bibitem{Bambi2005}
 C.Bambi, M.Giannotti and F.L. Villante, Phys.Rev.D., 71, 123524 (2005)

\bibitem{Zhu2015} Zhu, W.~W., and 19 colleagues 2015.\ Testing Theories of Gravitation Using 21-Year Timing of Pulsar Binary J1713+0747.\ The Astrophysical Journal 809, 41.

\bibitem{Anderson2015} J.D. Anderson, G. Schubert, V. Trimble, and M.R. Feldman, EPL 110, 1002 (2015)

\bibitem{Schlamminger2015} Schlamminger, S., Gundlach, J.~H., Newman, R.~D.\  Recent measurements of the gravitational constant as a function of time.\ Physical Review D 91, 121101 (2015)

\bibitem{fujii}
F. Wilczek, Phys. Rep. {\bf 104}, 143 (1984)

\bibitem{fujii1}
M. Ozer, M.O. Taha, Phys. Lett. B 171 (1986), 363; Nucl. Phys. B287 (1987) 776;

\bibitem{fujii2}
Y. Fujii, Astropart. Phys. { 5} (1996) 133

\bibitem{fujii3}
R.D. Peccei, J. Sola, C. Wetterich, Phys. Lett. B { 195} (1987) 183

\bibitem{fujii4}
L.H. Ford, Phys. Rev. D { 35} (1987) 2339

\bibitem{fujii5}
K. Freese, F.C. Adams, J.A. Frieman, E. Mottola, Nucl. Phys. B287 (1987) 797

\bibitem{fujii6}
M. Gasperini, Phys. Lett. B 194 (1987) 347

\bibitem{fujii7}
M. Reuter, C. Wetterich, Phys. Lett. B 188 (1987) 38

\bibitem{fujii8}
S.M. Barr, D. Hochberg, Phys. Lett. B  211  (1988) 49

\bibitem{fujii9}
N. Weiss Phys. Lett. B197 (1987) 42

\bibitem{fujii10}
Y. Fujii, T. Nishioka, Phys. Rev. D { 42} (1990) 361; Phys.Lett. B{ 254}  (1991) 347,
см. также приведенные там ссылки.

\bibitem{earlier-papers}
L.F. Abbott, Phys. Lett. 150B, 427 (1985)

\bibitem{earlier-papers1}
T. Banks, Nucl. Phys. B249, 332 (1985)

\bibitem{earlier-papers2}
O. Bertolami, Nuovo Cimento 93B, 36 (1986)

\bibitem{earlier-papers3}
S. M. Barr, Phys. Rev. D 36, 1691 (1987)

\bibitem{muk-rand}
S. Mukohyama, L. Randall, Phys. Rev. Lett. 92 (2004) 211302.

\bibitem{ad-kawa-2005}
A.D. Dolgov, M. Kawasaki, Phys. Atom. Nucl. {\bf 68}, 828; я‘ {\bf 68},
860; astro-ph/0307442,
Realistic cosmological model with dynamical cancellation of vacuum energy  (2005)

\bibitem{ad-kawa-2003}
A.D. Dolgov, M. Kawasaki,
astro-ph/0310822,
Stability of a cosmological model with dynamical cancellation of vacuum energy (2003)


\bibitem{two-scalars-fujii}
Y. Fujii, Phys. Rev. D {\bf 62},064004 (2000)

\bibitem{two-scalars-fujii1}
Y. Fujii, T. Nishioka, Phys. Lett. B {\bf 25}, 347 (1991)

\bibitem{two-scalars-fujii2}
Y. Fujii, Astropart Phys. {\bf 5}, 133 (1996).

\bibitem{two-scalars-rub}
V.A. Rubakov, Phys.Rev. D61 (2000) 061501, [hep-ph/9911305].

\bibitem{two-scalars-hebecker}
 A. Hebecker, C. Wetterich, Phys. Rev. Lett. {\bf 85}, 3339 (2000).

\bibitem{weinberg-2}
S. Weinberg, {\it The Quantum Theory of Fields}, v.1, sec. 5.9,
Cambridge University Press, 1995.

\bibitem{rub-G-of-t}
V.A. Rubakov, P.G. Tinyakov, Phys. Rev. D61 (2000) 087503.


\bibitem{notoph}
V. I. Ogievetsky,  I. V. Polubarinov, я‘ 4 (1966) 216; Sov. J. Nucl. Phys. 4 (1967) 156.
Stability

\bibitem{klink}
V.~Emelyanov, F.R.~Klinkhamer, Phys.Rev. D85 (2012) 063522

\bibitem{klink1}
V.~Emelyanov, F.R.~Klinkhamer, Int. J. Mod. Phys. D21 (2012) 1250025; arXiv:1108.1995

\bibitem{klink2}
V.~Emelyanov,  F.R.~Klinkhamer,  Phys.Rev. D85 (2012) 103508; arXiv:1109.4915

\bibitem{klink3}
V.~Emelyanov,  F.R.~Klinkhamer, Phys.Rev. D86 (2012) 027302.

\bibitem{2006Sci...312.1180S} Steinhardt, P.~J., Turok, N.\ 2006.\ Why the Cosmological Constant Is Small and Positive.\ Science 312, 1180-1183.

\bibitem{GreatDebate} Shapley, H., and Curtis, H.D., 1921. The scale of the
Universe. Bulletin of the National Research Council, 2, Part 3, No. 11, 171--217.

\bibitem{Boehle2016} Boehle, A., et al., Astrophysical Journal, \textbf{830} (2016) p. 17.

\bibitem{Hubble1929} Hubble, E.\ 1929.\ A Relation between Distance and Radial Velocity among Extra-Galactic Nebulae.\
Proceedings of the National Academy of Science 15, 168-173.

\bibitem{Sandage1995} Sandage, A.\ 1995.\ Practical Cosmology: Inventing the Past.\ Saas-Fee Advanced Course 23.~Lecture Notes
1993.~Swiss Society for Astrophysics and Astronomy, XIV, Springer-Verlag Berlin Heidelberg New York,
p.~1-232.

\bibitem{Wirtz1922} C.~Wirtz, Astronomische Nachrichten, {\bf 215}, 349 (1922).

\bibitem{Wirtz1936} C.~Wirtz, Zeitschrift f\"ur Astrophysik, {\bf 11}, 261  (1936).

\bibitem{SeitterDuerbeck1999} Seitter, W.~C., Duerbeck, H.~W.\ 1999.\ Carl Wilhelm Wirtz - Pioneer in Cosmic Dimensions.\ Harmonizing Cosmic Distance Scales in a Post-HIPPARCOS Era 167, 237-242.

\bibitem{vandenBergh2011} van den Bergh, S.\ 2011.\ Discovery of the Expansion of the Universe.\ Journal of the Royal Astronomical Society of Canada 105, 197.

\bibitem{Raifeartaigh2012} Raifeartaigh, C.~O 2012.\
The contribution of VM Slipher to the discovery of the expanding universe;
arXiv:1212.5499.

\bibitem{Humason1931} Humason, M.~L.\ 1931.\ Apparent Velocity-Shifts in the Spectra of Faint Nebulae.\ The Astrophysical Journal 74, 35.

\bibitem{Pruzh2015} ћ.¬.ѕружинска€, —.ћ.Ћисаков, 2015. ѕрирода  No. 12, 36

\bibitem{Pruzh2016} Pruzhinskaya, M.V., Lisakov, S.M., 2016. How supernovae became the basis of the observational cosmology. 2016, JAH\&H, 19(2), 203--215.

\bibitem{Lemaitre1927} Lema{\^i}tre, G. Un Univers homog\`ene de masse constante et de rayon croissant rendant compte de la vitesse radiale des n\'ebuleuses extra-galactiques. Annales de la Soci\'et\'e Scientifique de Bruxelles, A47 (1927) p. 49--59.

\bibitem{Lemaitre1931} Lema{\^i}tre, G.\ 1931.\ Expansion of the universe, A homogeneous universe of constant mass and increasing radius accounting for the radial velocity of extra-galactic nebulae.\ Monthly Notices of the Royal Astronomical Society 91, 483-490.

\bibitem{Hubble1926} Hubble, E.P. Extragalactic nebulae. Astrophysical Journal, \textbf{64} (1926) p. 321--369.

\bibitem{Shklovsky1967} Shklovsky, J.\ 1967.\ On the Nature of ``standard'' Absorption Spectrum
of the Quasi-Stellar Objects.\ The Astrophysical Journal 150, L1.

\bibitem{Kardashev1967} Kardashev, N.\ 1967.\ LEMA{\^I}TRE'S Universe and Observations.\
The Astrophysical Journal 150, L135.

\bibitem{GunnTinsley1975} Gunn, J.~E., Tinsley,
B.~M.\ 1975.\ An accelerating Universe.\ Nature 257, 454-457.

\bibitem{Tinsley1978} Tinsley, B.~M.\ 1978.\
Accelerating universe revisited.\ Nature 273, 208-211.

\bibitem{Fukugita1990} Fukugita, M.,
Yamashita, K., Takahara, F., Yoshii, Y.\ 1990.\ Test for the cosmological
constant with the number count of faint galaxies.\ The Astrophysical
Journal 361, L1-L4.

\bibitem{FukugitaHogan1990} Fukugita, M.,
Hogan, C.\ 1990.\ High H$_{0}$?.\ Nature 347, 120-121.

\bibitem{Fukugita1993} Fukugita, M., Hogan,
C.~J., Peebles, P.~J.~E.\ 1993.\ The cosmic distance scale and the Hubble
constant.\ Nature 366, 309-312.

\bibitem{Sandage1974}
   {{Sandage}, A. and {Tammann}, G.~A.},
  {\apj}, \textbf{194}, {559-568} (1974).

\bibitem{Weinberg1972}
S. Weinberg, Gravitation and Cosmology, 1972;
—. ¬ейнберг,  √равитаци€ и космологи€, ћир,  ћосква, 1975

\bibitem{Weinberg2008}
S. Weinberg, Cosmology, 2008;
—. ¬айнберг,   осмологи€, ”–——:  нижный дом ``Ћ»Ѕ–ќ ќћ'',  ћосква, 2013

\bibitem{Carroll2004}
S. Carroll, Spacetime and Geometry, An Introduction to General Relativity, Addison Wesley, 2004; arXiv:gr-qc/9712019

\bibitem{GorbunovRubakov2006}
ƒ.—.~√орбунов, ¬.ј.~–убаков, ¬ведение... (2006)
Ђ¬ведение в теорию ранней ¬селенной. “еори€ гор€чего Ѕольшого взрываї (ћ.: URSS, 2008)

\bibitem{KodamaSasaki1984} Kodama, H., Sasaki, M.\ 1984.\ Cosmological Perturbation Theory.\ Progress of Theoretical Physics Supplement 78, 1.

\bibitem{Challinor}
Challinor, Anthony
Constraining fundamental physics with the cosmic microwave background.
Invited review at the Workshop on Cosmology and Gravitational Physics, 15-16 December 2005, Thessaloniki, Greece;
arXiv:astro-ph/0606548


\bibitem{RubVlasov2012}
   Rubakov, V.~A., Vlasov, A.~D.,
    What do we learn from the CMB observations?,
  Physics of Atomic Nuclei, {\bf 75} (2012) 1123-1141; arXiv:1008.1704.

\bibitem{Keel2009}
{W.C.}~{{Keel}},  {{The Extragalactic Distance Scale}}
  (2009),
  \url{http://www.astr.ua.edu/keel/galaxies/distance.html}.

\bibitem{ChurazovEa2014} Churazov et al. \nat (2014).

\bibitem{Poznanski07} Poznanski, D., et al., \mnras, {\bf 382}, 1169 (2007)

\bibitem{Cooke09} Cooke, J., Sullivan, M., Barton, E.~J., Bullock, J.~S., Carlberg, R.~G., Gal-Yam, A.,
\& Tollerud, E., \nat, {\bf 460}, 237 (2009).

\bibitem{Jones2013} Jones, D.O., et al. The discovery of the most distant known Type Ia supernova at
redshift 1.914. Astrophysical Journal, \textbf{768} (2013) p. 166.

\bibitem{Schmidt98}
 B.P. Schmidt et al., \apj {\bf 507}, 46 (1998)

\bibitem{Riess98}
A.G. Riess et al.,  Astron.J. {\bf 116}, 1009 (1998).

\bibitem{Perlm99}
 S. Perlmutter et al., \apj {\bf 517}, 565 (1999).


\bibitem{Pskovskij77} {{Y.~P.} {{Pskovskii}}},   {\azh} \textbf{{54}}, {1188}
  ({1977}). {\sovast} \textbf{{21}}, {675} ({1977}).


\bibitem{Pskovskii1984} Pskovskii, Yu.P. {\azh} \textbf{{61}}, 1125 (1984).
Photometric classification and basic
parameters of type I supernovae. Soviet Astronomy, \textbf{28} (1984), p. 658-664.

\bibitem{Rust} Rust B.W. The use of supernovae light curves for testing the expansion hypothesis and
other cosmological relations // Univ. of Illinois, ORNL 4953, Ph.D. thesis, Oak Ridge National Lab., TN., 1974.

\bibitem{Phillips93} {{M.~M.} {{Phillips}}}, {\apjl} \textbf{{413}}, {L105} ({1993}).

\bibitem{Baade1938} Baade, W., 1938. The absolute photographic magni-
tude of supernovae. Astrophysical Journal,  \textbf{88} (1938) 285-304.

\bibitem{Wilson1939} Wilson, O.C., 1939. Possible applications of super-
novae to the study of the nebular red shifts. Astrophysical Journal, 90, 634-636.

\bibitem{Zwicky1939} Zwicky, F., 1939. On the theory and observation of
highly collapsed stars. Physical Review, 55, 726-743.

\bibitem{Pskovskii1967} Pskovskii, Yu.P., 1967. The photometric properties of
supernovae. Soviet Astronomy, \textbf{11}, 63-69. јстрон.∆. \textbf{44}, 82.

\bibitem{Phillips05}
{{M.~M.} {{Phillips}}}, in   \emph{{1604-2004: Supernovae as Cosmological
  Lighthouses}}, Ed. by  {{M.}{{Turatto}}}   {\emph{et~al.}} ({2005}), p. {211}.

\bibitem{Branch1992}
	Branch, David; Tammann, G. A.
Annual review of astronomy and astrophysics. \textbf{30}  (1992) p. 359-389.

\bibitem{Vaucouleurs1976}
	de Vaucouleurs, G.; Pence, W. D.
	Astrophysical Journal, \textbf{209} (1976) p. 687-692.

\bibitem{Folatelli2010}  G.{Folatelli},  M.M.{Phillips},  C.R.{Burns} et al.
  {\aj}, \textbf{139}, 120-144 (2010), eprint:{0910.3317}.

\bibitem{Astier2006}
{{P.}~{{Astier}}}
  {\emph{et~al.}}, {\aap}
  \textbf{{447}}, {31} ({2006}),
  eprint:{astro-ph/0510447}.

\bibitem{Betoule2014} Betoule, M., et al. Improved cosmological constraints from a joint analysis of the SDSS-II and SNLS supernova samples. A\&A, \textbf{568} (2014), p. 22,
 eprint:{arXiv:1401.4064}.

\bibitem{Riess2011ApJ...730..119R}
{Riess}, A.~G., {Macri}, L., {Casertano}, S., {et~al.} 2011, \apj, 730, 119


\bibitem{Scolnic2017} Scolnic, D. M., et al. The Complete Light-curve Sample of Spectroscopically
Confirmed Type Ia Supernovae from Pan-STARRS1 and Cosmological Constraints from The Combined Pantheon
Sample (2017); arXiv:1710.00845.

\bibitem{Dam2017}
Dam, L.~H., Heinesen, A., \& Wiltshire, D.~L.\  2017. Apparent cosmic acceleration from Type Ia supernovae.  \mnras, 472, 835

\bibitem{Racz2017} R{\'a}cz, G., Dobos, L., Beck, R., Szapudi, I., \& Csabai, I.\ 2017.
Concordance cosmology without dark energy. \mnras, 469, L1

\bibitem{Buchert2018} Buchert, T.\ 2018. On Backreaction in Newtonian cosmology. \mnras, 473, L46

\bibitem{DWeinberg2013}
{Weinberg}, D.~H. et al.
    Observational probes of cosmic acceleration,
   {\physrep} 530, 87-255 (2013); arXiv:1201.2434

\bibitem{BassettHlozek2009} Bassett, Hlozek
Baryon Acoustic Oscillations (2009); arXiv:0910.5224

\bibitem{Giovannini2005} Giovannini, M. {\it
International Journal of Modern Physics D} {\bf 14} (2005) 363; astro-ph/0412601

\bibitem{Giovannini2008}
	Giovannini, M. \ 2008,
	A Primer on the Physics of the Cosmic Microwave Background, by Massimo Giovannini. World Scientific Publishing Co., Singapore.

\bibitem{GorbunovRubakov2010}
√орбунов ƒ.—., –убаков ¬.ј. ¬ведение в теорию ранней ¬селенной.  осмологические возмущени€. »нфл€ционна€ теори€. 2010, 564 с



\bibitem{Komatsu2009} Komatsu, E., and 18 colleagues 2009.\ Five-Year Wilkinson
Microwave Anisotropy Probe Observations: Cosmological Interpretation.\ The
Astrophysical Journal Supplement Series 180, 330-376

\bibitem{HuSugiyama1996} Hu, W., Sugiyama, N.\ 1996.\ Small-Scale Cosmological
Perturbations: an Analytic Approach.\ The Astrophysical Journal 471, 542;
astro-ph/9510117


\bibitem{Dodelson2003} Dodelson, S.\ 2003.\ Modern cosmology. Amsterdam (Netherlands): Academic Press.~ISBN 0-12-219141-2, 2003, XIII + 440 p.\ .


\bibitem{Alam2016} Alam~Sh. et al.
	The clustering of galaxies in the completed SDSS-III Baryon Oscillation Spectroscopic Survey: cosmological analysis of the DR12 galaxy sample
	(2016); arXiv:1607.03155

\bibitem{Hong2015} Hong, Tao; Han, J. L.; Wen, Z. L.
	A detection of Baryon Acoustic Oscillations from the distribution of galaxy clusters.
	Astrophysical Journal, 826, article id. 154, 8 pp. (2016);
	arXiv:1511.00392.

\bibitem{Seo2016} Seo, H.-J., Beutler, F.,
Ross, A.~J., Saito, S.\ 2015.\ Modeling the reconstructed BAO in Fourier
space.
Monthly Notices of the Royal Astronomical Society, Volume 460, Issue 3, p.2453-2471 (2016); arXiv:1511.00663.

\bibitem{DR11} Anderson, L., Aubourg,
{\'E}., Bailey, S., et al.\ 2014, \mnras, 441, 24

\bibitem{Eisen1998} Eisenstein, D.~J.,Hu, W., \& Tegmark, M.\ 1998, \apj, 504, L57


\bibitem{Eisen2007} Eisenstein, D.~J., et al. 2007; arXiv:0607061

\bibitem{Padm2009} Padmanabhan N, White M,  Cohn J D  \textit{Physical Review D}, \textbf{79}, Issue 6, id. 063523  (2009); arXiv:0812.2905


\bibitem{DHT}
A. Dolgov, V. Halenka, I. Tkachev, 
 \ Power-law cosmology, SN Ia, and BAO.\ Journal of Cosmology and Astro-Particle
 Physics 10, 047 (2014); arXiv:1406.2445.

\bibitem{BAO}
  B.~A.~Bassett, R.~Hlozek,
  {\em Baryon Acoustic Oscillations}, in
  Dark Energy, Ed. P. Ruiz-Lapuente (2010);
  arXiv:0910.5224.

\bibitem{Tegmark:2006az}
  M.~Tegmark {\it et al.}  [SDSS Collaboration],
  Phys.\ Rev.\ D {\bf 74} (2006) 123507; astro-ph/0608632.

\bibitem{Ade:2013zuv}
  P.~A.~R.~Ade {\it et al.}  [Planck Collaboration],
  Astron.\ Astrophys.\  (2014); arXiv:1303.5076.


\bibitem{Beutler}
F. Beutler,  C. Blake,  M. Colless,  {\it et al.}, MNRAS, {\bf 416} (2011) 3017.

\bibitem{Padmanabhan}
N. Padmanabhan, X. Xu, D. Eisenstein,  {\it et al.}, MNRAS, {\bf 427} (2012) 2132.

\bibitem{Anderson}
L. Anderson, E. Aubourg, S. Bailey,  {\it et al.}, MNRAS, {\bf 427} (2012) 3435.

\bibitem{Xia}
Jun-Qing Xia, V. Vitagliano, S. Liberati, M. Viel, Phys. Rev. {\bf D85} (2012) 043520;
arXiv:1103.0378.

\bibitem{Percival:2009xn}
W.~J.~Percival {\it et al.}, MNRAS,  {\bf 401}  (2010) 2148.

\bibitem{Conley2011}
{{A.}~{{Conley}}},
  {{J.}~{{Guy}}},
  {{M.}~{{Sullivan}}},
  {\emph{et~al.}}, {\apjs}
  \textbf{{192}}, {1} ({2011}).

\bibitem{Hicken2009}
{{M.}~{{Hicken}}}   {\emph{et~al.}}, {\apj}   \textbf{{700}}, {1097} ({2009}); arXiv:0901.4804.

\bibitem{Kazin2014} Kazin, E.~A., and 28
colleagues 2014.\ The WiggleZ Dark Energy Survey: improved distance
measurements to z = 1 with reconstruction of the baryonic acoustic
feature.\ Monthly Notices of the Royal Astronomical Society \textbf{441}, 3524-3542

\bibitem{Guy2007} Guy, J., Astier, P., Baumont, S., Hardin, D., Pain, R., Regnault, N., Basa, S., Carl- berg, R.G., Conley, A., Fabbro, S., Fouchez, D., Hook, I.M., Howell, D.A., Perrett, K., Pritchet, C.J., Rich, J., Sullivan, M., Antilogus, P., Aubourg, E., Bazin, G., Bron- der, J., Filiol, M., Palanque-Delabrouille, N., Ripoche, P., Ruhlmann-Kleider, V., 2007.
SALT2: using distant supernovae to improve the use of Type Ia supernovae as distance indicators. A\&A
466, 11Ц21. doi: 10.1051/0 0 04-6361:20 066930.

\bibitem{Tripp} Tripp, R., Branch, D. Determination of the Hubble Constant Using a Two-Parameter
Luminosity Correction for Type IA Supernovae. The Astrophysical Journal, Volume 525, Issue 1, pp. 209-214
(1999).

\bibitem{Sullivan2003} Sullivan, M., Ellis, R.S., Aldering, et al., 2003. The hubble diagram of Type Ia
supernovae as a function of host galaxy morphology. MNRAS 340, 1057Ц1075. doi: 10.1046/j.1365-8711.2003. 06312.x.

\bibitem{Henne2017} Henne, V., Pruzhinskaya, M.V., Rosnet, P., et al. The influence of host galaxy
morphology on the properties of Type Ia supernovae from the JLA compilation. New Astronomy 51 (2017) 43--50.


\bibitem{Sullivan2010} Sullivan, M., Conley, A., Howell, et al., 2010. The dependence of Type Ia
supernovae luminosities on their host galaxies. MNRAS 406, 782--802. doi:
10.1111/j.1365-2966.2010.16731.x.

\bibitem{Johansson2013} Johansson, J., Thomas, D., Pforr, J., Maraston, C., Nichol, R.C., Smith, M.,
Lampeitl, H., Beifiori, A., Gupta, R.R., Schneider, D.P., 2013. SN Ia host galaxy properties from Sloan
Digital Sky Survey-II spectroscopy. MNRAS 435, 1680--1700. doi: 10.1093/ mnras/stt1408.

\bibitem{Sullivan2006} Sullivan, M., Le Borgne, D., Pritchet, C.J., et al., 2006. Rates and properties of
Type Ia supernovae as a function of mass and star formation in their host galaxies. ApJ 648, 868Ц883.
doi: 10.1086/506137.

\bibitem{Neill2009} Neill, J.D., Sullivan, M., Howell, D.A., Conley, A., et al., 2009. The local hosts of
type Ia supernovae. ApJ 707, 1449--1465. doi: 10.1088/0 0 04-637X/707/2/1449.

\bibitem{Rigault2015} Confirmation of a Star Formation Bias in Type Ia Supernova Distances and its Effect
on the Measurement of the Hubble Constant.
Astrophysical Journal, 802, article id. 20, 18 pp. (2015).


\bibitem{Folatelli2013} Folatelli, G., and 29 colleagues 2013.\ Spectroscopy of Type Ia Supernovae by the Carnegie Supernova Project.\ The Astrophysical Journal 773, 53, eprint:1305.6997

\bibitem{Bogomazov2011} Bogomazov, A.I, Tutukov, A.V. Type Ia supernovae: Non-standard candles of the universe. Astronomy Reports, \textbf{55}, Issue 6 (2011), pp. 497--504.


\bibitem{Li2003}
{{W.}~{{Li}}},
  {{A.~V.} {{Filippenko}}},
  {{R.}~{{Chornock}}},
  {\emph{et~al.}}, {\pasp}
  \textbf{{115}}, {453} ({2003}),
  eprint:{astro-ph/0301428}.

\bibitem{Phillips2007}  {Phillips}, M.M., {Li}, W., {Frieman}, J. ., {Blinnikov}, S.I.
et al.
 {\pasp}, \textbf{119}, {360-387}  (2007); astro-ph/0611295.

\bibitem{Fakhouri2015} Fakhouri et al.
Improving Cosmological Distance Measurements Using Twin Type Ia Supernovae,
ApJ 815, 58 (2015);
arXiv:1511.01102

\bibitem{2007JCAP...02..011A} Alam, U., Sahni, V., Starobinsky, A.~A.\ 2007.\ Exploring the properties of
dark energy using type-Ia supernovae and other datasets.\ Journal of Cosmology and Astro-Particle Physics
2, 011.

\bibitem{2009PhRvD..80j1301S} Shafieloo, A., Sahni, V., Starobinsky, A.~A.\ 2009.\ Is cosmic acceleration
slowing down?.\ Physical Review D 80, 101301.

\bibitem{SIB-h}
Blinnikov, S., Potashov, M., Baklanov, P., Dolgov, A.\ 2012.\ Direct determination
of the Hubble parameter using type IIn supernovae.\ Journal of Experimental and Theoretical Physics Letters 96, 153-157.

\bibitem{SIB-h1}
Potashov, M., Blinnikov, S., Baklanov, P., Dolgov, A.\ 2013.\ Direct distance measurements to SN 2009ip.\ Monthly Notices of the Royal Astronomical Society 431, L98-L101.

\bibitem{SIB-h2}
Baklanov, P.~V., Blinnikov, S.~I., Potashov, M.~S., Dolgov, A.~D.\ 2013.\ Study of supernovae important for cosmology.\ Journal of Experimental and Theoretical Physics Letters 98, 432-439.

\bibitem{2016PhRvL.116f1102A} Abbott, B.~P., and 1012 colleagues 2016.\ Observation of Gravitational Waves from a Binary Black Hole Merger.\ Physical Review Letters 116, 061102.

\bibitem{2017PhRvL.119p1101A} Abbott, B.~P., and 1125 colleagues 2017.\ GW170817: Observation of Gravitational Waves from a Binary Neutron Star Inspiral.\ Physical Review Letters 119, 161101.

\bibitem{2017ApJ...848L..13A} Abbott, B.~P., and 1155 colleagues 2017.\ Gravitational Waves and Gamma-Rays from a Binary Neutron Star Merger: GW170817 and GRB 170817A.\ The Astrophysical Journal 848, L13.

\bibitem{2017ApJ...848L..15S} Savchenko, V., and 22 colleagues 2017.\ INTEGRAL Detection of the First Prompt Gamma-Ray Signal Coincident with the Gravitational-wave Event GW170817.\ The Astrophysical Journal 848, L15.

\bibitem{2017ApJ...850L...1L} Lipunov, V.~M., and 32 colleagues 2017.\ MASTER Optical Detection of the First LIGO/Virgo Neutron Star Binary Merger GW170817.\ The Astrophysical Journal 850, L1.

\bibitem{1984PAZh...10..422B} Blinnikov, S.~I., Novikov, I.~D., Perevodchikova, T.~V., Polnarev, A.~G.\ 1984.\ Exploding neutron stars in close binaries.\ Pisma v Astronomicheskii Zhurnal 10, 422-428.

\bibitem{1984SvAL...10..177B} Blinnikov, S.~I., Novikov, I.~D., Perevodchikova, T.~V., Polnarev, A.~G.\ 1984.\ Exploding Neutron Stars in Close Binaries.\ Soviet Astronomy Letters 10, 177-179, ArXiv:1808.05287.

\bibitem{1976ApJ...210..549L} Lattimer, J.~M., Schramm, D.~N.\ 1976.\ The tidal disruption of neutron stars by black holes in close binaries.\ The Astrophysical Journal 210, 549-567.

\bibitem{1977ApJ...215..311C} Clark, J.~P.~A., Eardley, D.~M.\ 1977.\ Evolution of close neutron star binaries.\ The Astrophysical Journal 215, 311-322.

\bibitem{1990AZh....67.1181B} Blinnikov, S.~I., Imshennik, V.~S., Nadezhin, D.~K., Novikov, I.~D., Perevodchikova, T.~V., Polnarev, A.~G.\ 1990.\ Explosion of a low mass neutron star.\ Astronomicheskii Zhurnal 67, 1181-1194.

\bibitem{1990SvA....34..595B} Blinnikov, S.~I., Imshennik, V.~S., Nadezhin, D.~K., Novikov, I.~D., Perevodchikova, T.~V., Polnarev, A.~G.\ 1990.\ Explosion of a Low-Mass Neutron Star.\ Soviet Astronomy 34, 595.

\bibitem{1986Natur.323..310S} Schutz, B.~F.\ 1986.\ Determining the Hubble constant from gravitational wave observations.\ Nature 323, 310.

\bibitem{2017Natur.551...85A} Abbott, B.~P., and 1313 colleagues (The LIGO Scientific Collaboration,
The Virgo Collaboration, The 1M2H Collaboration, The
Dark Energy Camera GW-EM Collaboration and the
DES Collaboration, The DLT40 Collaboration, The Las
Cumbres Observatory Collaboration, The VINRO UGE
Collaboration \& The MASTER Collaboration) 2017.\ A gravitational-wave standard siren measurement of the Hubble constant.\ Nature 551, 85-88.

\bibitem{2016ApJ...826...56R} Riess, A.~G., and 14 colleagues 2016.\ A 2.4\% Determination of the Local Value of the Hubble Constant.\ The Astrophysical Journal 826, 56.

\bibitem{2018ApJ...861..126R} Riess, A.~G., and 11 colleagues 2018.\ Milky Way Cepheid Standards for Measuring Cosmic Distances and Application to Gaia DR2: Implications for the Hubble Constant.\ The Astrophysical Journal 861, 126.

 \bibitem{PlanckCollaboration2016} Planck Collaboration 2016.\ Planck 2015 results. XIII. Cosmological parameters.\ Astronomy and Astrophysics 594, A13.

\bibitem{2018arXiv180706209P} Planck Collaboration, and 178 colleagues 2018.\ Planck 2018 results. VI. Cosmological parameters.\ ArXiv e-prints arXiv:1807.06209.

\bibitem{Leibundgut01}
B.{Leibundgut}.
\newblock {\em Nuclear Physics A}, 688:1--8, May 2001.

\bibitem{KK1974}
R.~P. {Kirshner} and J.~{Kwan}.
 {\apj}, 193:27--36, October 1974.

\bibitem{BaronSEAM}
E.~{Baron}, P.~E. {Nugent}, D.~{Branch}, and P.~H. {Hauschildt}.
\newblock {\em \apjl}, 616:L91--L94, December 2004.

\bibitem{Bartel2007}
N.~{Bartel}, M.~F. {Bietenholz}, M.~P. {Rupen}, and V.~V. {Dwarkadas}.
\newblock {\em \apj}, 668:924--940, October 2007.

\bibitem{Taddia2013} Taddia, F., Stritzinger, M.D., Sollerman, J., et al.
 {\it Astron. Astroph.}  {\bf 555} (2013) A10.


\bibitem{Grasberg1971} Grassberg, E.~K., Imshennik, V.~S., Nadyozhin, D.~K.\ 1971.\ On the Theory of the Light Curves of Supernovae.\ Astrophysics and Space Science 10, 28-51.

\bibitem{Grasberg1986} Grasberg, E.~K.,
Nadezhin, D.~K.\ 1986.\ Type-II Supernovae - Two Successive Explosions.\
Soviet Astronomy Letters 12, 68-70.

\bibitem{ChugaiEa04}
N.~N. {Chugai}, S.~I. {Blinnikov}, R.~J. {Cumming}, et al. 
\newblock {\em \mnras}, 352:1213--1231, August 2004.

\bibitem{WooBliHeg2007}
S.~E. {Woosley}, S.~{Blinnikov}, and A.~{Heger}.
\newblock {\em \nat}, 450:390--392, November 2007.

\bibitem{Baade1926}
W.~{Baade}.
\newblock {\em Astronomische Nachrichten}, 228:359, October 1926.

\bibitem{Wesselink1946}
A.~J. {Wesselink}.
\newblock {\em Bulletin of the Astronomical Institutes of the Netherlands},
  10:91, January 1946.


\bibitem{phantom}
R. R. Caldwell, Phys. Lett. B 545 (2002) 23.


\bibitem{ratra-peebles}
P.J.E. Peebles, B. Ratra, Astrophys J., 325 (1988) L17; \\
B. Ratra, P.J.E. Peebles, Phys. Rev. D37 (1988) 3406.

\bibitem{wetter}
C. Wetterich, Nucl. Phys. B302 (1988) 668;
Nucl. Phys. B302 (1988) 645;


\bibitem{quintessence-0}
I. Zlatev, L. Wang, P.J. Steinhardt, Phys. Rev. Lett. 82 (1999) 896
\bibitem{quintessence-1}
A.R. Liddle, R.J. Scherrer, Phys. Rev. D59 023509(1999)
\bibitem{quintessence-2}
T. Chiba, N. Sugiyama, T. Nakamura, MNRAS 289 (1997)  L5
\bibitem{quintessence-3}
R.R. Caldwell, R. Dave, P.J. Steinhardt, Phys. Rev. Lett. 80 (1998) 1582
\bibitem{quintessence-4}
W. Hu, D.J. Eisenstein and M. Tegmark, Phys. Rev. D 59, 023512(1998)
\bibitem{quintessence-5}
G. Huey, L. Wang, R.R. Caldwell, P.J. Steinhardt, Phys.Rev. D59 (1999) 063005
\bibitem{quintessence-6}
 P.G. Ferreria, M. Joyce, Phys. Rev. Lett. 79 (1997) 4740; Phys. Rev. D58, 023503 (1998)
\bibitem{quintessence-7}
K. Coble, S. Dodelson, J.A. Frieman, Phys. Rev. D 55 (1997) 1851
\bibitem{quintessence-8}
M.S. Turner, M. White, Phys. Rev. D 56 (1997) 4439
\bibitem{quintessence-9}
J. Frieman, I. Waga, Phys. Rev. D57 (1998) 4642
\bibitem{quintessence-10}
S.M. Carroll, Phys. Rev. Lett. 81 (1998) 3067;
  Quintessence and the Rest of the World;  astroph/9806099
\bibitem{quintessence-11}
P.J. Steinhardt, L. Wang, I. Zlatev, Phys. Rev. 59 (1999) 123504
\bibitem{quintessence-12}
T. Chiba,  Phys. Rev. D60 (1999) 083508;
gr-qc/9903094; Quintessence, the Gravitational Constant, and Gravity
\bibitem{quintessence-13}
M.C.Bento, O. Bertolami, Gen.Rel.Grav. 31 (1999) 1461;
gr-qc/9905075; Compactification, Vacuum Energy and Quintessence
\bibitem{quintessence-14}
F. Perrotta, C. Baccigalupi,  S. Matarrese, Phys. Rev. D61 (2000) 023507;
 astro-ph/9906066; Extended Quintessence
\bibitem{quintessence-15}
J. Garriaga, M. Livio, A. Vilenkin, Phys. Rev. D61 (2000) 023503;
 astro-ph/9906210. The cosmological constant and the time of its dominance
\bibitem{quintessence-16}
N. Bartolo,  M. Pietroni, Phys. Rev. D61 (2000) 023518;
hep-ph/9908521; Scalar-Tensor Gravity and Quintessence
\bibitem{quintessence-17}
Y. Fujii, Phys. Rev. D62 (2000) 044011.
Quintessence, scalar-tensor theories and non-Newtonian gravity


\bibitem{rev-quint}
E. J. Copeland, M. Sami, S. Tsujikawa, Int. J. Mod. Phys. D15, 1753 (2006);
\bibitem{rev-quint1}
R. Bean,  TASI Lectures on Cosmic Acceleration; arXiv:1003.4468
\bibitem{rev-quint2}
M.J. Mortonson, D.H. Weinberg, M. White,
to appear as Chapter 25 of Particle Data Group 2014 Review of Particle Physics;
arXiv:1401.0046



\bibitem{born-infeld}
M.~Born and L.~Infeld,
Proc.\ Roy.\ Soc.\ Lond.\ A { 144} (1934) 425.

\bibitem{det-gravity}
S.~Deser and G.~W.~Gibbons,
Class.\ Quant.\ Grav.\  {\bf 15}, L35 (1998); \\
D. Comelli, A. Dolgov, JHEP 0411 (2004) 062.


\bibitem{exp-U}
F. Lucchin, S. Matarrese,  Phys. Rev. D {\bf 32}, 1316 (1985).

\bibitem{exp-U2}
V. Sahni, H. Feldman, A. Stebbins, Astrophys. J. 385, 1 (1992).


\bibitem{grav-mdf}
S. Capozziello, S. Carloni, A. Troisi,
{\it RecentRes. Dev. Astron. Astrophys.} 1, 625 (2003); astro-ph/0303041

\bibitem{grav-mdf1}
S.M. Carroll, V. Duvvuri, M. Trodden, M.S. Turner,
{\it Phys.Rev.} {\bf D} 70, 043528 (2004); astro-ph/0306438.

\bibitem{DolgKaw}
A.D. Dolgov, M.Kawasaki, {\it Phys. Lett.} {\bf B} 573, 1 (2003).

\bibitem{mdf-mdf}
A.A. Starobinsky, JETP Lett. 86 (2007) 157

\bibitem{mdf-mdf1}
W. Hu, I. Sawicki, Phys. Rev. D 76 (2007) 064004

\bibitem{mdf-mdf2}
A. Appleby, R. Battye, Phys. Lett. B 654 (2007) 7.

\bibitem{odin-rev}
S. Nojiri, S. Odintsov, Phys.Rept. 505 (2011) 59.

\bibitem{appl-bat-star}
S.A. Appleby, R.A. Battye, A.A. Starobinsky, JCAP 1006 (2010) 005.

\bibitem{frolov}
A.V. Frolov, Phys. Rev. Lett. 101 (2008) 061103.

\bibitem{ea-ad}
E.V. Arbuzova, A.D. Dolgov,  Phys.Lett. B700 (2011) 289.

\bibitem{lr-sing}
L. Reverberi, Phys. Rev. D 87 (2013) 084005.

\bibitem{aas-R2}
V.Ts. Gurovich, A.A. Starobinsky, Sov. Phys. JETP 50 (1979) 844;
[∆Ё“‘ 77 (1979) 1683]

\bibitem{aas-R2a}
A.A. Starobinsky, JETP Lett. 30 (1979) 682; [ѕисьма ∆Ё“‘  30 (1979) 719]

\bibitem{aas-R2b}
A. A. Starobinsky, ``Nonsingular model of the Universe with the quantum- gravitational de Sitter stage and its observational consequences'',
in Proc. of the Sec- ond Seminar ``Quantum Theory of Gravity'' (Moscow, 13-15 Oct. 1981), INR Press, Moscow , 1982, pp. 58-72; reprinted in: Quantum Gravity, eds. M. A. Markov and P. C. West. Plenum Publ. Co., N.Y., 1984, pp. 103-128.

\bibitem{aas-infl}
A. A. Starobinsky, Phys. Lett. B91, 99 (1980).

\bibitem{pp-R2}
S. G. Mamaev, V. M. Mostepanenko and A. A. Starobinsky, JETP 43, 823 (1976)

\bibitem{pp-R2a}
Ya. B. Zeldovich, A. A. Starobinsky, JETP Lett. 26 (1977) 252

\bibitem{pp-R2b}
A. Vilenkin, Phys. Rev. D32 (1985)  2511

\bibitem{pp-R2c}
M. B. Mijic, M. S. Morris, Wai-Mo Suen, Phys. Rev. D34 (1986) 2934

\bibitem{pp-R2d}
Wai-Mo Suen, P. R. Anderson, Phys. Rev. D35 (1987) 2940.

\bibitem{GP}
D. S. Gorbunov and A. G. Panin, Phys. Lett. B700, 157 (2011); arXiv:1009.2448 [hep-ph].

\bibitem{ea-ad-lr}
E.V. Arbuzova, A.D. Dolgov, L. Reverberi, JCAP 1202 (2012) 049;  arXiv:1112.4995.

\bibitem{ea-ad-lr-2}
E.V. Arbuzova, A.D. Dolgov, L. Reverberi,   Eur. Phys .J. C72 (2012) 2247

\bibitem{ea-ad-lr-2a}
 E.V. Arbuzova, A.D. Dolgov, L. Reverberi, Phys. Rev. D88 (2013) 024035.

\bibitem{GT}
D. Gorbunov and A. Tokareva, J. Exp. Theor. Phys. 120, 528 (2015); arXiv:1412.3413

\bibitem{GTa}
D. Gorbunov and A. Tokareva, Phys. Rev. D96, 103527 (2017); arXiv:1412.3770.


\bibitem{adr-reply}
E.V. Arbuzova, A.D. Dolgov, L. Reverberi; arXiv:1707.02541

\bibitem{adr-anti}
E.V. Arbuzova, A.D. Dolgov, L. Reverberi, Astroparticle Physics 54 (2014) 44.

\bibitem{JB-theorem}
S. Capozziello, M. De Laurentis, Phys. Rep. 509 (2011) 167

\bibitem{JB-theorem1}
S. Capozziello, A. Stabile, A. Troisi, Phys. Rev. D 76 (2007) 1040.

\bibitem{sch-dombriz}
A. de la Cruz-Dombriz, A. Dobado, A. L. Maroto, Phys. Rev. {\bf D80}, 124011 (2009).

\bibitem{cembranos}
J.A.R. Cembranos, A. de la Cruz-Dombriz, B. Montes Nunez, JCAP 1204 (2012) 021.

\bibitem{Fock}
 ¬.ј. ‘ок. “еори€ пространства, времени и т€готени€. ћ.: √»““Ћ, 1955. - 504 с.

\bibitem{Wald}
R.M.Wald. General Relativity, University of Chicago Press, 506 pages,  1984.

–оберт ћ. ”олд
ќбща€ теори€ относительности
–ос. ун-т дружбы народов, 2008 - 692~c.


\bibitem{Poisson04}
{{E.}~{{Poisson}}},
  \emph{{{A relativist's toolkit : the mathematics of black-hole
  mechanics}}} ({Cambridge, Cambridge University Press},
  {2004}).

\bibitem{KolbTurner}
Edward W. Kolb, Michael S. Turner
 The Early Universe. Addison-Wesley (1990).

\bibitem{GibH1977}
{{G.~W.} {Gibbons}} {and}
  {{S.~W.} {Hawking}},
  {Phys. Rev. D} \textbf{{15}},
  {2752} ({1977}).

\bibitem{York1972}
{{J.~W.} {York}},
  {Phys. Rev. Lett.} \textbf{{28}},
  {1082} ({1972}).

\bibitem{HawkHor1996}
{{S.~W.} {{Hawking}}}
  {and} {{G.~T.}
  {{Horowitz}}}, {Classical and Quantum Gravity}
  \textbf{{13}}, {1487} ({1996}),
  eprint:{gr-qc/9501014}.

\bibitem{MTW}
{{C.~W.}~{{Misner}}},
  {{K.~S.}~{{Thorne}}}
  {and} {{J.~A.}~{{Wheeler}}},
  \emph{{{Gravitation}}}
  ({W.~H.~Freeman and Company}, {1973}).
–усский перевод в 3х томах.

\bibitem{LL}
Ћандау Ћ.ƒ. и Ћифшиц ≈.ћ.,
“еори€ пол€.Ч 8-е изд., стереот., ћ.: ‘»«ћј“Ћ»“, 2003.-534 с.
%
{{L.~D.} {{Landau}}} {and}
  {{E.~M.} {{Lifshitz}}},
  \emph{{{The Classical Theory of Fields}}}
  ({Oxford, Pergamon Press}, {1975}).

\bibitem{pdg}
M. Tanabashi et al. (Particle Data Group),
Review of Particle Physics,
Phys. Rev. D 98, 030001 (2018)


\bibitem{planck-h}
P.A.R. Ade et al. (Planck Collab. 2013 XVI); arXiv:1303.5076v1

\bibitem{h-astro}
A.G. Riess et al., Astrophys. J. 730, 119 (2011).

\bibitem{BDT}
Z. Berezhiani, A.D. Dolgov, I. I Tkahev, Phys.Rev. D92 (2015) No.6; arXiv:061303

\bibitem{BDT1}
A. Chudaykin, D. Gorbunov, I. Tkachev,  Phys.Rev. D94 (2016) 023528 .



\bibitem{w-combine}
S. Alam et al., MNRAS 470, 2617 (2017)

\bibitem{w-of-time}
V. Sahni, A. Shafieloo, A.A. Starobinsky.
Model-independent Evidence for Dark Energy Evolution from Baryon Acoustic Oscillations.\ The
Astrophysical Journal 793, L40 (2014);  arXiv:1406.2209

\bibitem{2017PhR...692....1N} Nojiri, S., Odintsov, S.~D., Oikonomou, V.~K.\ 2017.\ Modified gravity theories on a nutshell: Inflation, bounce and late-time evolution.\ Physics Reports 692, 1-104; arXiv:1705.11098 [gr-qc]


\bibitem{2018PhLB..779..425N} Nojiri, S., Odintsov, S.~D.\ 2018.\ Cosmological bound from the neutron star merger GW170817 in scalar-tensor and F(R) gravity theories.\ Physics Letters B 779, 425-429;  arXiv:1711.00492 [astro-ph.CO]

\bibitem{Capozziello2018} Salvatore Capozziello, Carlo Alberto Mantica, Luca Guido Molinari.\ Cosmological perfect-fluids in $f(R)$ gravity; arXiv:1810.03204 [gr-qc] (2018)

\bibitem{2016ARNPS..66...95J} Joyce, A., Lombriser, L., Schmidt, F.\ 2016.\ Dark Energy Versus Modified Gravity.\ Annual Review of Nuclear and Particle Science 66, 95-122; arXiv:1601.06133

\bibitem{2018RPPh...81a6902B} Brax, P.\ 2018.\ What makes the Universe accelerate? A review on what dark energy could be and how to test it.\ Reports on Progress in Physics 81, 016902.


\bibitem{2016ApJ...833L..30R} Rubin, D., Hayden, B.\ 2016.\ Is the Expansion of the Universe Accelerating? All Signs Point to Yes.\ The Astrophysical Journal 833, L30; arXiv:1610.08972


\bibitem{WrightLi2017} Wright, B.~S., Li, B.\ 2017.\ Type Ia supernovae, standardisable candles, and gravity.\ Phys. Rev. D 97, 083505 (2018); arXiv:1710.07018

\bibitem{Brennan2017} Brennan, T.~D., Carta, F., Vafa, C.\ 2017.\ The String Landscape, the Swampland, and the Missing Corner;\ arXiv:1711.00864

\bibitem{Obied2018} Obied, G., Ooguri, H., Spodyneiko, L., Vafa, C.\ 2018.\ De Sitter Space and the Swampland;\ arXiv:1806.08362

\bibitem{2018PhLB..784..271A} Agrawal, P., Obied, G., Steinhardt, P.~J., Vafa, C.\ 2018.\ On the cosmological implications of the string Swampland.\ Physics Letters B 784, 271-276; arXiv:1806.09718

\bibitem{Akrami2018} Akrami, Y., Kallosh, R., Linde, A., Vardanyan, V.\ 2018.\ The landscape, the swampland and the era of precision cosmology.\ ArXiv e-prints arXiv:1808.09440.




%
%
%
%
%
%
%
%
%
%




\end{thebibliography}
\end{document}